\documentclass[
aps,
pra,
showpacs,
amsmath,
amssymb,
preprint,
floatfix,
lengthcheck, 
superscriptaddress,
longbibliography
]{revtex4-1}

\usepackage{graphicx}
\usepackage{mathtools}
\usepackage{braket}
\usepackage{hyperref}
\usepackage{bbold}
\usepackage{calrsfs}
\usepackage{dsfont}
\usepackage{cancel}
\usepackage{mathrsfs}
\usepackage{multirow}
\usepackage{latexsym}
\usepackage[normalem]{ulem}
\usepackage{float}

\usepackage{color}

\DeclareMathAlphabet{\pazocal}{OMS}{zplm}{m}{n}

\usepackage{bm}

\begin{document}


\title{Down-conversion processes in ab-initio non-relativistic quantum electrodynamics}

\author{Davis M. Welakuh}
\email[Electronic address:\;]{davis.welakuh@mpsd.mpg.de}
\affiliation{Max Planck Institute for the Structure and Dynamics of Matter and Center for Free-Electron Laser Science \& Department of Physics, Luruper Chaussee 149, 22761 Hamburg, Germany}

\author{Michael Ruggenthaler}
\email[Electronic address:\;]{michael.ruggenthaler@mpsd.mpg.de}
\affiliation{Max Planck Institute for the Structure and Dynamics of Matter and Center for Free-Electron Laser Science \& Department of Physics, Luruper Chaussee 149, 22761 Hamburg, Germany}
\affiliation{The Hamburg Center for Ultrafast Imaging, Luruper Chaussee 149, 22761 Hamburg, Germany.}

\author{Mary-Leena M. Tchenkoue}
\email[Electronic address:\;]{mary-Leena.tchenkoue@mpsd.mpg.de}
\affiliation{Max Planck Institute for the Structure and Dynamics of Matter and Center for Free-Electron Laser Science \& Department of Physics, Luruper Chaussee 149, 22761 Hamburg, Germany}

\author{Heiko Appel}
\email[Electronic address:\;]{heiko.appel@mpsd.mpg.de}
\affiliation{Max Planck Institute for the Structure and Dynamics of Matter and Center for Free-Electron Laser Science \& Department of Physics, Luruper Chaussee 149, 22761 Hamburg, Germany}

\author{Angel Rubio}
\email[Electronic address:\;]{angel.rubio@mpsd.mpg.de}
\affiliation{Max Planck Institute for the Structure and Dynamics of Matter and Center for Free-Electron Laser Science \& Department of Physics, Luruper Chaussee 149, 22761 Hamburg, Germany}
\affiliation{Center for Computational Quantum Physics, Flatiron Institute, 162 5th Avenue, New York, NY 10010, USA}
\affiliation{The Hamburg Center for Ultrafast Imaging, Luruper Chaussee 149, 22761 Hamburg, Germany.}


\begin{abstract}
The availability of efficient photon sources with specific properties is important for quantum-technological applications. However, the realization of such photon sources is often challenging and hence alternative perspectives that suggest new means to enhance desired properties while suppressing detrimental processes are valuable. In this work we highlight that ab-initio simulations of coupled light-matter systems can provide such new avenues. We show for a simple model of a quantum ring that by treating light and matter on equal footing we can create and enhance novel pathways for down-conversion processes. By changing the matter subsystem as well as the photonic environment in experimentally feasible ways, we can engineer hybrid light-matter states that enhance at the same time the efficiency of the down-conversion process and the non-classicality of the created photons. Furthermore we show that this also leads to a faster down-conversion, potentially avoiding detrimental decoherence effects.    
 \end{abstract}


\maketitle


\section{Introduction}

Quantum-light sources exhibiting two-photon emission, especially entangled photon pairs, are crucial building blocks for quantum-information processing protocols~\cite{obrien2009,zheng2000}, cryptography~\cite{jennewein2000}, or teleportation~\cite{boschi1998}. As opposed to the characterization of single-photon sources, or "photon guns", which have sub-Poissonian statistics~\cite{stevenson2012}, the characterization of two-photon processes shows statistics that vary from non-classical (sub-Poissonian) to even chaotic (super-Poissonian) behavior~\cite{callsen2013}. Spontaneous parametric down-conversion or parametric down-conversion (PDC) is the coherent generation of a pair of photons with lower frequency (signal photons) by injecting a higher-frequency photonic field (pump photon) into a nonlinear medium~\cite{boyd1992}. Three-wave mixing has been mostly used for the generation of entangled photon pairs~\cite{ou1999,dousse2010,burillo2016,chang2016}. The need for on-demand efficient two-photon sources expanded PDC from using nonlinear crystals with picoseconds (ps) pulsed lasers~\cite{evans2010,bruno2014} to atom-like systems coupled to cavities~\cite{law1996,thompson2006}, photon-pair generation using the biexciton-exciton cascade in quantum dots~\cite{akopian2006,pathak2011,dousse2010,muller2014} and the nascent fields of polaritonic chemistry~\cite{juan2020} and circuit quantum electrodynamics (QED)~\cite{abdo2013,kamal2014}. The most widely used of these methods suffer from low photon emission rates and limited scalability. For example, an average of one in every $10^{12}$ photons is down-converted when using a bulk material~\cite{burnham1970}. Also, it was shown that entanglement between emitted photon pairs could be degraded due to the presence of energy-level splitting of the intermediate excitonic states in, e.g., quantum dots~\cite{akopian2006,stevenson2006}. Semiconductor quantum rings, where the charge carriers are confined in the radial direction, have been demonstrated to be excellent quantum emitters which generate single-photon states with strong antibunching under certain conditions~\cite{abbarchi2009}. Here the quantum confinement of the charge carriers together with modified ring geometries led to the observation of geometry-dependent photon antibunching. This makes quantum rings a suitable medium for photon generation since the electronic properties of the quantum ring can be modified by varying the geometric confinement parameters which in turn changes the optical spectrum~\cite{fomin2018}.

In order to investigate such down-conversion processes theoretically one usually assumes that light and matter can be separated and treated differently. The non-linear optics approach, for instance, uses non-linear response functions and susceptibilities of matter-only quantum mechanics to characterize such a process and connects them with a classical description of the light field~\cite{couteau2018,evans2010,bruno2014}. The quantum-optics approach, on the other hand, treats the light field quantized and couples the resulting photons to a simplified few-level description of the matter~\cite{burillo2016,chang2016,kockum2017}. In the latter approach one often even gets rid of the matter part altogether by defining effective photon-only Hamiltonians which model the photon-photon interaction due to the matter system~\cite{villas2003,villas2005,couteau2018}. 
In both approaches the efficiency and properties of the down-conversion process in question are usually determined by dipole-transition elements which depends on the symmetry the matter subsystem possesses. For instance, since PDC is a three-wave mixing process, the second-order non-linear susceptibility is the dominant contribution. For systems that possess a single symmetry (like the model considered below) this quantity is negligible~\cite{duque2012} resulting in an inefficient or even impossible PDC process. For this reason, one would conventionally break the symmetry of the quantum ring by some external classical field to engineer appropriate dipole-allowed transitions~\cite{rasanen2007,fomin2018} or use double-ring structures~\cite{abbarchi2009}. However, besides this more conventional perspective, we here highlight the possibilities that arise if we do not make the initial assumption to treat light and matter separately. Instead, by perform numerical simulations of non-relativistic QED, where light and matter are treated on equal quantized footing, we show how novel hybrid light-matter states (polaritons) are created that can act as intermediators for a photon down-conversion process. By making use of the full flexibility of the matter and the photon subsystems we can achieve a high-degree of control.

Our ab-initio simulations in the following are based on the Pauli-Fierz Hamiltonian of non-relativistic QED~\cite{spohn2004,ruggenthaler2017b,Jestaedt2020} in the long-wavelength limit~\cite{flick2017,rokaj2017,schaefer2018,schaefer2020}. As a simple yet illustrative example system we consider the aforementioned GaAs semiconductor quantum ring featuring a single effective electron coupled to a photonic environment. Varying the anharmonicity of the quantum ring we change the character of the matter subsystem continuously from harmonic to strongly anharmonic without breaking the radial symmetry. We, however, allow changes in the photonic environment, e.g., the geometry of a multi-mode cavity. This provides means to control the basic coupling parameters and can hence interpolate between free-space-like situations (weak coupling with strong dissipation) and ultra-strong coupling situations. Since the coupling between light and matter is fully quantized, also for the quantum ring (non-dipole allowed or virtual-state) down-conversion processes take place. We then investigate how specific properties and the efficiency of such a process can be optimized without relying on (experimentally challenging) specialized input fields but instead making use of the hybrid nature of coupled light-matter systems. For the case of degenerate (both output modes have the same frequency) down-conversion we see that by coupling the photons strongly to a virtual transition while optimizing the configuration of the coupled system, we can enhance the efficiency as well as the non-classicality of the created photons. Just increasing the intensity of the input field does not make the down-conversion process more efficient. Furthermore, increasing the coupling between light and matter shifts the down-conversion process to earlier times. This temporal control potentially allows to suppress the influence of dissipation in desired features like non-classicality and entanglement. All these results for a simple system highlight the possibilities that become available for designing novel photon sources if light and matter are treated on equal footing in an ab-initio description.

\section{Ab-initio description of down-conversion process}
~\label{sec:theory}

The starting point of our investigation is very general and we only make the assumption that our bound matter system is small compared to the wavelength of the relevant photon modes. In this case we  can consider the Pauli-Fierz Hamiltonian in dipole approximation~\cite{schaefer2020}
\begin{align}
\hat{H}_{\text{PF}} = \hat{H}_{el} - \frac{e}{m}\hat{\textbf{A}}\cdot\hat{\textbf{p}} + \frac{e^{2}}{2m} \hat{\textbf{A}}^{2}  + \sum_{\alpha=1}^{M} \hat{H}_{\alpha} . \label{coupled}
\end{align}
Here, $\hat{H}_{el}$ is the bare Hamiltonian of the matter subsystem in Coulomb gauge and $\hat{\textbf{p}}$ is the corresponding momentum. The charge and mass are given by $e$ and $m$, respectively. The operator $\hat{A} = \sum_{\alpha=1}^{M} \lambda_{\alpha} \textbf{e}_{\alpha} \hat{q}_{\alpha}$ is the vector potential with the dipole coupling strength $\lambda_{\alpha}$ determined by the value of the respective mode function (of frequency $\omega_{\alpha}$) at the position of the matter system (center of charge) and $\textbf{e}_{\alpha}$ the polarization vector~\cite{schaefer2020} of $M$ photon modes. Here, $\hat{q}_{\alpha} = \sqrt{\frac{\hbar}{2\omega_{\alpha}}}\left(\hat{a}_{\alpha} + \hat{a}_{\alpha}^{\dagger}\right)$ is the canonical photon coordinate that satisfies the commutation relation $\left[\hat{q}_{\alpha}, \hat{p}_{\alpha'}\right] = i \hbar \delta_{\alpha, \alpha'}$ and $\hat{a}_{\alpha}$, $\hat{a}_{\alpha}^{\dagger}$ are the usual annihilation and creation operators, respectively. The photon Hamiltonian is given by $\hat{H}_{\alpha} = \hbar \omega_{\alpha} (\hat{a}_{\alpha}^{\dagger} \hat{a}_{\alpha} + 1/2) = \frac{1}{2} \left(\hat{p}_{\alpha}^{2} + \omega_{\alpha}^{2}\hat{q}_{\alpha}^{2} \right)$. In the case that we keep many modes (up to a certain physical cutoff) to simulate the continuum, we need to be aware that this might make it necessary to use the bare mass of the charged particles instead of the already renormalized (physical) mass~\cite{spohn2004,flick2019,schaefer2018,rokaj2020}. Here, we make the usual assumption that only a small part of the photon continuum is changed with respect to the free-space case due to a cavity or photonic nanostructure, and we treat only these changes explicitly. The rest of the continuum of modes is subsumed in the usual renormalized (physical) mass of the charged particles~\cite{spohn2004}. Thus in practice, out of $M$-modes of the electromagnetic field, we just keep a few effective modes that are relevant. We note that this change of the photon modes also modifies the free-space Coulomb interaction~\cite{greiner1996,Jestaedt2020}. So care has to be taken when assuming that $\hat{H}_{el}$ is just due to a free-space matter system as described by the standard Schr\"odinger equation once the interaction with the photon field is taken into account explicitly. These cautionary remarks translate equivalently to the usual few-level models, where $\hat{H}_{el}$ is replaced typically by a few states from free-space calculations, or when making the mean-field assumption for the coupling between light and matter which leads to the Maxwell-Schr\"odinger equation of non-linear optics~\cite{ruggenthaler2017b}.

\subsection{Semiconductor quantum ring in multi-mode photonic environment}

\begin{figure}[t] 
\centerline{\includegraphics[width=0.5\textwidth]{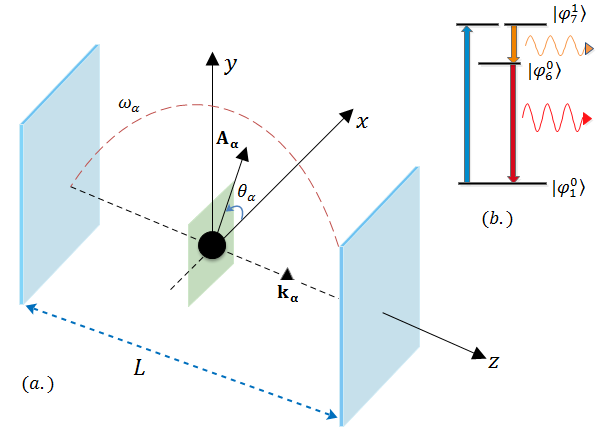}}
\caption{(a.) Setup for investigating down-conversion processes featuring a 2D GaAs quantum ring at the center of a multi-mode photonic environment. The fields are linearly polarized in the $x,y-$ plane where the effective quantum ring particle is trapped in a binding potential. The fields $\hat{\textbf{A}}_{\alpha}$ are polarized perpendicular to the propagation vectors $\textbf{k}_{\alpha}$. (b.) Simplified energy-level scheme indicating the relevant transitions in a non-degenerate down-conversion process. The simulations take the full level structure into account (see App.~\ref{app:quantum-ring} for details)}
\label{fig:setup-spdc}
\end{figure}

For investigating the down-conversion process we need to make a specific choice for the matter system and the photonic environment. We choose for the matter subsystem that mediates the down-conversion process a two-dimensional (2D) semiconductor GaAs quantum ring~\cite{rasanen2007,flick2015}. We describe the quantum ring by a single effective electron confined in a 2D mexican-hat potential (see Fig.~\ref{fig:potential} in App.~\ref{app:quantum-ring}). A more realistic description with many interacting electrons and possibly even phononic excitations would be possible if we employ more efficient first-principle methods such as quantum-electrodynamical density-functional theory (QEDFT)~\cite{ruggenthaler2014, pellegrini2015,flick2017c,Jestaedt2020} or polaritonic coupled cluster theory~\cite{mordovina2020polaritonic,haugland2020coupled}. Nevertheless, the current level of description already suffices to demonstrate the novel pathways for down-conversion that become accessible with an ab-initio light-matter description. Moreover, by changing the anharmonicity of the GaAs quantum ring we obtain control over the electronic level structure. We can use this control, which is also experimentally realizable~\cite{fomin2018}, to optimize the down-conversion process without breaking the symmetry of the matter subsystem. Finally, due to the rotational symmetry~\cite{hartmann2019,vinasco2018,fomin2018}, a simple matter-only analysis would indicate that no (efficient) down-conversion takes place. Due to the rotational symmetry of the eigenstates, only transitions that change the angular momentum by one (see also a detailed discussion in App.~\ref{app:quantum-ring}) are dipole allowed, and hence a process as indicated in Fig.~\ref{fig:setup-spdc} b.) is not dipole allowed. This becomes evident also in the second-order non-linear susceptibility, which is negligible in this case~\cite{duque2012}. Yet, since we do not decouple light and matter, we will still find a certain probability that such a process takes place. The efficiency and the details of the ensuing process will then depend crucially on the details of the photonic environment.

For the photonic subsystem we consider a multi-mode environment. To be precise, we assume that we can manipulate three of these modes, i.e., $\hat{\textbf{A}}_{1}$, $\hat{\textbf{A}}_{2}$ and $\hat{\textbf{A}}_{3}$, at will, while the rest of the continuum of modes stays largely unaffected. We consider $\hat{\textbf{A}}_{1}$ as the input (pump) mode with frequency $\omega_1$, polarization direction $\textbf{e}_{1}$ and coupling strength $\lambda_1$, and $\hat{\textbf{A}}_{2}$ and $\hat{\textbf{A}}_{3}$ as the output (signal) modes with corresponding frequencies, polarizations and coupling strengths.  

We arrange the matter and the photon system as depicted in Fig.~\ref{fig:setup-spdc}, such that only the $x$ and $y$ polarization directions are relevant and couple to the quantum ring. We then choose a coordinate system such that $\hat{\textbf{A}}_{1} = \hat{A}_{1}\textbf{e}_{x}$, $\hat{\textbf{A}}_{2} = \hat{A}_{2}(-\sin\theta_{2}\textbf{e}_{x} + \cos\theta_{2}\textbf{e}_{y} )$ and $\hat{\textbf{A}}_{3} = \hat{A}_{3}(\sin\theta_{3}\textbf{e}_{x} + \cos\theta_{3}\textbf{e}_{y} )$, where $\theta_{2}$ and $\theta_{3}$ are the angles of the respective polarization vectors. With this sketch of a possible setup we can connect the change in coupling strength to a change in the length of the cavity $L$, i.e., $\lambda = \sqrt{2/\epsilon L }$ where the dielectric permittivity of the GaAs quantum ring is $\epsilon = 12.7\epsilon_{0}$. For simplicity we assume in the following that all three modes have the same coupling strength $\lambda$. The rest of the continuum of modes is assumed to be subsumed, on the one hand, in the effective mass of the quantum ring particle (in accordance to the case of no nanophotonic environment we take $m=0.067m_e$, see also App.~\ref{app:quantum-ring}) while we treat a finite number of modes that constitute the linewidth of the enhanced modes explicitly in order to account for dissipation and decoherence in the electron-photon coupled system~\cite{flick2019}. The Hamiltonian for this description is given in the following form
\begin{align}
\hat{H} = \hat{H}_{S} + \hat{H}_{B} + \hat{H}_{SB} \label{cspdc-hamiltonian-SB}
\end{align}
where the internal system Hamiltonian is
\begin{align}
\hat{H}_{S} &= \hat{H}_{el} + \hat{H}_{1} + \hat{H}_{2} + \hat{H}_{3} - \frac{e}{m}\hat{A}_{1}\hat{p}_{x} \label{cspdc-hamiltonian} \\
& \quad  - \frac{e}{m}\left[\hat{A}_{2}\left(-\hat{p}_{x}\sin(\theta_{2}) + \hat{p}_{y}\cos(\theta_{2})\right) \right. \nonumber \\
& \left. \qquad\qquad   + \hat{A}_{3}\left(\hat{p}_{x}\sin(\theta_{3}) + \hat{p}_{y}\cos(\theta_{3})\right)\right]  \nonumber\\
& \quad + \frac{e^{2}}{2m}\left[\hat{A}_{1}^{2} + \hat{A}_{2}^{2} + \hat{A}_{3}^{2}  -2\hat{A}_{2}\hat{A}_{1}\sin(\theta_{2})  \right. \nonumber\\ 
& \left. \qquad\qquad\quad - 2\hat{A}_{3}\hat{A}_{1} \sin(\theta_{3}) + 2\hat{A}_{3}\hat{A}_{2}\cos(\theta_{2}+\theta_{3})\right]\nonumber  
\end{align}
and the bath and system-bath coupling that constitute the rest of the $(M-3)$ modes in the full Hamiltonian is  
\begin{align}
\hat{H}_{B} &=\!\!\sum_{\alpha=4}^{M} \! \hbar\omega_{\alpha} \left(\hat{a}_{\alpha}^{\dagger} \hat{a}_{\alpha} + \frac{1}{2}\right) , \nonumber \\
\hat{H}_{SB} &=\!\!\sum_{\alpha=4}^{M} \! \left[\! - \!\frac{e}{m}\hat{\textbf{A}}_{\alpha} \! \cdot \!\hat{\textbf{p}} \!+ \! \frac{e^{2}}{2m}\!\left(\!2 \hat{\textbf{A}}_{1} \!+ \! 2\hat{\textbf{A}}_{2} \!+ \! 2\hat{\textbf{A}}_{3} \! + \!\sum_{\beta=4}^{M}\!\! \hat{\textbf{A}}_{\beta} \! \right)\! \!\! \cdot\! \hat{\textbf{A}}_{\alpha} \right], \nonumber 
\end{align}
where $\hat{\textbf{p}} = \hat{p}_{x}\textbf{e}_{x} + \hat{p}_{y} \textbf{e}_{y}$. For our further considerations the last two terms of Eq.~\eqref{cspdc-hamiltonian-SB} constitutes the active photonic bath and its coupling to the system. For the photonic bath, we consider $(M-3)=M_{70}=70$ bath modes that are treated in photon Fock number states together with the three relevant modes. The vector potential of the bath modes is $\hat{\textbf{A}}_{\alpha} = \lambda_{\alpha}' \textbf{e}_{\alpha} \sqrt{\frac{\hbar}{2\omega_{\alpha}}}\left(\hat{a}_{\alpha} + \hat{a}_{\alpha}^{\dagger}\right)$ where $\lambda_{\alpha}'$ is the coupling of the bath modes.

\subsection{Time-evolution of the coupled system}
~\label{subsec:evolution}

To investigate the down-conversion process in detail we perform the time evolution of different initial states $|\Psi_{in} (0)\rangle$ of the coupled matter-photon system. To do so we explicitly propagate the time-dependent Schr\"{o}dinger equation $i\hbar \frac{\partial}{\partial t} |\Psi (t)\rangle = \hat{H}|\Psi (t)\rangle$ with the Hamiltonian of Eq.~\eqref{cspdc-hamiltonian-SB}. The initial states that we consider in the following are factorizable product states of the form $|\Psi_{in} (0)\rangle = |\varphi_{0}^{1}\rangle|\phi_{1}\rangle|0_{2}\rangle |0_{3}\rangle \dots | 0_{M}\rangle$, where $|\varphi_{0}^{1}\rangle$ is the ground state of the uncoupled quantum ring and $|0_{\alpha}\rangle$ is the zero-photon state of mode $\alpha$. The pump mode $\alpha=1$ will take different initial states. The simplest choice is that $\ket{\phi_1} = \ket{1_1}$ is just a single-photon Fock state. In most cases we will consider $|\phi_{1}\rangle=|\xi_{1}\rangle= e^{-|\xi_{1}|^{2}/2} \sum_{n_{1}=0}^{\infty}  \left(\xi_{1}^{n_{1}}/\sqrt{n_{1}!}\right)|n_{1}\rangle$ where $\xi_{1}$ is the amplitude and $|n_{1}\rangle$ the Fock states of mode 1. This implies that we have on average $|\xi_1 |^2$ photons at the beginning in the input mode. If we increase $|\xi|^2 \gg 1$ we approach a classical laser field (see App.~\ref{app:external-pump} for details). We note that the here chosen factorizable initial state is not special, and we find a similar behavior also with different choices. This is discussed in more detail in Sec.~\ref{sec:methods}.


We solve the time-dependent Schr\"odinger equation of the coupled light-matter system with a Lanczos propagation scheme~\cite{hochbruck1997}. We represent the matter Hamiltonian on a two-dimensional uniform real-space grid of $N_{x}=N_{y}=127$ grid points (implying $127^2$ states are taken into account) with spacing $\Delta x = \Delta y =0.7052$ nm while applying an eighth-order finite-difference scheme for the momentum operator and Laplacian. At several points (for comparison or for numerical efficiency) we instead of the real-space grid represent the matter subsystem by its truncated uncoupled eigenstate basis. This amounts for only a few states to the usual few-level approximation (see App.~\ref{app:few-levels} for further details). The photon modes are represented in a basis of Fock number states as discussed in App.~\ref{app:numerical-details} for the different input fields and descriptions. In the combined electron-photon space, we explicitly construct matrix representations for all operators. The expectation value for observables are computed for a time step of $\Delta t = 0.029$ fs of the time-evolved wave function.

\subsection{Characterization of down-converted photons}
~\label{subsec:characterization}

To characterize the down-converted photons, we compute the mean photon occupation $n_{\alpha} = \langle \hat{a}_{\alpha}^{\dagger}\hat{a}_{\alpha}\rangle$ for the pump ($\alpha=1$) and down-converted signal modes ($\alpha=2,3$). The photon occupation $n_{1} = \langle \hat{a}_{1}^{\dagger}\hat{a}_{1}\rangle$ is computed to contrast the amount of photonic occupation in the down-converted mode $n_{2} = \langle \hat{a}_{2}^{\dagger}\hat{a}_{2}\rangle$ and mode $n_{3} = \langle \hat{a}_{3}^{\dagger}\hat{a}_{3}\rangle$. Furthermore we consider the population of the single-, two- and three-photon Fock state of the relevant modes, i.e., $|\braket{1_{\alpha}|\Psi(t)}|^2$, $|\braket{2_{\alpha}|\Psi(t)}|^2$ and $|\braket{3_{\alpha}|\Psi(t)}|^2$ for $\alpha = 1,2,3$ of the down-conversion. This will allow us in simple cases to identify the standard PDC process. 

Of importance is the character of the down-converted photons, which can be determined by computing the Mandel $Q_{\alpha}$ parameter~\cite{mandel1979} defined as
\begin{align}
Q_{\alpha} =  \frac{\langle \hat{a}_{\alpha}^{\dagger}\hat{a}_{\alpha}^{\dagger}\hat{a}_{\alpha}\hat{a}_{\alpha}\rangle - \langle \hat{a}_{\alpha}^{\dagger}\hat{a}_{\alpha}\rangle^{2} }{\langle \hat{a}_{\alpha}^{\dagger}\hat{a}_{\alpha}\rangle} .  \label{mandel-q}
\end{align}
The Mandel $Q_{\alpha}$ measures the deviation of the photon statistics from a Poisson distribution and thus is a measure for the non-classicality. For a field with non-classical properties, the range of values lies between $-1 \leq Q_{\alpha}< 0$ which corresponds to sub-Poissonian statistics (anti-bunching behavior). Fields with super-Poissonian statistics (bunching behavior) have $Q_{\alpha} > 0$ and for a coherent state with Poissonian statistics we have $Q_{\alpha} = 0$ \cite{loudon2000}. 

In addition, we compute the second-order cross-correlation function (intensity correlations) of the photon field as a measure of correlation between the modes. The multi-mode intensity correlation function is defined by
\begin{align}
g_{\alpha\beta}^{(2)} =  \frac{\langle \hat{a}_{\alpha}^{\dagger}\hat{a}_{\alpha}\hat{a}_{\beta}^{\dagger}\hat{a}_{\beta}\rangle  }{\langle \hat{a}_{\alpha}^{\dagger}\hat{a}_{\alpha}\rangle\langle \hat{a}_{\beta}^{\dagger}\hat{a}_{\beta}\rangle} . \label{g-2-coherence}
\end{align}
The correlation function takes values greater than one for correlated modes. For uncorrelated modes, it is equal to one and it takes values smaller than one if the modes are anti-correlated~\cite{loudon2000,gerry2005,kalaga2016}. 

Also, we compute the purity $\gamma_{\alpha}$ with $\alpha=1,2,3$ as a measure for entanglement. The purity is obtained by tracing over the square of the one-body reduced density matrix of the respective subsystem (see App.~\ref{app:density-matrix} for details). If the purity is equal to one, the system can be expressed as a factorizable state of the subsystem (modes $\alpha=1,2,3$) and the rest of the coupled matter-photon system. If the purity is smaller than one, we have a non-factorizable state, which indicates correlation between the subsystems.

\section{Single input-photon down-conversion and temporal control}
~\label{sec:fock-initial-state}

We start by considering the case were the input mode 1 is occupied by just one photon, the coupling between light and matter is weak such that the usual non-linear optics considerations apply, and we have a strongly anharmonic quantum ring such that a few-level approximation is reasonable as well. We therefore choose $\ket{\phi_1} = \ket{1_1}$, fix a small coupling strength of $\lambda = 0.014$ and use an anharmonicity of $V_0 = 200$ meV (see App.~\ref{app:quantum-ring} for details on the quantum ring). By selecting the frequency of mode 1 in resonance with the dipole-allowed transition between the ground- and eleventh-excited state ($\ket{\varphi_1^0} \leftrightarrow |\varphi_{7}^{1}\rangle$) of the bare quantum ring, which is $\hbar\omega_{1} =24.65$ meV, and by choosing the signal mode 2 with energy $\hbar\omega_{2} = 1.36$ meV which is resonant with the tenth- and eleventh-excited states (($|\varphi_{7}^{1}\rangle\leftrightarrow|\varphi_{6}^{0}\rangle$) as well as signal mode 3 with energy $\hbar\omega_{3} = 23.29$ meV resonant with the ground- and tenth-excited state ($|\varphi_{6}^{0}\rangle\leftrightarrow|\varphi_{1}^{0}\rangle$), a few-level picture is appropriate (see also Fig.~\ref{fig:setup-spdc} b). We will, however, not a priori restrict the number of electronic states involved. That means we will use all electronic states or else have checked for convergence with respect to the number of states in all considered observables. Only in Sec.~\ref{subsec:few-levels} we make a comparison with an a priori restriction to only a few levels as well as with a semi-classical treatment, which shows how certain observables are not well captured. We further note that the superindex of the matter-only eigenstates refers to their angular momentum. Thus the transition in resonance with mode 3 is not dipole allowed, since only states that differ by exactly one in their angular quantum number have a non-zero dipole transition element (see App.~\ref{app:quantum-ring} for details). Due to treating light and matter fully coupled we will, however, still find a down-conversion process in our ab-initio simulation.

Finally, we need to fix the polarization directions of the input and output modes and also define the bath modes. In order to maximize the photon-pair generation, we choose the mixing angles $\theta_{2}=\theta_{3}=90^{\circ}$ such that both fields of the signal modes are horizontally polarized as $\hat{\textbf{A}}_{2}=-\hat{A}_{2}\textbf{e}_{x}$ and $\hat{\textbf{A}}_{3}=\hat{A}_{3}\textbf{e}_{x}$. This choice results in maximization of the interference terms of Eq.~(\ref{cspdc-hamiltonian}) since the sines and cosine of the mixing angles become one (see the detailed investigation in Sec.~\ref{subsec:polarization}). With these choices the down-conversion process obeys the energy and momentum conservation $\hbar\omega_{1} = \hbar\omega_{2} + \hbar\omega_{3}$ and $\hbar\textbf{k}_{1} = \hbar\textbf{k}_{2} + \hbar\textbf{k}_{3}$, respectively. The polarizations of the $M_{70}$ bath modes are chosen equally aligned as the modes 1,2,3 (since else the bath modes would just couple more weakly). We sample equally-spaced energy ranges around the main modes of $\hbar\omega_{1}$, $\hbar\omega_{2}$, and $\hbar\omega_{3}$. For $\hbar\omega_{2}$ the energy range is $\hbar\omega_{B_{2}} = [0.113,4.521]$ meV, which is sampled with 20 bath modes, and for $\hbar\omega_{2}$ and $ \hbar\omega_{3}$, 50 bath modes are employed to sample for the combined energy range $\hbar\omega_{B_{13}} = [11.303,27.128]$ meV with equal spacing $\hbar\Delta \omega=0.25$ meV. The coupling strength of the bath modes is chosen to be $\lambda' = 0.007$. For the combined system-bath Hamiltonian of Eq.~(\ref{cspdc-hamiltonian-SB}), we treat the $M_{70}$ and three relevant modes by including three photon Fock states (vacuum, one-photon and two-photon states) in each of these sampled modes (see App.~\ref{app:numerical-details} for details). In the following, we present the time evolution for the combined system and bath (i.e. Eq.~(\ref{cspdc-hamiltonian-SB})) and system only (i.e. Eq.~(\ref{cspdc-hamiltonian})). For the simulation including the bath, we compute observables only for the relevant three modes while the bath modes serves just as dissipation and decoherence channels. 

\subsection{Dissipation and coherence time}
~\label{sec:dissipation-and-coherence}

\begin{figure}[bth]
\centerline{\includegraphics[width=0.5\textwidth]{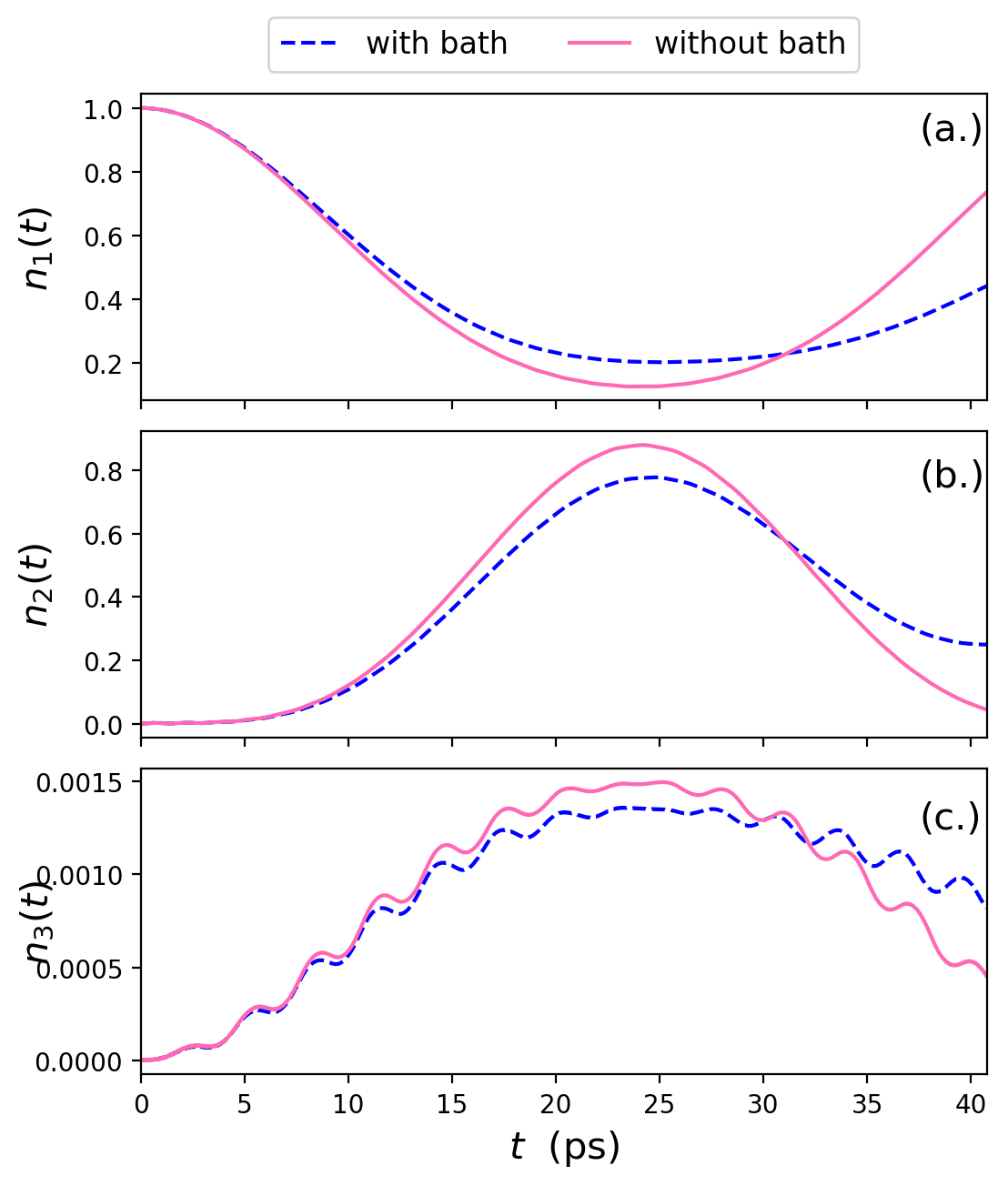}}
\caption{Real-time photon occupations of the pump mode (a.) and signal modes (b. and c.) for weak coupling $\lambda=0.014$ and bath coupling $\lambda'=0.007$. Qualitatively the coherent simulation without the bath modes (solid pink line) agrees with the simulation including the bath modes (dashed blue line) and only towards the end of the simulation, i.e., the end of the coherence time, the differences become large.}
\label{fig:with-without-bath-ns}
\end{figure}

We first focus on the influence of the bath modes on the down-conversion by performing two different simulations: one with the bath modes (i.e. time-propagate Eq.~(\ref{cspdc-hamiltonian-SB})) and one without the bath modes (i.e. time-propagate only Eq.~(\ref{cspdc-hamiltonian})). In both cases, since the effective coupling strength $g_{\alpha}=\lambda\sqrt{\hbar/2\omega_{\alpha}}$ of the transition $|\varphi_{7}^{1}\rangle\leftrightarrow|\varphi_{1}^{0}\rangle$ is small when compared to the coupling of the other main transitions (see discussion in Sec.~\ref{sec:input-fock-state} for further details), the effective electron preferably relaxes by cascaded emission into modes 2 and 3. This is inferred as the coupling strength is proportional to the square root of the spontaneous decay rate (see App.~(\ref{app:few-levels})). We therefore expect that mode 2 and 3 will become strongly populated once the initial photon in the pump mode 1 interacts with the matter subsystem. Thus in the ab-initio description of the PDC process the effective coupling strengths $g_\alpha$ become decisive and the bare dipole-transition elements are no longer the only important quantity to consider. And indeed we find in Fig.~\ref{fig:with-without-bath-ns}, where we show the mode occupations for the input and two signal modes, that this holds true for both situations. Qualitatively both simulations show the same behavior, with the main difference that the down-conversion from mode 1 to the signal modes is less effective when the bath modes are included and the maximum of the down-converted number of photons appears slightly later.
If we would simulate for longer times, the differences would become more pronounced and (provided we do not wait too long such that we hit the unphysical revival time due to having only finitely many modes~\cite{PhysRev.107.337}) the photon numbers would relax to the ones of the coupled matter-photon ground state. For the weak-coupling case this number is effectively zero. We would therefore consider the photons emitted. Since we cannot simulate the full emission process in this numerical setup, we instead make the simple assumption that what appears as the maximum amount of photons in each of the signal modes corresponds to what would be detected outside of the system. And since the bath modes do not seem to change these numbers strongly, even for this weakly coupled case where the bath contributes considerably, the bath-free simulation is a good approximation as long as we do not go beyond the coherence time of roughly 40 picoseconds (ps). By coherence time we mean the time interval in which the bath-free (fully coherent) simulation is a good approximation to the simulation including a bath. If we increased the coupling to the bath modes (stronger dissipation) the coherence time would be shorter. On the other hand, if we consider a stronger coupling to the input and signal modes, while keeping the bath modes fixed (essentially increasing the finesse of the cavity), the coherence times would be longer. It is not surprising that simulating 70 modes coupled to the electronic system is numerically very expensive and would not allow for all the different cases of down-conversion processes that we investigate in this work. In order to compare all these different cases as unbiased as possible we therefore consider them all with a bath-free simulation of a coherence time of about 40 ps. This is not a completely arbitrary number, since quantum rings are known to have long coherence times on the order of ps, before other dissipation channels destroy the coherence~\cite{fomin2018}. Thus this analysis allows us to avoid the use of (numerically expensive) non-unitary master-equation approaches to approximately treat the effect of the bath on the system~\cite{breuer2007} or to keep the bath explicitly in our following considerations.

\begin{figure}[bth]
\centerline{\includegraphics[width=0.5\textwidth]{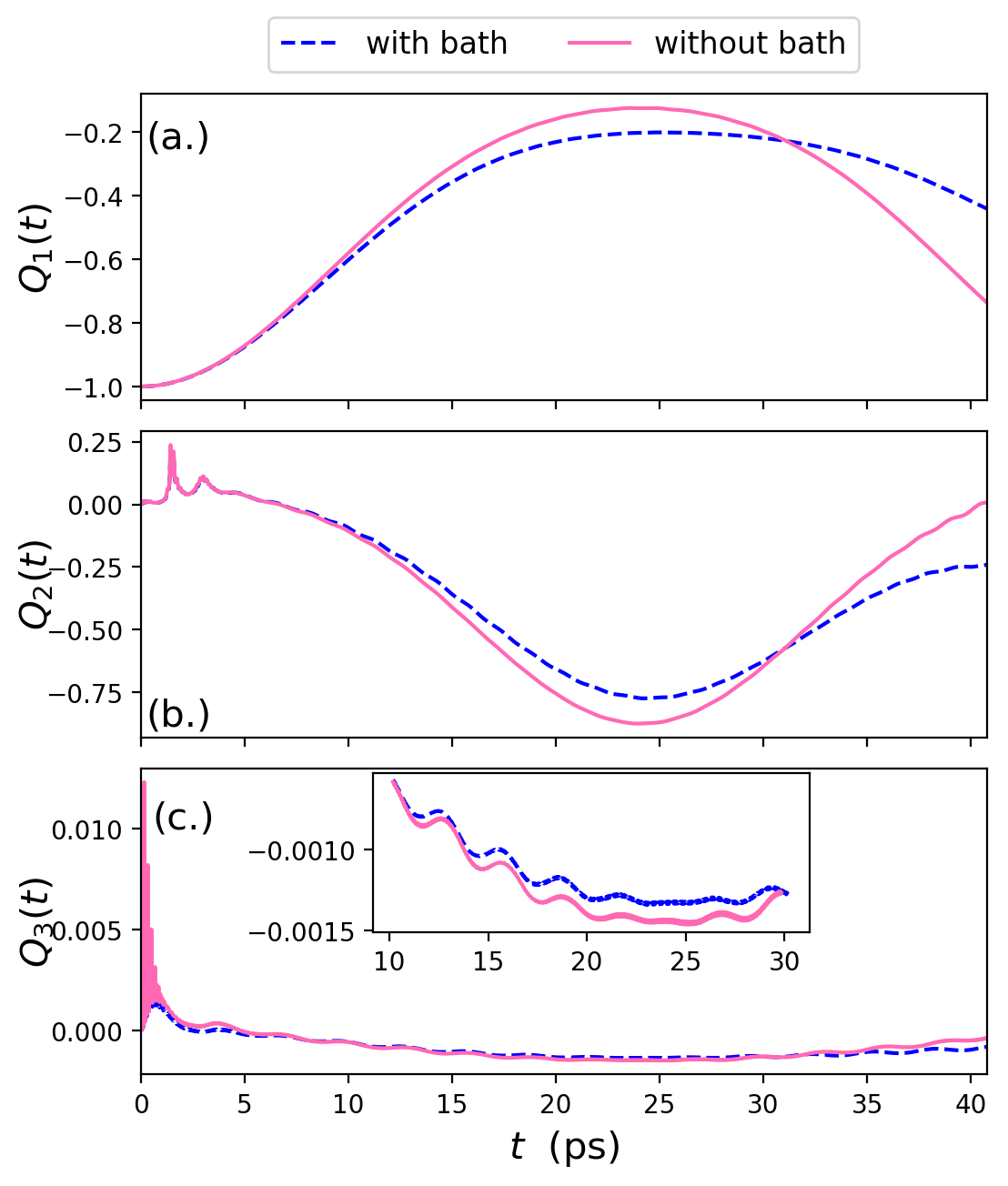}}
\caption{Real-time Mandel $Q$ parameter of the pump mode (a.) and signal modes (b. and c.) for weak coupling $\lambda=0.014$ and bath coupling $\lambda'=0.007$. Again the coherent simulation (solid line) agrees qualitatively with the simulation including the bath modes (dashed blue line).}
\label{fig:with-without-bath-Qs}
\end{figure}

Since we will consider also other quantities, let us check whether the coherent simulation faithfully reproduces these observables as well. The most important quantity in the following will be the Mandel $Q$ parameter to determine the statistics of the down converted photons. We find in Fig.~\ref{fig:with-without-bath-Qs} that again both simulations agree qualitatively and remain close over a long period of time.       
This demonstrates that the bath-free simulations capture at least qualitatively also the more complex properties of the created photon pair. Since we will be focused on an enhanced coupling to the main modes, the influence in these simulations of the bath modes will be even less important. The same holds true if we consider many input photons and classical external pumping.

\subsection{Temporal control of down-converted photons}
~\label{sec:input-fock-state}

\begin{figure}[t] 
	\centerline{\includegraphics[width=0.5\textwidth]{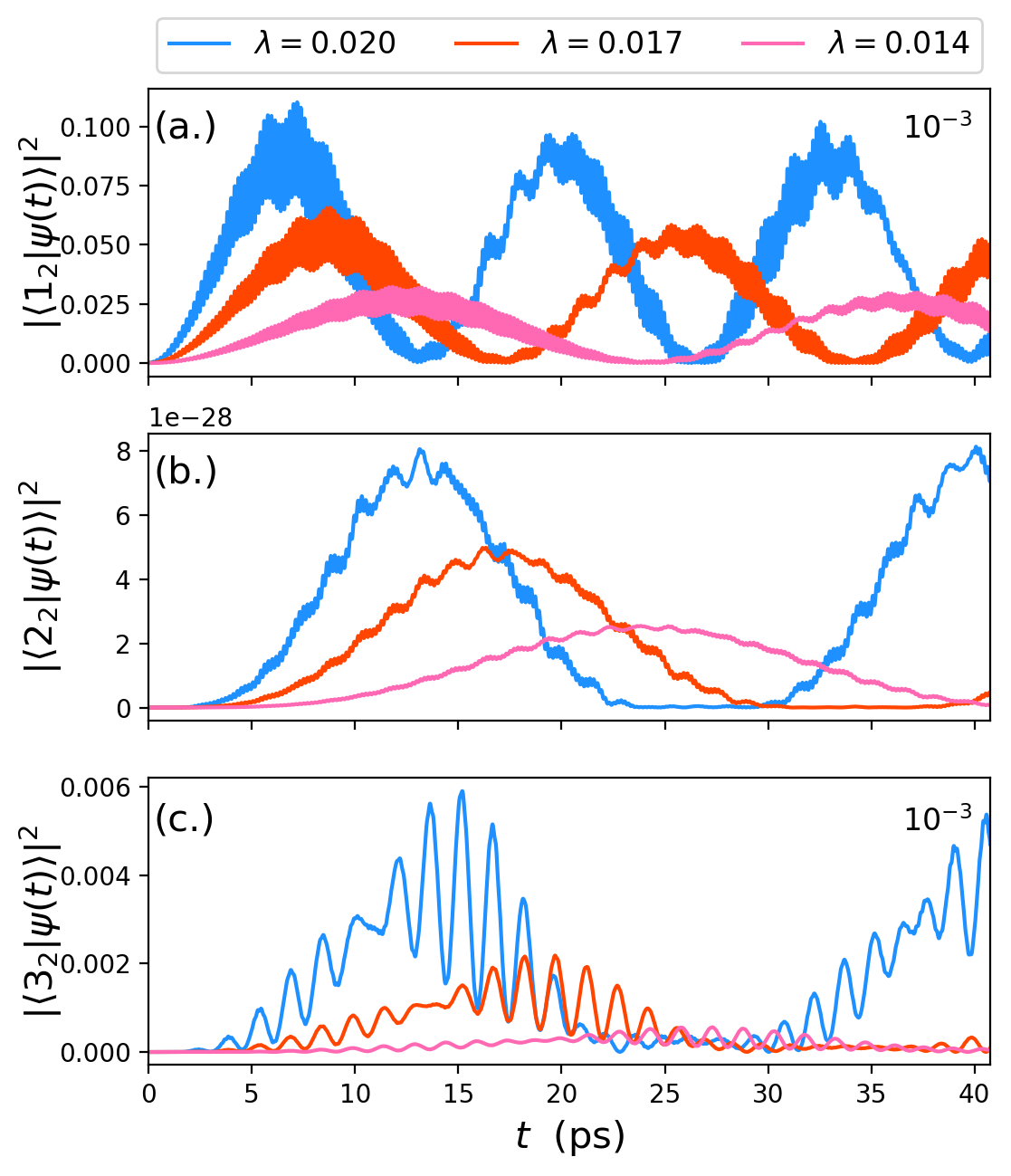}}
	\caption{Real-time Fock state populations of signal mode 2 from weak to ultra-strong coupling. Panels (a.), (b.), (c.) shows the one-, two- and three-photon Fock states, respectively. The one-photon Fock state dominates for a single photon in the input mode.}
	\label{fig:fock-occupation-mode-2}
\end{figure}

\begin{figure}[t] 
	\centerline{\includegraphics[width=0.5\textwidth]{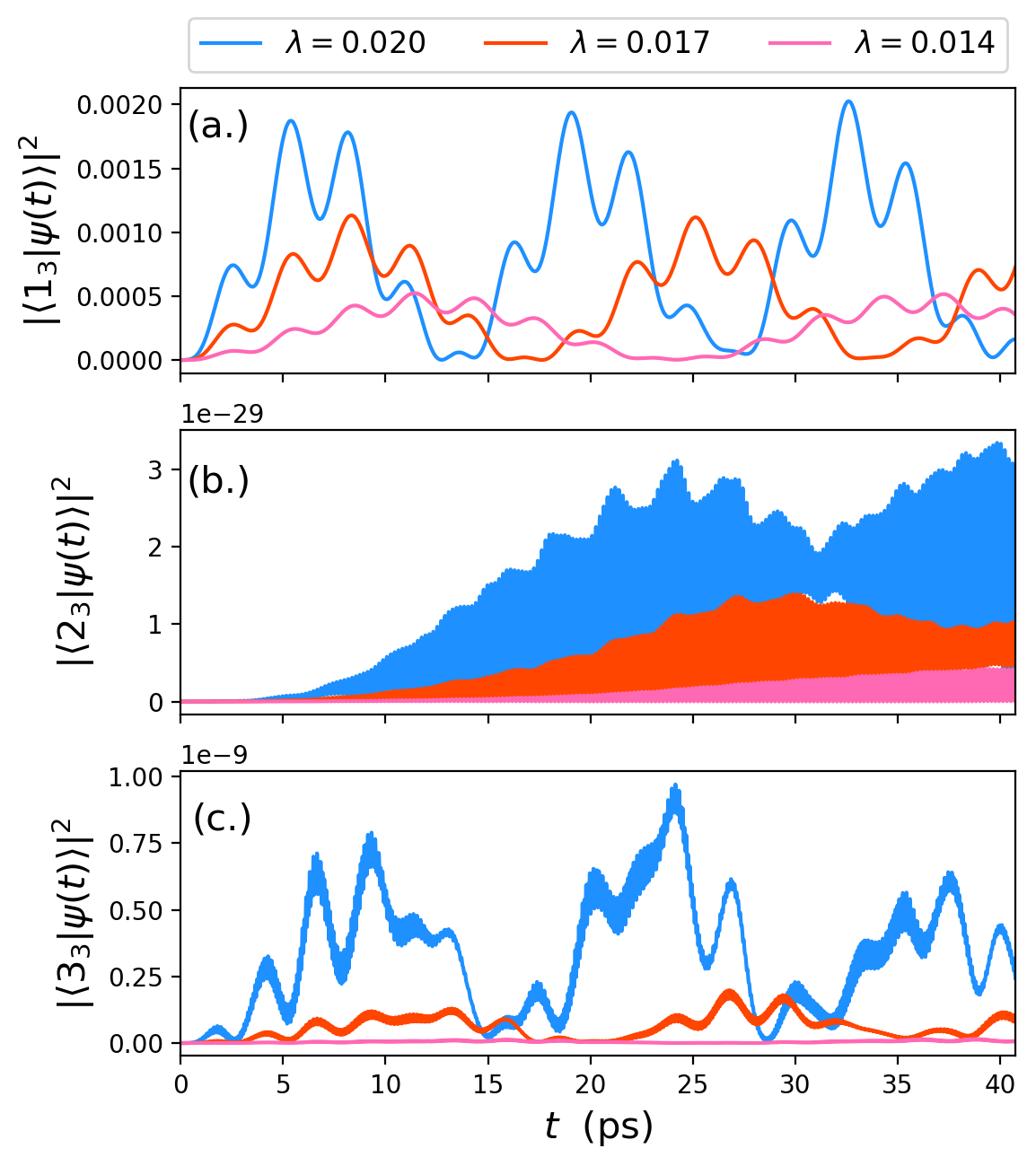}}
	\caption{Real-time Fock state populations of signal mode 3 from weak to ultra-strong coupling. Panels (a.), (b.), (c.) shows the one-, two- and three-photon Fock states, respectively. The one-photon Fock state dominates for a single photon in the input mode also in this case.}
	\label{fig:fock-occupation-mode-3}
\end{figure}

\begin{figure}[t] 
\centerline{\includegraphics[width=0.5\textwidth]{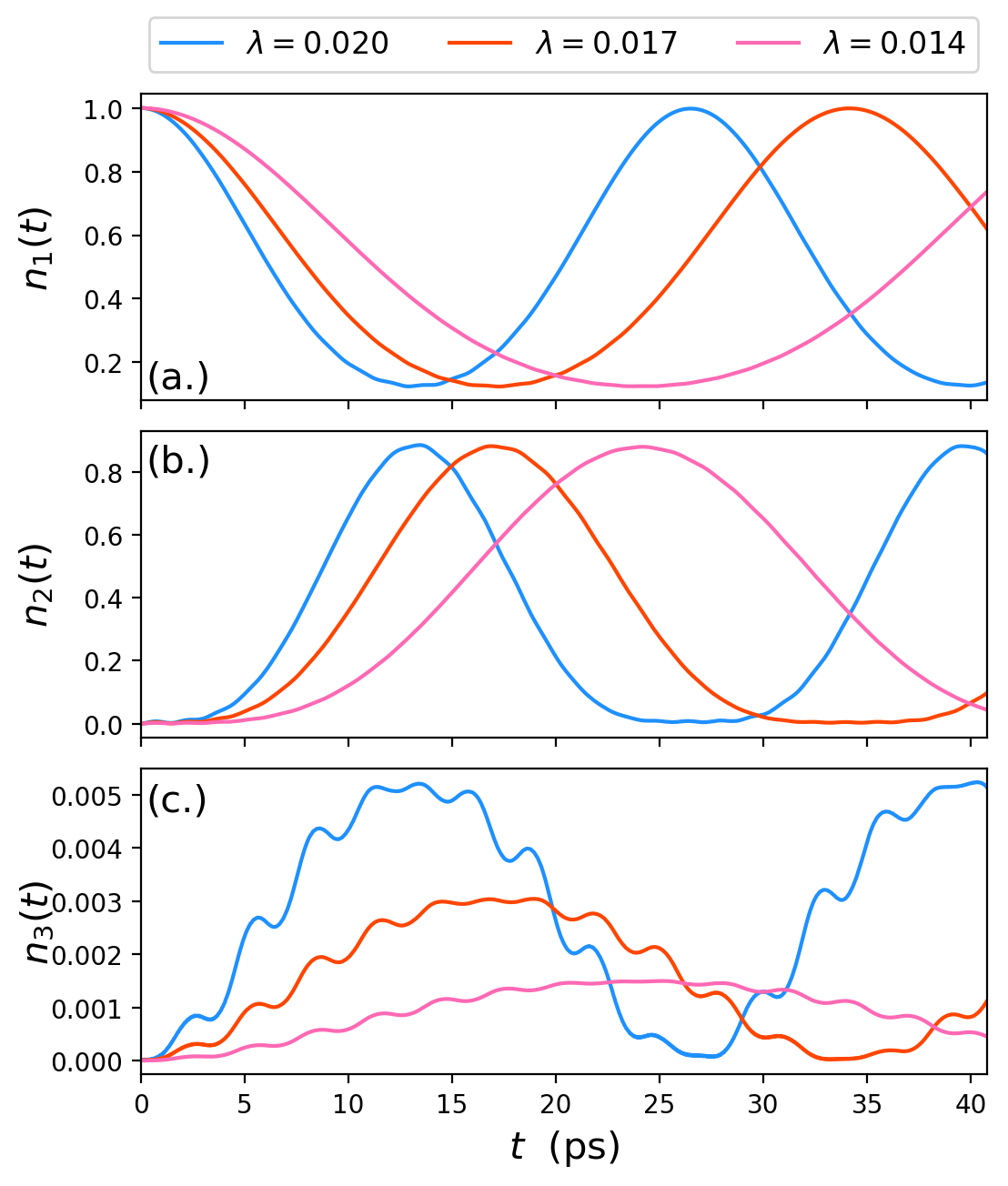}}
\caption{Real-time photon occupations of the input mode $n_1(t)$ initially in a single-photon Fock state, and the photon occupations of the down-converted signal photons $n_2(t)$ and $n_3(t)$ from weak (pink solid line) to ultra-strong coupling (blue line).}
\label{fig:fock-occupations}
\end{figure}

While we have above shown the mode occupations and we see a drop in the input mode and an increase in the signal modes, this does not imply that we have indeed converted one photon of $\omega_1$ into a photon with $\omega_2$ and one with $\omega_3$. For this we need to consider the populations of the different Fock number states in each mode. While in mode 1 we know by construction that we have only one photon at $t=0$ and zero in the other modes, it could easily be that we also populate 2 and 3 photon states in the down-conversion process. This would imply also higher-order process play a role and we could not identify it in a simple manner as a two-photon down-conversion process. However, in Figs.~\ref{fig:fock-occupation-mode-2} and \ref{fig:fock-occupation-mode-3} and focusing on the pink solid lines, we find that our intuition about the PDC process was correct. Only the one-photon states have significant population throughout the simulation. 
If we increase the coupling strength and go from the weak to the ultra-strong coupling domain it is not clear a priori that we still only have single-photon processes and no higher-order (multi-level/multi-photon) contributions become important. We vary the coupling $\lambda$ by varying the effective cavity length $L$ (see also Fig.~\ref{fig:setup-spdc}). This also leads to modified effective coupling strengths $g_{\alpha}=\lambda\sqrt{\hbar/2\omega_{\alpha}}$ (see Tab.~\ref{tab:couplings} for detail values), while the bare dipole transition elements stay unchanged and are given in scaled effective atomic units in App.~(\ref{app:quantum-ring}). 
\begin{table}
\begin{center}
\begin{tabular}{ | c | c | c | c | }
	\hline
	Coupling      &   weak    &  strong     & ultra-strong  \\\hline \hline
	$L$ ($\mu$m)  &   100     &    50       &      30       \\\hline
	$\lambda$     &   0.014   &   0.020     &     0.026     \\\hline
	$g_{1}$       &  0.00675  &   0.00954   &   0.01232     \\\hline
	$g_{2}$       &  0.02875  &   0.04065   &   0.05249     \\\hline
	$g_{3}$       &  0.00694  &   0.00981   &   0.01267     \\\hline
	$g_{1}*g_{2}$ &  0.00019  &   0.00039   &   0.00065     \\\hline
	$g_{1}*g_{3}$ &  0.00005  &   0.00009   &   0.00016     \\\hline
	$g_{2}*g_{3}$ &  0.00020  &   0.00040   &   0.00067     \\\hline
\end{tabular}
\end{center}
\caption{Electron-photon coupling strengths by varying the cavity length/mode volume. The coupling strengths  $g_{\alpha} = \lambda\sqrt{\hbar/2\omega_{\alpha}}$ are different for coupling to different different modes of frequency $\omega_{\alpha}$. Decreasing the cavity length $L$ increases the coupling strengths $g_{\alpha}$ and their respective products.}
\label{tab:couplings}
\end{table}
From Figs.~\ref{fig:fock-occupation-mode-2} and \ref{fig:fock-occupation-mode-3} we see that also for the strong coupling calculations the PDC remains qualitatively similar to the weak coupling case. Yet if we look for the total amount of photons in each mode we find in Fig.~\ref{fig:fock-occupations}
that we can increase the amount of photons in the signal modes considerably. This demonstrate that the effective coupling strength becomes a decisive quantity for the efficiency of the process in an ab-initio description. It indicates how strongly light and matter mix and how strongly the matter-only eigenstates are modified and turn into hybrid light-matter states potentially beneficial for the PDC process. Indeed, as has been shown in several works~\cite{flick2015,schaefer2018}, the vacuum field of the modes can break the rotational symmetry and therefore $g_\alpha$ can lead to increased dipole transition elements. Although the dipole elements are no longer a good descriptor for the interaction between light and matter in this regime, they nevertheless provide an intuitive picture for the PDC process. And they also highlight that by engineering the vacuum of the electromagnetic field, the properties of the coupled system can be very different from the individual uncoupled systems. For example, changing the photonic environment by decreasing $L$ increases the coupling $g_\alpha$ even in the case of a weak dipole moment. Also, the photon-photon couplings $g_{1}*g_{2},g_{1}*g_{3},g_{2}*g_{3}$ between the modes increases with increasing couplings. We note that these terms arise due to the induced diamagnetic currents and are an often overlooked yet important contribution in many light-matter phenomena~\cite{rokaj2017,schaefer2020,rokaj2020}.        

\begin{figure}[t] 
	\centerline{\includegraphics[width=0.5\textwidth]{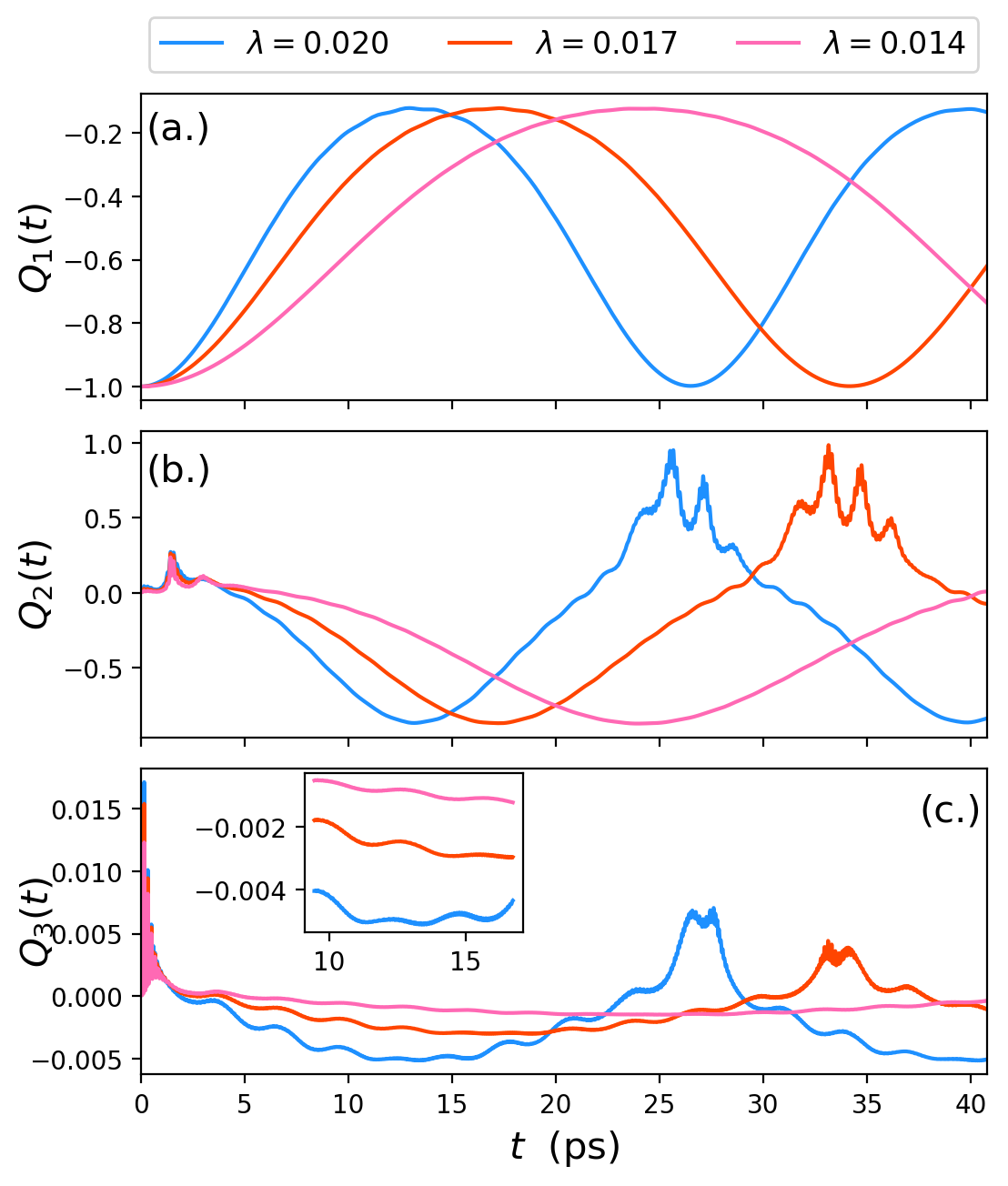}}
	\caption{Real-time photon statistics from weak to ultra-strong coupling for the input single-photon Fock state and down-converted signal photons. (a.) The pump mode $1$ has sub-Poissonian photon statistics for the entire evolution for all three coupling strengths. (b.) Strong anti-bunching as well as bunching features for different couplings of signal mode $2$. (c.) The mode 3 stays close to a coherent state throughout the PDC process. In all cases the coupling strength shifts the appearance of the different features to earlier times.}
	\label{fig:fock-mandels}
\end{figure}
\begin{figure}[t] 
	\centerline{\includegraphics[width=0.5\textwidth]{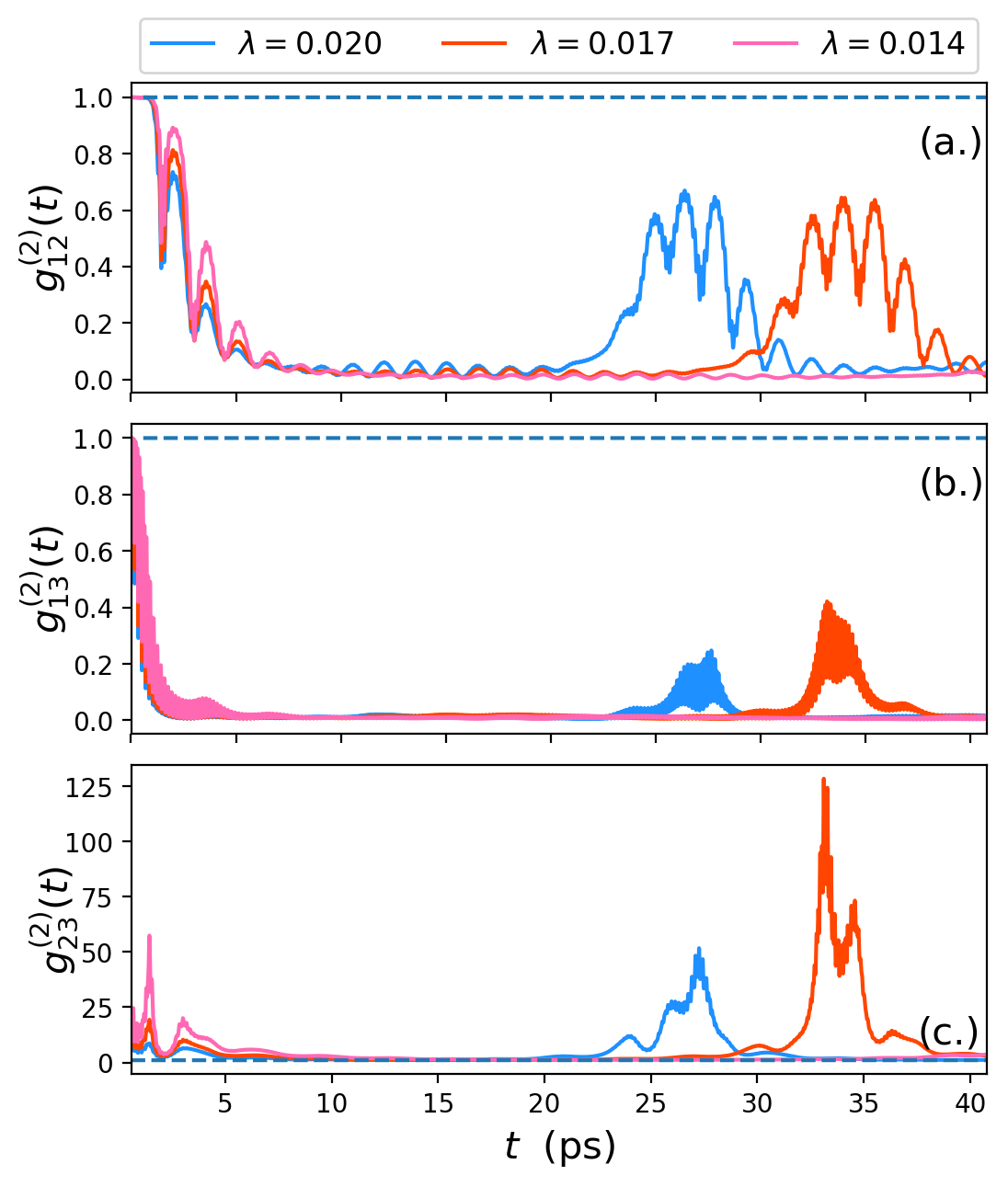}}
	\caption[]{Real-time cross-correlation between the pump and signal modes for the input single-photon Fock state. The dashed line indicates when the modes are correlated (above the dashed line) or anti-correlated (below the dashed line). In panels (a.) and (b.), the pump and signal modes 2 and 3 are anti-correlated. (c.)  The down-converted photons of modes 2 and 3 are correlated for the whole evolution.}
	\label{fig:fock-coherences}
\end{figure}

In the remainder of this section we focus on a different effect. In the Figs.~\ref{fig:fock-occupation-mode-2}, \ref{fig:fock-occupation-mode-3} and \ref{fig:fock-occupations} we consistently find that the stronger the coupling, the earlier the down-conversion of photons happens. Having the coherence time in mind, we find that we can potentially beat the undesired dissipative processes by strong and ultra-strong coupling. That is, by shifting the creation of the down-converted photons to earlier times, while the coupled system is still behaving coherently, we can hope for attaining also special, coherence-driven features of the PDC process such as non-classicality and entanglement of the photons. Considering the Mandel $Q$ parameter of the different photons (see Fig.~\ref{fig:fock-mandels}), we observe that the time of appearance of non-classical features can be controlled by the coupling strength as well.  
The overall shape of the time-evolution of the different Mandel $Q$ parameters stays relatively rigid. The same holds true for the intensity correlation functions (see Fig.~\ref{fig:fock-coherences}).
They nicely show how the photon in mode 1 is anti-correlated with respect to the signal modes, while the signal modes are strongly correlated.

While we do not model the emission of the down-converted photons from the photonic environment, we believe that controlling the timing of the creation process of the down-converted photons by varying the coupling strength will have a direct impact on the features of the emitted photons. At least for a simple description of an instantaneous emission, the features of the created signal photons are carried over to the emitted signal photons. It remains, however, unclear whether these features are only there for the special and in practice highly-demanding choice of a single photon in mode 1 and whether also the temporal control of these features is lost with another initial state or when the pump mode is replaced by a classical external pump field, i.e., a laser.

\section{Input mode in a coherent state}
~\label{sec:ab-initio}

The generation of a single photon in a specific mode is highly challenging, and due to the usually low efficiency of the PDC process it is also not easy to observe such a process. So in practice one has to increase the number of photons to observe any down-converted photons. That means we will have more than just one photon in mode 1 at the beginning. It is, however, not clear how this will change the main features observed for the case above. Let us therefore here consider how a change in the initial state influences the different observables.

\begin{figure}[t] 
\centerline{\includegraphics[width=0.5\textwidth]{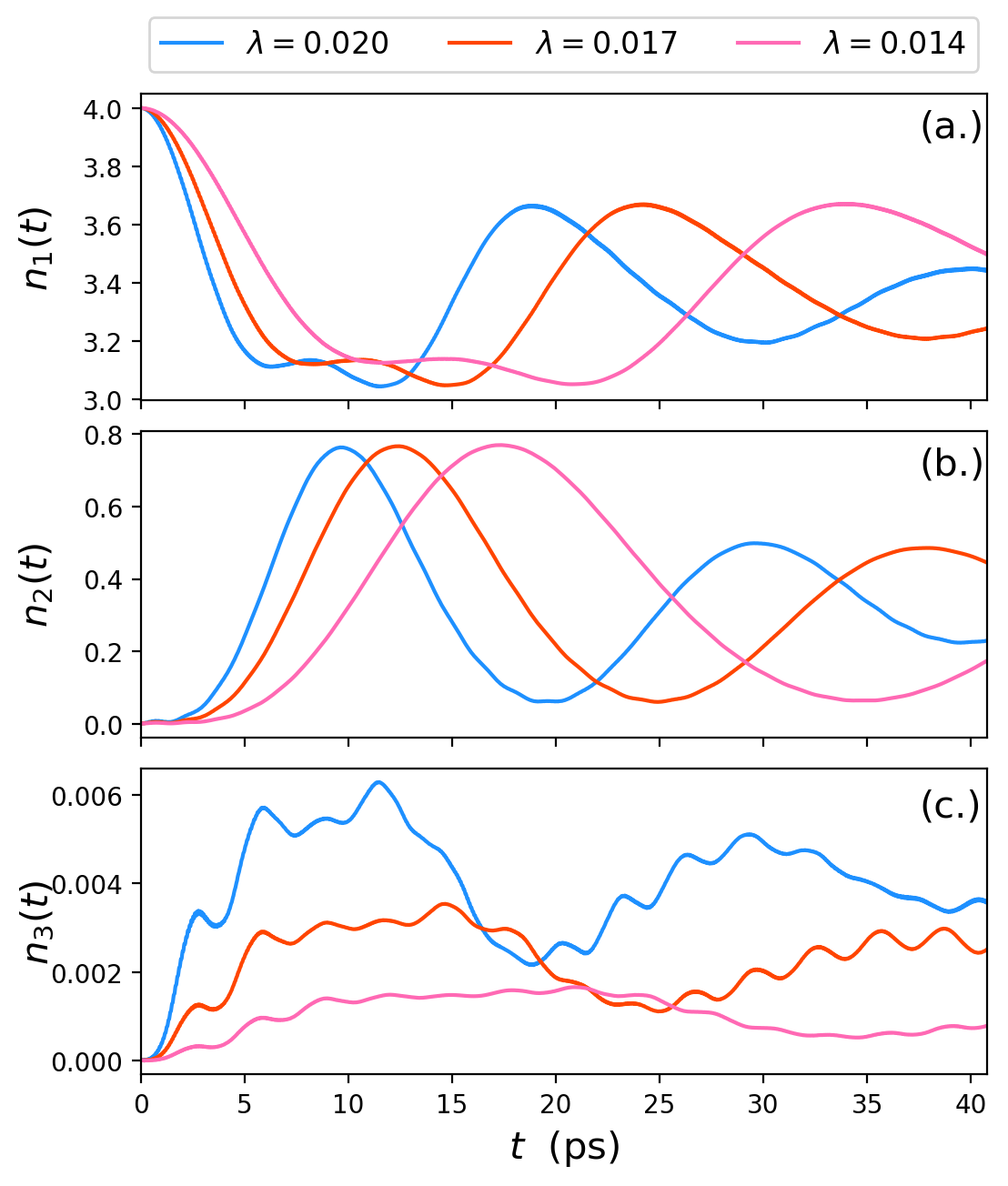}}
\caption{Real-time photon occupations of the input mode $n_1(t)$ initially in a coherent state, and the photon occupations of the down-converted signal photons $n_2(t)$ and $n_3(t)$ from weak to ultra-strong coupling.}
\label{fig:occupations}
\end{figure}

\begin{figure}[t] 
\centerline{\includegraphics[width=0.5\textwidth]{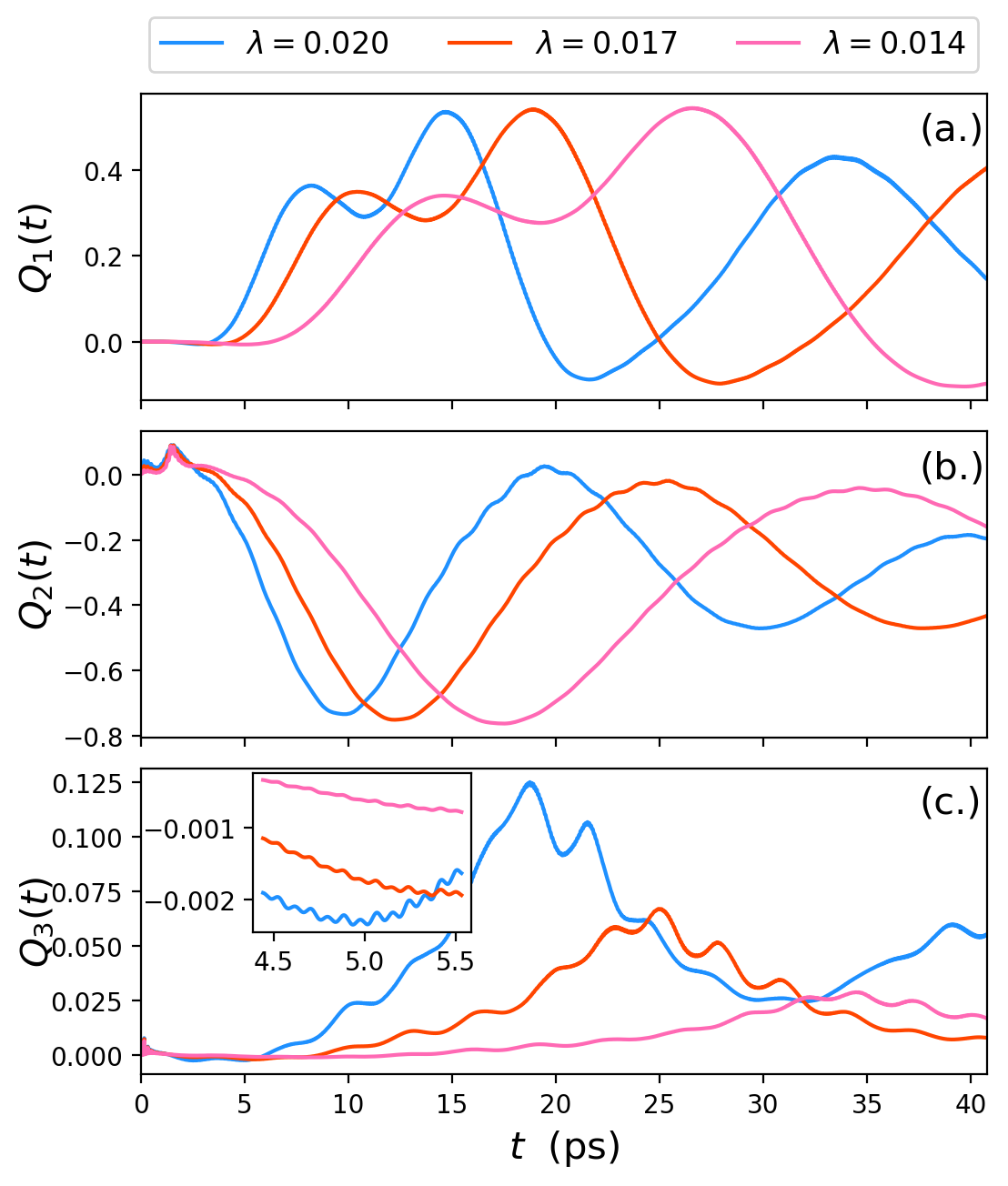}}
\caption{Real-time photon statistics from weak to ultra-strong coupling for an input coherent state and down-converted signal photons. (a.) The pump mode $1$ varies from fields with Poissonian, super-Poissonian and sub-Poissonain statistics. (b.) Strong anti-bunching feature for different couplings regimes of mode $2$. (c.) The emitted photon in mode 3 is non-classical for a brief time interval with maximum non-classicality of $Q_{3}=-0.0025$ at $t=4.93$ (ultra-strong coupling) ps as shown in the inset. In all cases the coupling strength shifts the appearance of the different features to earlier times.}
\label{fig:mandels}
\end{figure}

\begin{figure}[t] 
\centerline{\includegraphics[width=0.5\textwidth]{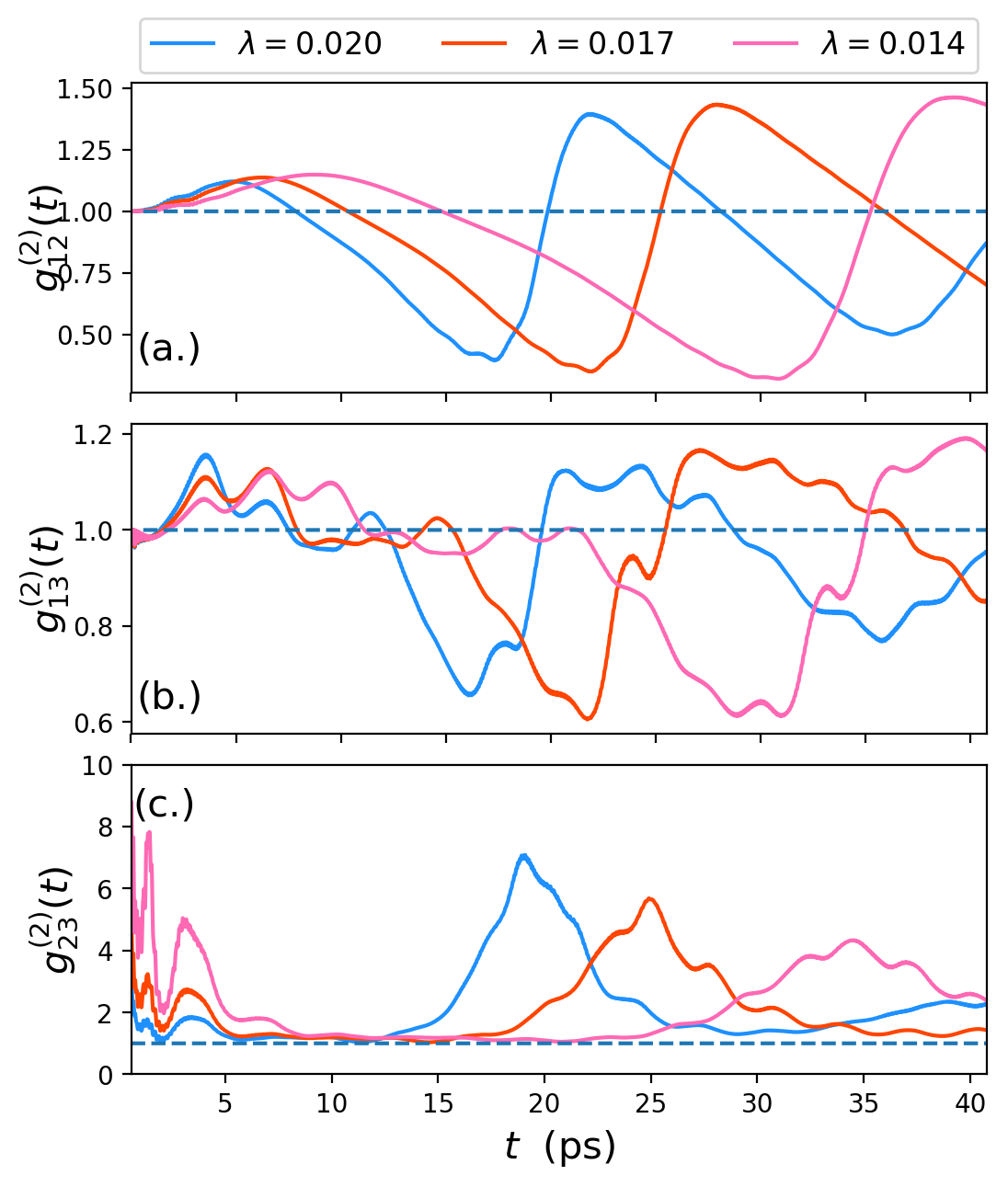}}
\caption[]{Real-time cross-correlation between the pump and signal modes. The dashed line indicates when the modes are correlated (above the dashed line)  and anti-correlated (below the dashed line). In panels (a.) and (b.) we have correlation and anti-correlation between the pump and signal photons at different times. (c.) The generated photon pairs of modes 2 and 3 are time-entangled for the whole evolution. In (c.), we disregard the time interval between $t=0$ and $t=0.58$ ps due to finite numerical precision~\cite{figcomment}}
\label{fig:coherences}
\end{figure}

\begin{figure}[t] 
\centerline{\includegraphics[width=0.5\textwidth]{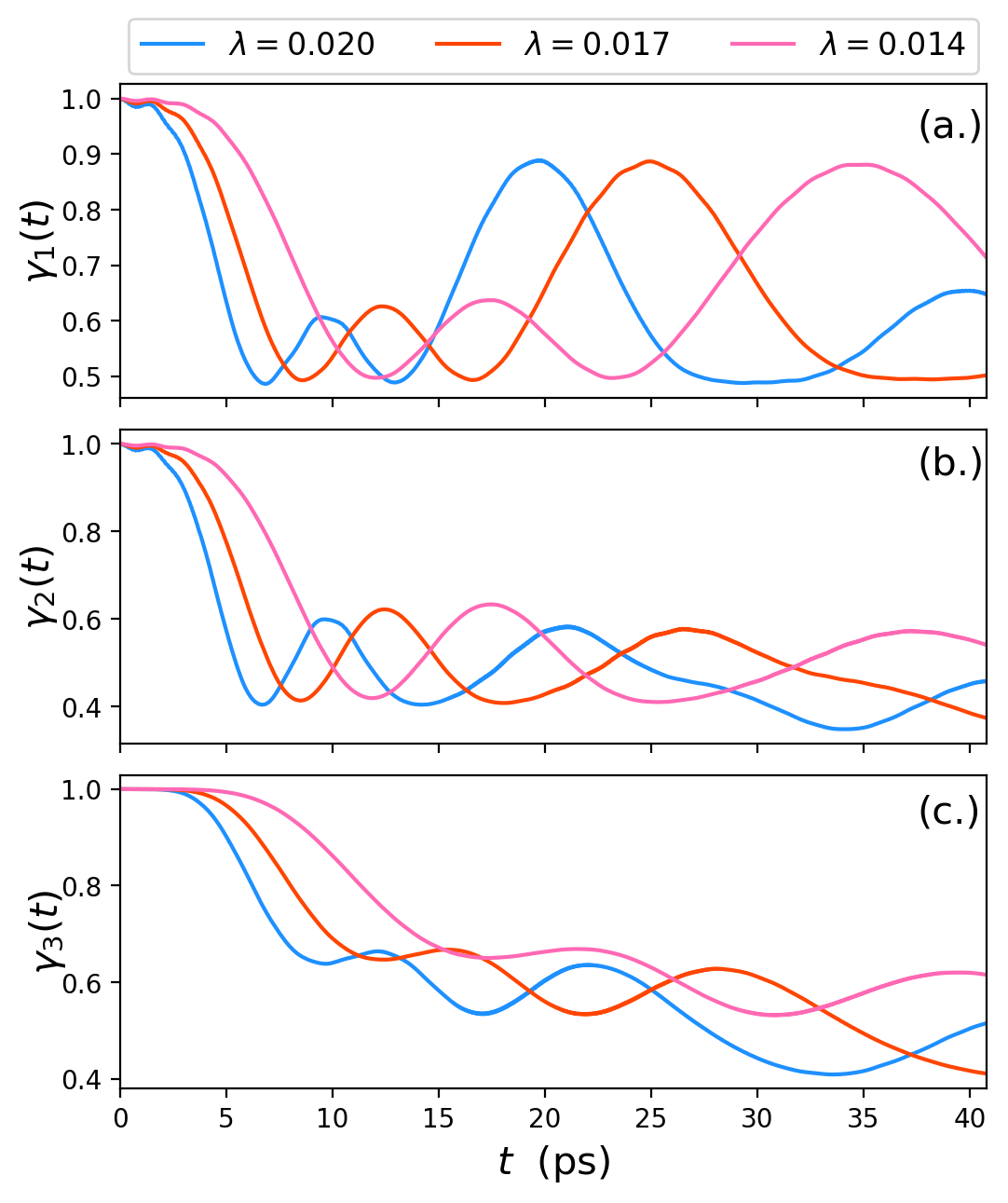}}
\caption{Measure of entanglement for the pump and signal modes. For the three modes in panels (a.), (b.) and (c.), signal mode 2 is most entangled, signal mode 3 is intermediate, and pump mode 1 is least entangled. In addition, by increasing the coupling from weak to ultra-strong not only changes the entanglement profile but also makes the modes more entangled.}
\label{fig:purity}
\end{figure}

The pump mode is now initially prepared in a coherent state $\ket{\phi_1} = |\xi_{1}\rangle$ and thus, its vector potential has the strength $\langle\xi_{1}|\hat{A}_{1}|\xi_{1}\rangle=\lambda\sqrt{2\hbar/\omega_{1}}|\xi_{1}|$ while that of the signal modes are zero at the beginning. We choose $\xi_{1}$ such that the mean photon number $\langle\hat{a}_{1}^{\dagger}\hat{a}_{1}\rangle = |\xi_{1}|^{2} =4$. We have thereby just changed the occupation slightly. However, this is already enough to no longer be able to identify in a simple manner the usual PDC as one photon being down-converted to two photons, since now many photon states mix (see also App.~\ref{app:fock-coherent-occup}).
\begin{figure}[t] 
\centerline{\includegraphics[width=0.5\textwidth]{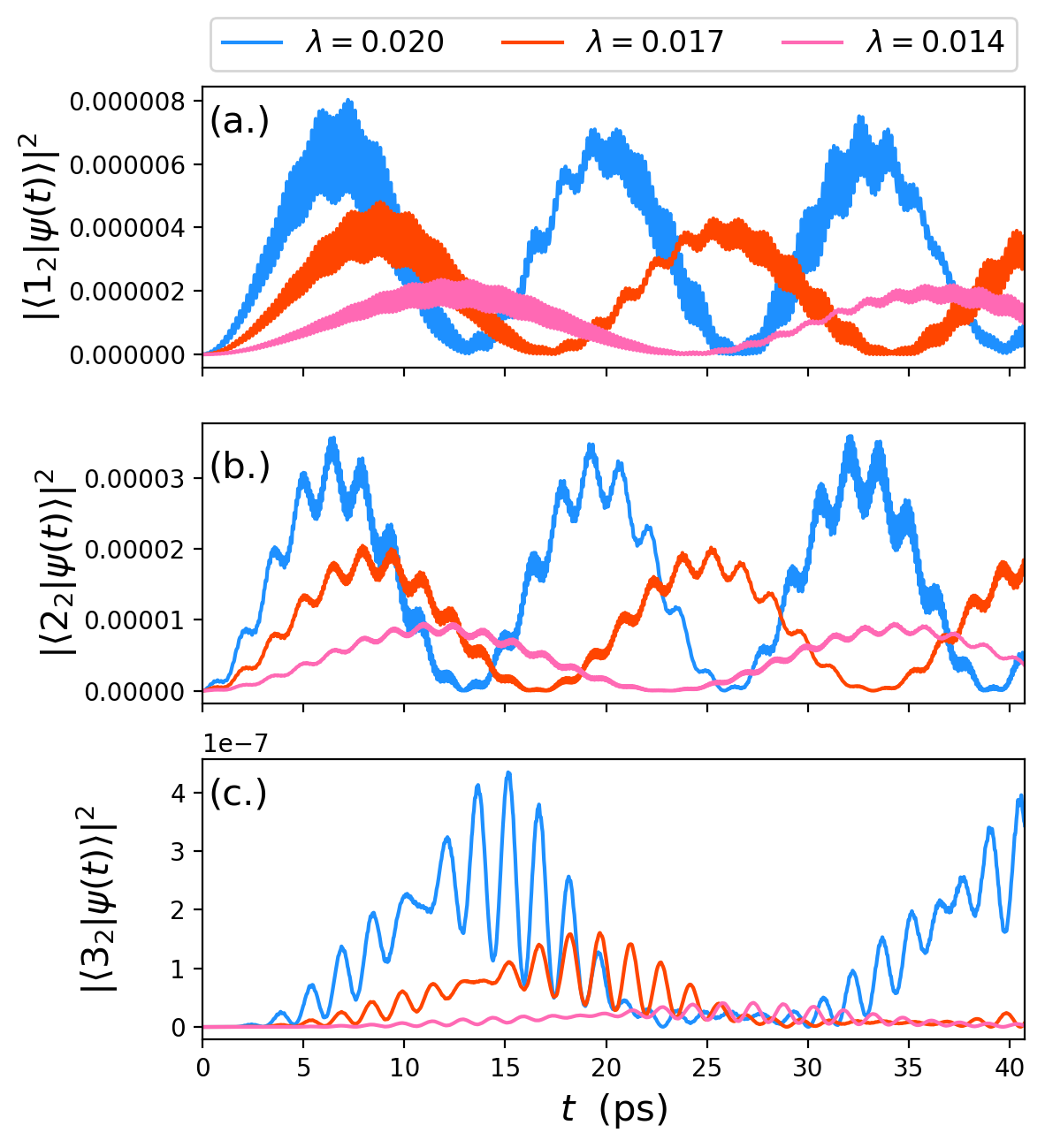}}
\caption{Real-time Fock state populations of the signal mode 2 from weak to ultra-strong coupling for the input coherent state. The one-photon Fock state in (a.) is mostly populated in comparison to the two- and three-photon Fock states in (b.) and (c.), respectively.}
\label{fig:coherent-fock-occupation-mode-2}
\end{figure}
We now find that the two- and three-photon states are also relatively strongly occupied. This becomes evident in Fig.~\ref{fig:coherent-fock-occupation-mode-2}, where the second-Fock state populations are an order of magnitude larger than the ones of the single-photon Fock states. This also changes the total mode occupations as can be seen in Fig.~\ref{fig:occupations}. 
Qualitatively the mode occupations remain relatively close to the one-photon case and also the feature of faster down-conversion in the strong coupling limit is preserved. Some details (especially for the input mode 1) are changed in comparison with the single-photon case when considering the non-classicality of the down-converted photons. As can be seen in Fig.~\ref{fig:mandels}, in mode 2 the anti-bunching behavior is increased and mode 3 becomes now more non-classical as well.
The cross-correlation features of the different modes also change strongly. While before the input and the different signal modes where anti-correlated all the time, indicating that the input photon is annihilated and the signal photons created instead, we now have a richer behavior. In Fig.~\ref{fig:coherences} (a.) and (b.) we see that there is a switching between anti-correlation and correlation in time.
This change is not so surprising because we have now several photons in the initial state and the simple picture of one photon annihilated in mode 1 and one photon created in mode 2 and 3 respectively, is not so straightforward anymore. However, the main feature of correlated down-converted photons persists also for this initial state (see  Fig.~\ref{fig:coherences} (c.)). Conservation of energy and momentum in this cascaded process gives rise to the correlations, implying energy-time entanglement~\cite{muthukrishnan2004} which is important for on-demand generation of entangled photon pairs~\cite{muller2014,dousse2010}. This interpretation is strengthened if we furthermore consider the purity $\gamma_{\alpha}$ to be the entanglement measure for the present case. As shown in Fig.~\ref{fig:purity} 
the different modes become entangled over time. The entanglement and its time profile can be controlled by the coupling strength and pushed to earlier times as well. Treating the pump and signal fields as quantized modes, the three fields are shown to be entangled (see Fig.~\ref{fig:purity}) and have varying correlations (see Fig.~\ref{fig:coherences}) which allows for applications in quantum information networks~\cite{villar2006}.

\section{Classical input fields}
~\label{sec:methods} 

While a coherent initial state is already closer to classical input field, we want to investigate whether a description based on a genuine classical external pumping changes the picture fundamentally. Although it is to be expected that a classical description of the input field for the relatively few photons we considered so far is not perfect, it should show qualitatively similar behavior at least for certain observables. We would then expect that these features are relatively unaffected by the details of the input field, i.e., how we populate the input mode or alternatively how we drive the electronic system. This is a very desirable feature since it would indicate that our findings are robust and do not depend on miniscule details of the setup.

At this point we face a further choice. Pumping the system with a classical field is possible in two distinct ways in an ab-initio description. Firstly we can prescribe an external electromagnetic field that couples to the matter system and its induced currents generate photons in the respective output modes. Alternatively we could prescribe an external current that couples to the input mode and thus we pump this mode directly. Physically these options should not be too different (by Maxwell's equation both can be connected to each other, see also App.~\ref{app:external-pump}). To find the influence of these two options we in the following consider both cases. 
\begin{figure}[t] 
\centerline{\includegraphics[width=0.5\textwidth]{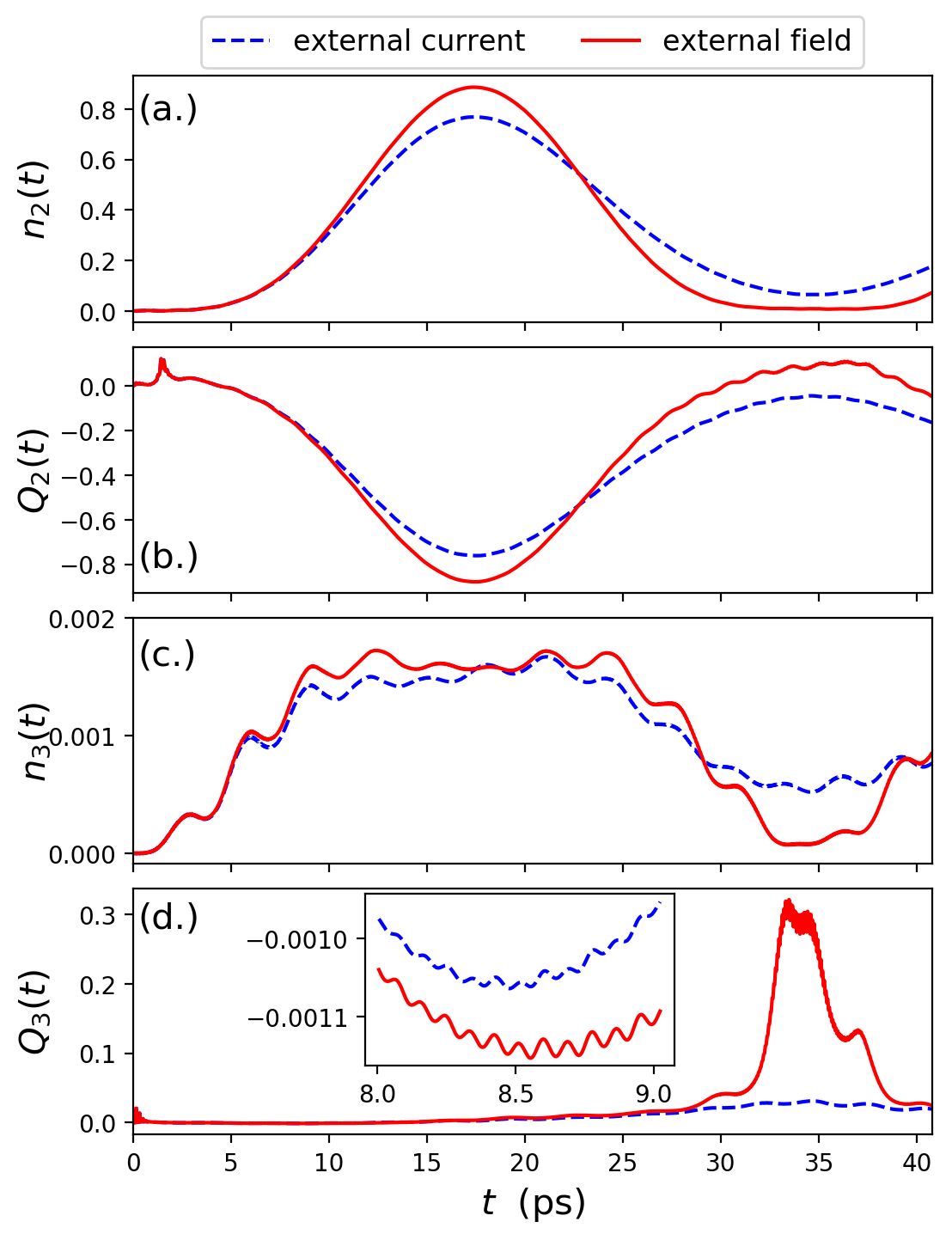}}
\caption{Comparison of the two different external drivings for weak matter-photon coupling $\lambda = 0.014$. In red we use an external laser pulse, and in blue we use an external current to pump mode 1 directly. We have chosen the pump pulse and the pump current to be connected via Maxwell's equations such that for weak coupling both lead to similar effects.}
\label{fig:cspdc-vs-spdc}
\end{figure}
First, we consider the more traditional option, that is, we choose some external electromagnetic field that drives the matter subsystem. Such a situation is often considered to avoid the need to have three explicit photon modes~\cite{chang2016,liu2005,liu2014}. In practice this means we replace input mode 1 by a classical external field and keep the rest of the system's Hamiltonian the same, i.e., we have 
\begin{align}
\hat{H}_{S}'(t) = \hat{H}_{S}' + \hat{H}_{ext}(t) . \label{external-PDC}
\end{align}
Here, the time-independent Hamiltonian $\hat{H}_{S}'$ and external driving term $\hat{H}_{ext}(t)$ are given explicitly in App.~\ref{app:external-pump}. The external pulse used corresponds to the classical field induced by the below defined external source term for the photon field in the case that mode 1 would be uncoupled. To facilitate a simple comparison we therefore select the weak coupling regime of Tab.~(\ref{tab:couplings}) in the following. Else we would need to adopt the form of the external pulse to expect a reasonable agreement. We note that as a result of discarding mode 1 the corresponding observables of this mode become inaccessible in a direct manner.

This is not the case for the second approach, where we pump mode 1 by an external current, i.e., a source term for the photons. In this case we can still investigate properties of mode 1. The corresponding Hamiltonian becomes
\begin{align}
\hat{H}_{S}(t) = \hat{H}_{S} + \hat{A}_{1}\cdot j_{1}(t) . \label{external-cPDC}
\end{align}
The external current is an envelope Gaussian of the form $j_{1}(t)=j_{1}\exp\left(-(t-t_{0})^{2}/\tau^{2}\right)\sin(\omega_{1}t)$. The parameters of the Gaussian pulse is chosen such that at time $t=0.23$ ps, the pump mode is driven to an excited state with on average $n_{1}(0)=4$ photons. Thus the vector potential of the pump mode is the same as that of the coherent state at the initial time (discussed in Sec.~\ref{sec:ab-initio}). For both cases, we choose the initial state to be the non-factorizable ground-state of $\hat{H}_{S}$ and $\hat{H}_{S}'$, respectively.

\begin{figure}[t] 
\centerline{\includegraphics[width=0.5\textwidth]{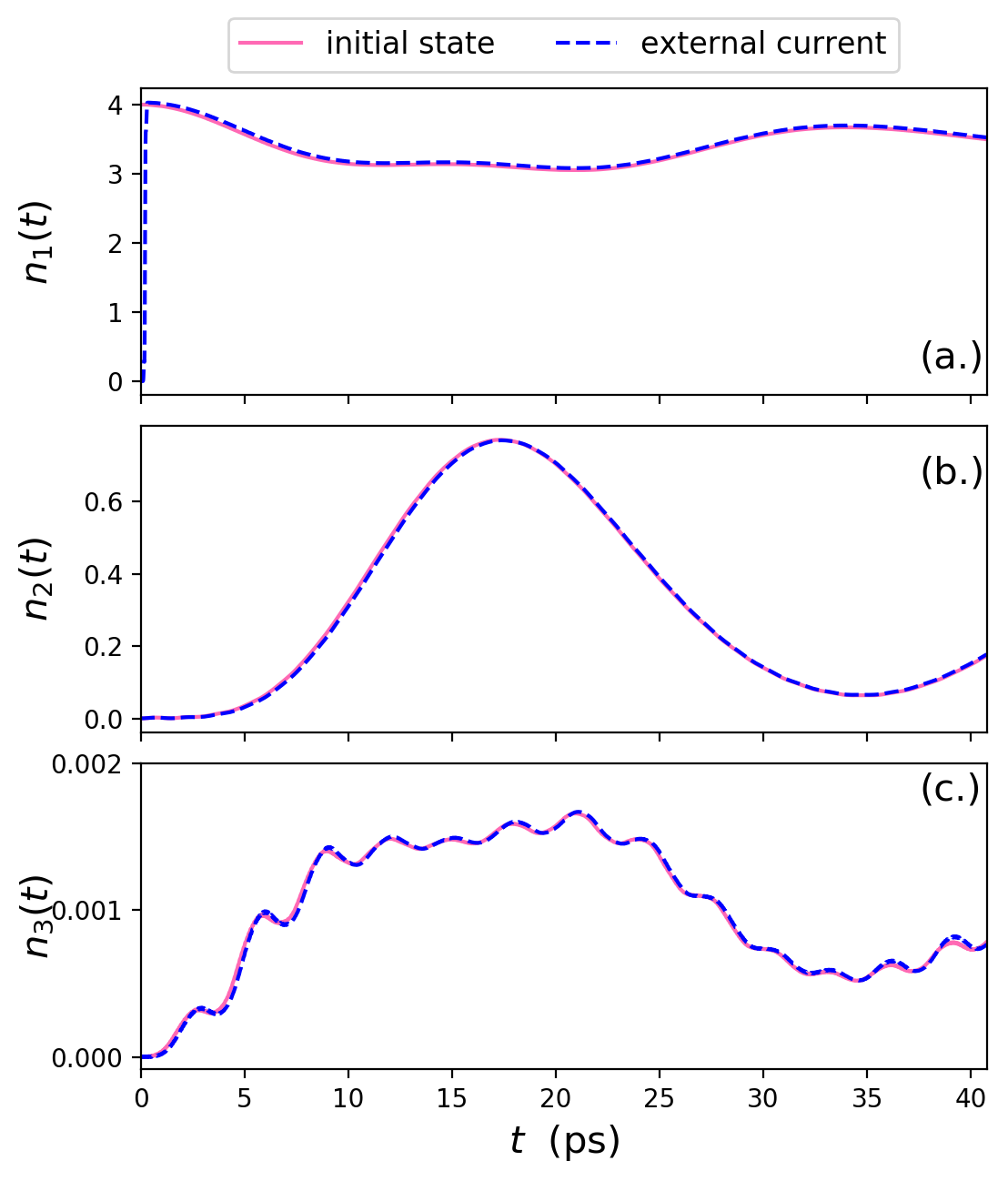}}
\caption{Comparison between an initial coherent state with on average of 4 photons (discussed in detail in Sec.~\ref{sec:ab-initio}) and an external current that simulates the creation of these 4 photons for weak coupling $\lambda = 0.014$. (a.) At $t=0.23$ ps, 4 photons are readily excited by the external current $j_{1}(t)$. The external current and initial state are qualitatively the same for the entire profile of the mode occupations.}
\label{fig:ini_vs_ext_ns}
\end{figure}

Comparing now both cases of external driving in Fig.~\ref{fig:cspdc-vs-spdc}, we see that they agree qualitatively for the mode occupations and for the Mandel $Q$ parameters. We find, however, that at later times, e.g., for $Q_3$, they can differ strongly. If we then compare to the case of no external driving and having 4 photons in mode 1 at $t=0$ instead (discussed in Sec.~\ref{sec:ab-initio}), we find that the chosen external fields qualitatively reproduce this case (see Fig.~\ref{fig:ini_vs_ext_ns})

\section{Comparison to standard approximations}
~\label{subsec:few-levels}

\begin{figure}[t] 
\centerline{\includegraphics[width=0.5\textwidth]{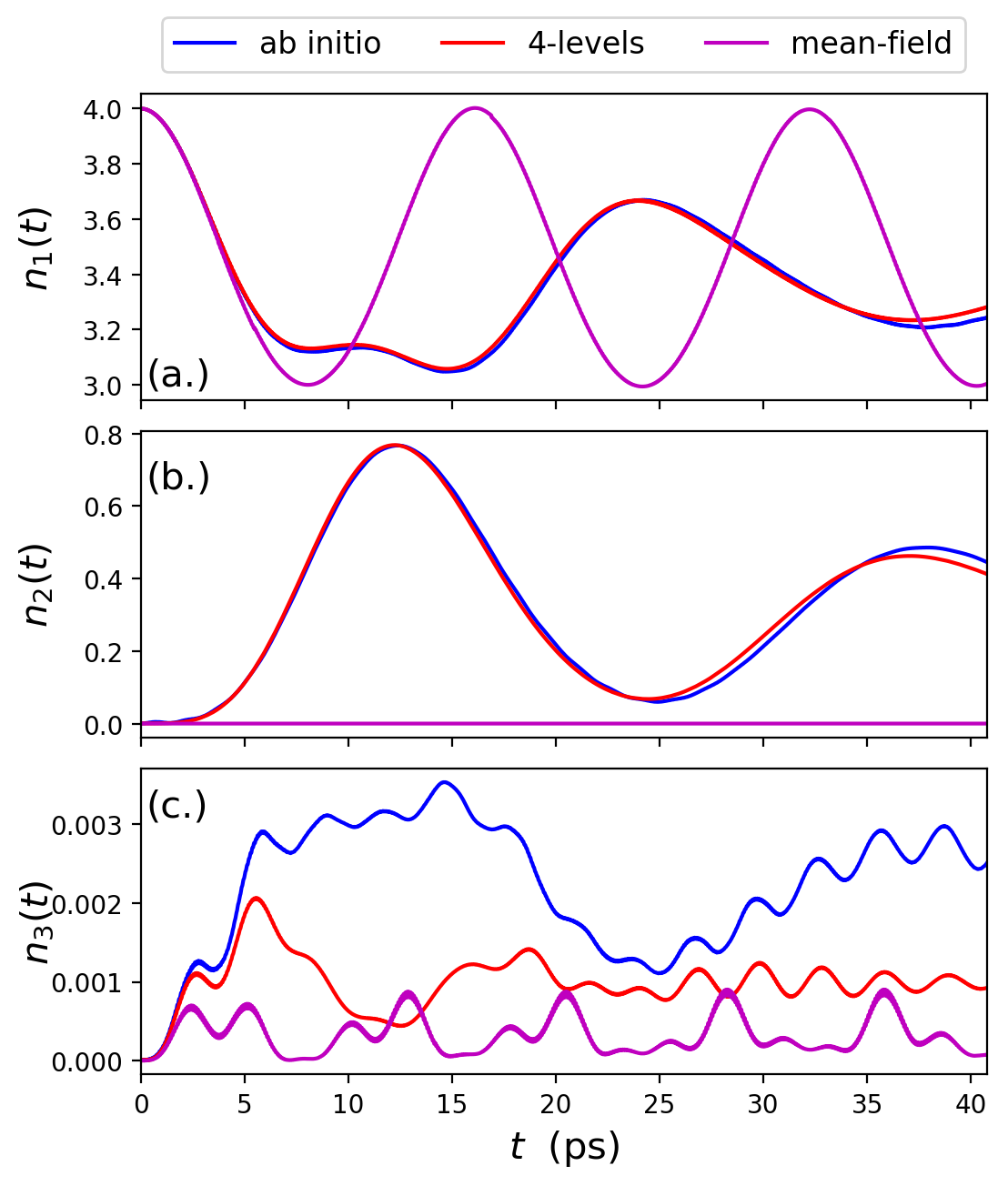}}
\caption{Performance of few-level and Maxwell-Schr\"odinger approximation. In (a.) and (b.) the few-level approximation is relatively accurate while the Maxwell-Schr\"odinger approximation is qualitatively correct only up to $t=5$ ps. In (c.) both approximations are qualitatively off and do not capture the occupation of mode 3.}
\label{fig:real-down}
\end{figure}

\begin{figure}[t] 
\centerline{\includegraphics[width=0.5\textwidth]{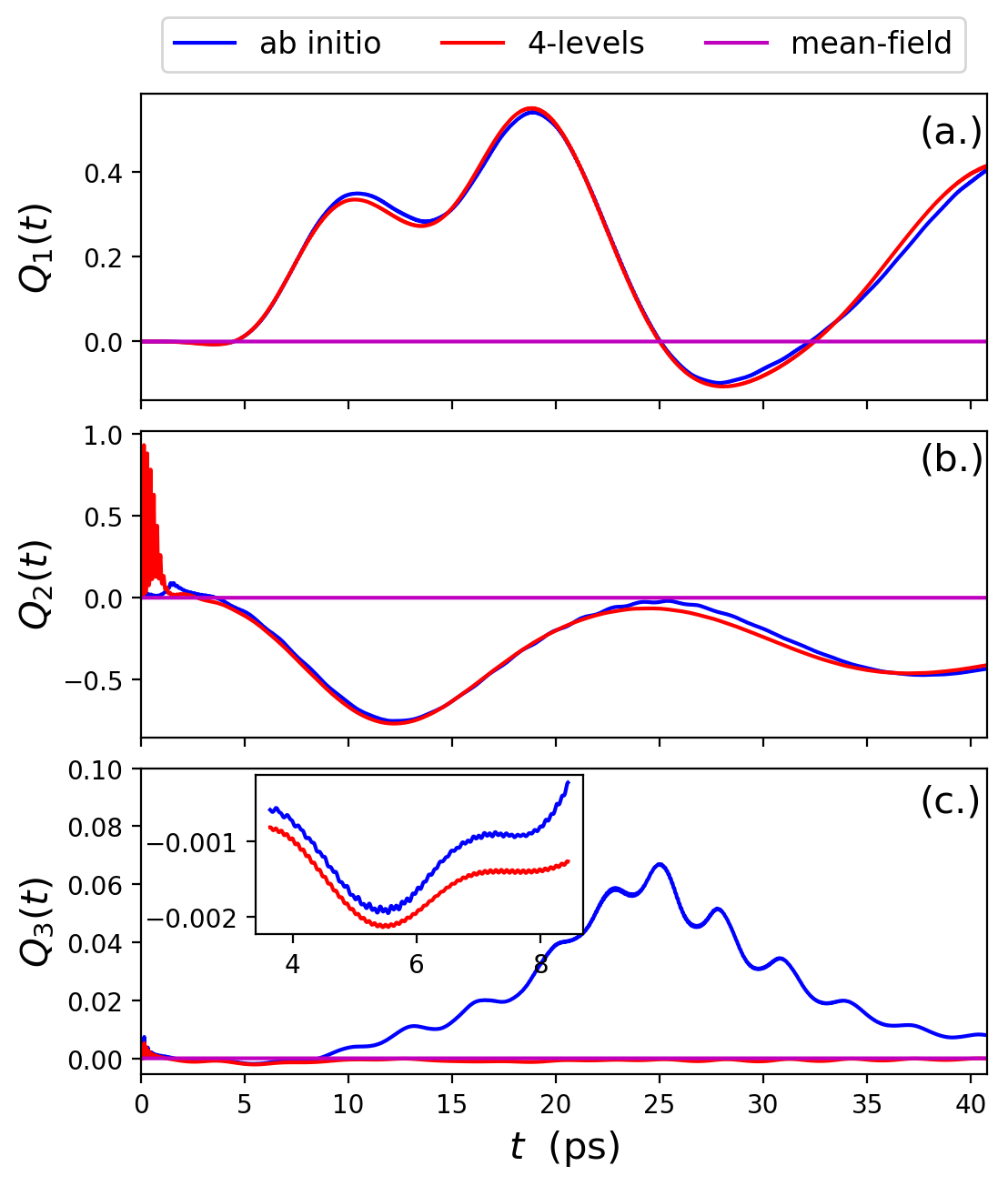}}
\caption{Performance of few-level and Maxwell-Schr\"odinger approximation. While the few-level approximation captures the Mandel $Q_1$ in (a.) quite well, it is wrong initially for $Q_2$ in (b.) and wrong for $Q_3$ in (c.). As expected, the Maxwell-Schr\"odinger approximation does not capture the quantum features of the photon field in (a.) - (c.).}
\label{fig:real-mandel}
\end{figure}

We have found so far that while treating the pump mode differently has an influence on certain observables and that it can make the interpretation of the down-conversion process more involved, other observables stay relatively unchanged. Yet, can we capture these effects also with standard approaches and avoid a full ab-initio treatment? By this we mean on the one hand quantum-optical approaches based on the few-level approximation and on the other hand semi-classical approaches which are employed in non-linear optics.     

The few-level approximation takes only a few matter states into account in contrast to the large amount of states that are considered in ab-initio simulations~\cite{stefano2019}. In ab-initio simulations one increases the amount of states until either the observables of interest or even the full wave function does no longer change. This basis-set-limit procedure is common place to ensure that the obtained results are not pathologically influenced by a cut off in the matter-photon basis. Disregarding the issue of basis-set convergence, we can instead choose to consider only a few ''relevant'' matter states, which leads to a Hamiltonian of the form given in App.~\ref{app:few-levels}. In this section we choose four levels (as in Fig.~(\ref{fig:setup-spdc}b) or the highlighted energy levels in Fig.~(\ref{fig:spectrum-GaAs}a)) which by simple energy arguments are the most relevant states (for details see App.~\ref{app:few-levels}). By simplifying the matter subsystem in this way discards the possibilities of the existence of other hybridized polariton and virtual states that occur in an ab-initio treatment when the quantized field is coupled to the full matter subsystem. In the few-levels approximation considered here, we include the mode-mode interactions that arises from the diamagnetic term as opposed to cases were they are dropped out~\cite{kockum2017,stefano2019}. Ignoring these terms can lead to results that will differ considerably. 

In non-linear optics one would usually use a semi-classical approximation and consider, for instance, non-linear matter-only response functions. However, in the case of the quantum ring these would just be zero at the employed frequencies~\cite{duque2012} and thus no further conclusions could be drawn. We therefore go one rung higher and employ an adopted Maxwell-Schr\"odinger approximation~\cite{Jestaedt2020}, which is a further common approximation used in the field of non-linear optics. It replaces the quantized photon field by its mean-field expression and hence leads to a set of coupled non-linear equations
\begin{align}
&i\hbar \frac{\partial}{\partial t} \varphi(\textbf{r}, t) = \hat{H}_{\text{MS}}([q_{\alpha}],t) \varphi(\textbf{r},t) , \label{schroed} \\
&\left(\frac{d^{2}}{dt^{2}} + \omega_{\alpha}^{2}\right)q_{\alpha}(t) = j_{\alpha}([\textbf{p}, q_{\alpha}],t) . \label{maxwell}
\end{align}
Here, $\varphi(\textbf{r}, t)$ is the wave function of the matter subsystem and $j_{\alpha}(t)$ is the current that self-consistently couples the mode-resolved Maxwell fields to the electronic subsystem (see App.~\ref{app:maxwell-schroedinger} for details). We note that we include in the Maxwell-Schr\"odinger approximation also the mean-field mode-mode interactions (and with this the diamagnetic current). Discarding these terms, as commonly done, will make the Maxwell-Schr\"odinger approximation less accurate. While the PDC process will be well described by this level of theory for the weak-coupling situation, in the strong-coupling case where hybrid light-matter states emerge, it is expected to be less reliable.

In Figs.~\ref{fig:real-down} and \ref{fig:real-mandel} we display characteristic observables of the down-conversion process in the strong coupling regime (i.e. $\lambda = 0.017$). The Maxwell-Schr\"odinger approximation fails, as expected, in the few-photon strong-coupling limit for all observables. In this case the quantum features of the electromagnetic field are essential and hence the mean-field-type approach of the Maxwell-Schr\"odinger theory is inadequate. Nevertheless, once we go to many photons, it becomes much more precise such that in the limit of arbitrarily many photons the full ab-initio theory becomes essentially equivalent to the Maxwell-Schr\"odinger theory (see also App.~\ref{app:maxwell-schroedinger}). Therefore we can use this level of theory to assess how increasing the number of input photons strongly, affects the down-conversion process (see Sec.~\ref{subsec:strong}). Beside the inadequacy of the semi-classical approximation for describing the down-conversion in the few-photon limit, we also find in Figs.~\ref{fig:real-down} and \ref{fig:real-mandel} that the standard few-level approximation is not completely reliable as well. The main reason is that since the transition that is in resonance with mode 3 is not dipole allowed, many more levels have a similar contribution. This highlights that such a reduction to merely a few states can become inaccurate once several observables are considered at once. Only upon having converged the full wave function to an appropriate accuracy can we hope to have access to all possible observables.

Consequently, in the case that light and matter couple strongly and hybrid-light matter states mediate the down-conversion process, an ab-initio light-matter description becomes necessary.  


\section{Optimization of the down-conversion: the degenerate case}
~\label{sec:iHG}

\begin{figure}[t] 
\centerline{\includegraphics[width=0.5\textwidth]{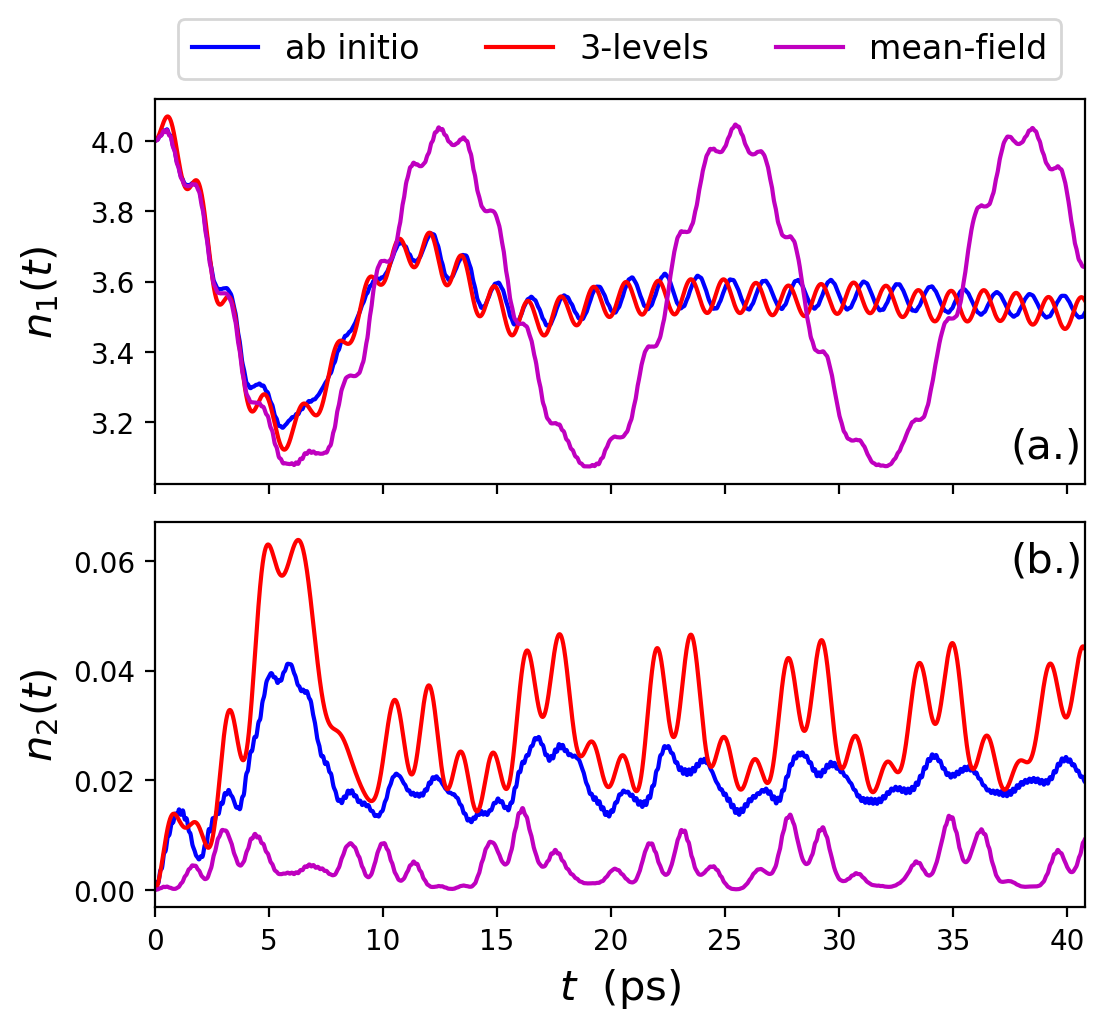}}
\caption{Example of the degenerate down-conversion process in the strong coupling regime ($\lambda=0.017$)  and for a mixing angle of $\theta_{1}=60^{\circ}$. (a.) The 3-level approximation qualitatively captures the full dynamics of $n_{1}(t)$, while the Maxwell-Schr\"odinger description deviates around $t=5$ ps for the entire dynamics. (b.) Associated $n_{2}(t)$ of down-converted photons where both approximations are off from the ab-initio result.}
\label{fig:iHG-n1-n2}
\end{figure}

\begin{figure}[t] 
\centerline{\includegraphics[width=0.5\textwidth]{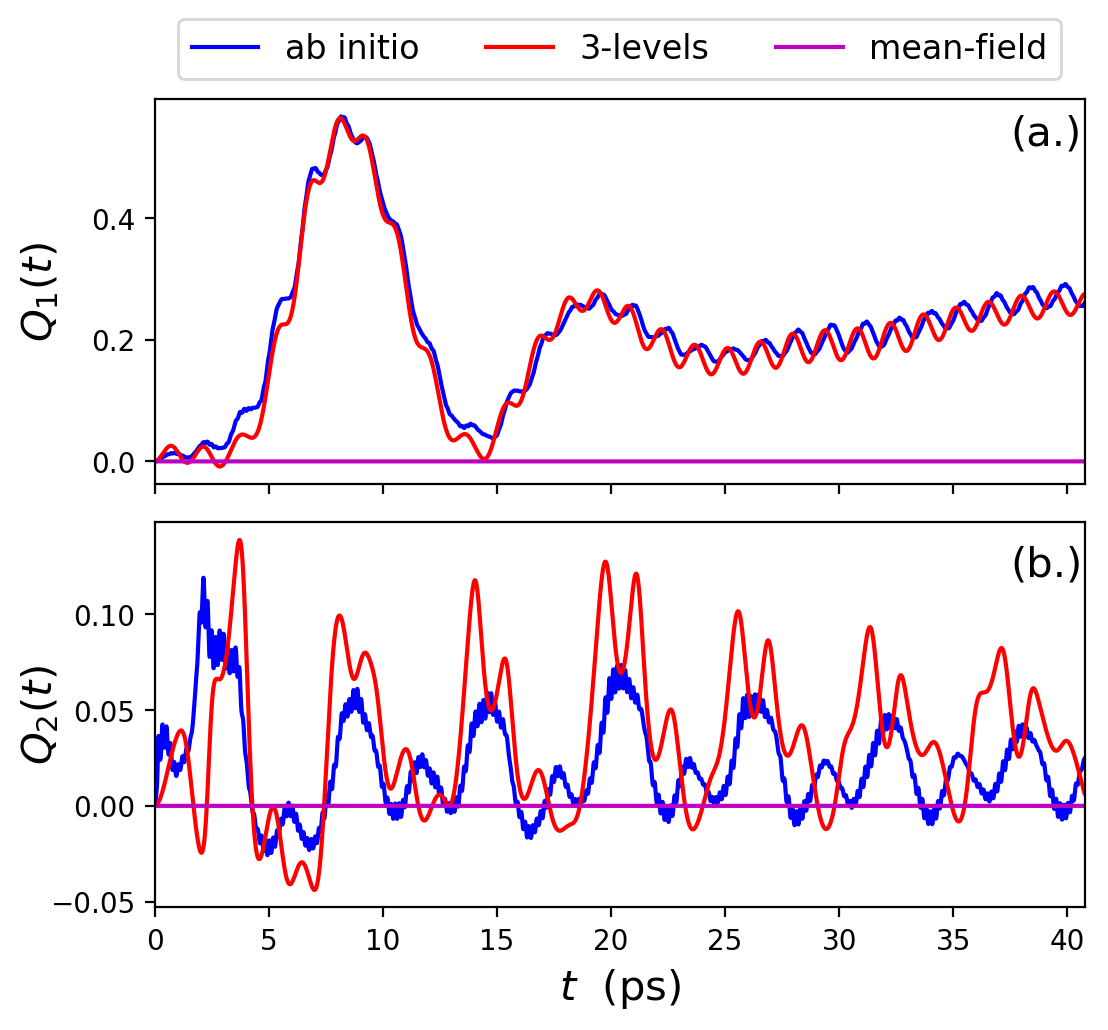}}
\caption{Example of the degenerate down-conversion process in the strong coupling regime ($\lambda=0.017$)  and for a mixing angle of $\theta_{1}=60^{\circ}$. (a.) The 3-level approximation qualitatively captures the full dynamics of $Q_{1}(t)$, while the mean-field is zero for the full simulation by construction. (b.) The 3-levels and Maxwell-Schr\"odinger approximations are off for the complete time evolution.}
\label{fig:iHG-Q1-Q2}
\end{figure}

In this section we want to find the optimal setting for having, on the one hand, an efficient polariton-mediated down-conversion process, and on the other hand, non-classical and controlled down-converted photons. In our current context we mean by non-classical the minimum Mandel $Q$ parameter, i.e., anti-bunching behavior. To make the space of parameters as small as possible we will here consider the degenerate case, i.e., mode 2 and 3 have the same frequency. As a result of the degeneracy in energy in both signal modes 2 and 3, we decouple the signal mode 3 from Eq.~(\ref{cspdc-hamiltonian}) such that the down-converted photons populate only mode 2. We fix the polarization of signal mode 2 as $\hat{\textbf{A}}_{2}=\hat{A}_{2}\textbf{e}_{y}$ by choosing $\theta_{2} = 0^{\circ}$ and change the polarization of the pump field to $\hat{\textbf{A}}_{1} = \hat{A}_{1}(\cos\theta_{1}\textbf{e}_{x} + \sin\theta_{1}\textbf{e}_{y} )$ where $\theta_{1}$ is the mixing angle between the photonic and electronic subsystems. In this case the Hamiltonian of Eq.~(\ref{cspdc-hamiltonian}) reduces to the two-mode electron-photon Hamiltonian
\begin{align}
\hat{H}_{D} &= \hat{H}_{el} + \hat{H}_{1} + \hat{H}_{2} - \frac{e}{m}\hat{A}_{1}\left(\hat{p}_{x} \cos\theta_{1} + \hat{p}_{y} \sin\theta_{1} \right)  \nonumber\\
& \quad - \frac{e}{m}\hat{A}_{2}\hat{p}_{y} + \frac{e^{2}}{2m}\left[\hat{A}_{1}^{2} + \hat{A}_{2}^{2} +  2\hat{A}_{1}\hat{A}_{2} \sin(\theta_{1})\right] . \label{ihg-hamiltonian}
\end{align}

As increasing the coupling strength pushes the down-conversion process to earlier times, here we consider its maximal efficiency and minimum Mandel-$Q$ parameter in mode 2. We will vary the polarization directions (see Sec.~\ref{subsec:polarization}), the anharmonicity of the binding potential of the GaAs quantum ring (see Sec.~\ref{subsec:anharmonicty}), and the coupling strength as well as the number of input photons (see Sec.~\ref{subsec:strong}). To judge the efficiency we consider the maximal amount of mode occupation $n_2$ (except of in Sec.~\ref{subsec:strong} where we take a more general definition) over the range of the first 40 ps. For the non-classicality we determine the minimal amount of $Q_2$ over the same time interval.

We simulate the time-evolution dynamics of the ab-initio system, the 3-level approximation (due to having a degenerate down-conversion process) and the self-consistent Maxwell-Schr\"{o}dinger approximation in the strong-coupling regime ($\lambda =0.017$) starting from a coherent state in mode 1 with 4 photons ($\xi_{1}=2$). For degenerate two-photon generation, the pump field with energy $\hbar\omega_{1} =1.413$ meV drives resonantly the transition between the ground- and first-excited state $|\varphi_{1}^{0}\rangle\leftrightarrow|\varphi_{2}^{1}\rangle$, thereby populating the state $|\varphi_{2}^{1}\rangle$.  The signal mode has half the energy $\hbar\omega_{2} = 0.706$ meV of the $|\varphi_{1}^{0}\rangle\leftrightarrow|\varphi_{2}^{1}\rangle$ transition corresponding to a two-photon process. Since the coupling $g_{1}=0.0398$ of the pump mode is less than the coupling $g_{2}=0.0563$ of the signal mode, the effective electron in state $|\varphi_{2}^{1}\rangle$ is likely to relax to the ground-state through a two-photon emission channel via a virtual state. From a standard perspective this process should be again very inefficient. Yet within an ab-initio light-matter treatment tuning the photonic environment to a virtual state will in the strong and ultra-strong coupling regime lead to the creation of a hybrid state that can efficiently mediate the down-conversion of the input photons.

In Figs.~(\ref{fig:iHG-n1-n2}) and (\ref{fig:iHG-Q1-Q2}), we show the real-time photon occupations and Mandel $Q_{\alpha}$ parameters of the incoming pump photon at the mode-mixing angle $\theta_{1}=60^{\circ}$ and the degenerate down-converted photons. For the ab-initio results, the profile of $n_{1}(t)$ shows that at $t=5.61$ ps, 0.82 of a photon is annihilated by absorption to promote the electron to the state $|\varphi_{2}^{1}\rangle$. At $t=5.84$ ps, the particle subsequently relaxes to the ground state by emitting photons with maximum mode occupation $n_{2}=0.041$. The apparent weak profile for $n_{2}(t)$ is due to emission via virtual states. The photon statistics of the pump field starts out in a coherent state and leads to a field with super-Poissonian statistics while the photon statistics of the generated photons varies between a field with bunching and anti-bunching features at different times, with the minimal value of $Q_{2}=-0.0199$ at $t=4.94$ ps. The few-levels approximation (in this case we only take three states into account, see also App.~\ref{app:quantum-ring}) in Figs.~\ref{fig:iHG-n1-n2} (a.) and \ref{fig:iHG-Q1-Q2}(a.) is relatively accurate, while the Maxwell-Schr\"{o}dinger performs is qualitatively correct only up $t=5$ ps and by construction it remains a coherent field for the entire evolution with constant $Q_{1}=0$. Both approximations deviate from the full solution in Figs.~\ref{fig:iHG-n1-n2} (b.) and \ref{fig:iHG-Q1-Q2} (b.). While the three-level approximation consistently overestimates the down-conversion efficiency as well as the quantumness of the photons in mode 2 (because it restricts the matter system too strongly), the Maxwell-Schr\"odinger approximation under-estimates these quantities. For the Mandel $Q$ parameter this is by design and for the down-conversion efficiency the Maxwell-Schr\"odinger discards all the correlation between light and matter that becomes beneficial in the strong coupling case. 

We note that this down-conversion scheme is indeed an inverse second-harmonic generation made possible by coupling the signal mode to a virtual state at half the energy of the first degenerate excited state $|\varphi_{2}^{1}\rangle$. This will, however, not be the limiting case since the same can be done for the third-, fourth- and $N$th-photon generation, thereby defining an inverse harmonic generation for realizing an $N$-photon photon gun. By tuning the system in an experimentally realizable way, the setup can potentially be used an $N$-photon source with highly non-classical properties. Such an optimization will be discussed below for the inverse second-harmonic generation process.

\subsection{Optimization of field polarization}
~\label{subsec:polarization}

\begin{figure}[t] 
\centerline{\includegraphics[width=0.5\textwidth]{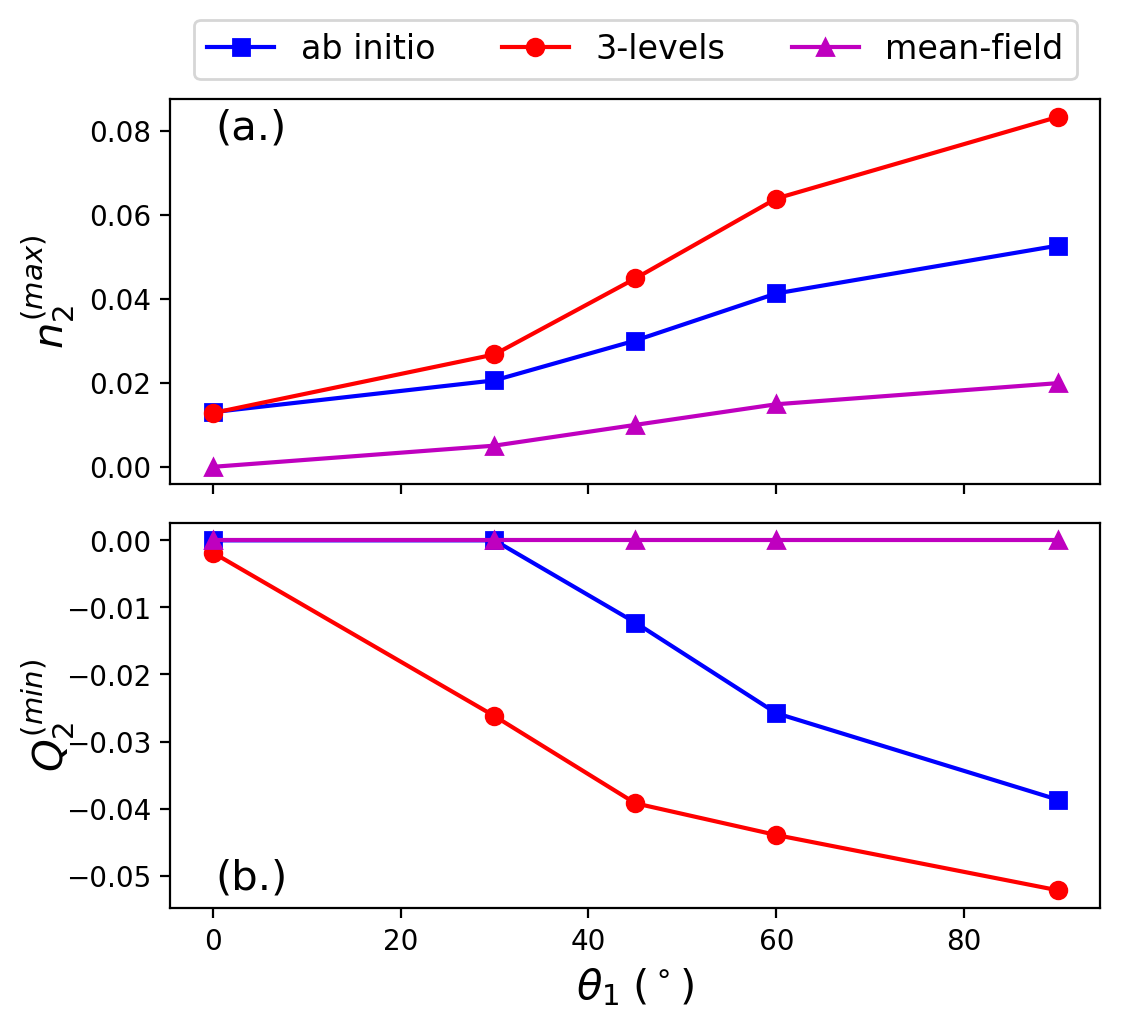}}
\caption{(a.) The influence of the interference term in the down-conversion process by varying the mixing angle $\theta_{1}$. Increasing $\theta_{1}$ increases the photon occupation. The 3-levels and Maxwell-Schr\"odinger approximations are off from the exact results. (b.) Increasing the mixing angle $\theta_{1}$ results in increasing sub-Poissonian statistics (anti-bunching) of the down-converted photons.}
\label{fig:iHG-n2-max}
\end{figure}

Now we consider which relative polarization is most efficient and at the same time minimizes $Q_2$. In our setup this means that we consider the mixing angle $\theta_{1}$ and the contribution of the interference term $2\hat{A}_{1}\hat{A}_{2}\sin(\theta_{1})$. For this we perform the time-propagation for different mixing angles $\theta_{1} = 0^{\circ}, 30^{\circ}, 45^{\circ}, 60^{\circ}, 90^{\circ}$ for strong coupling ($\lambda=0.017$) and choose as the input mode 1 in a coherent state with $\xi_{1}=2$ and have $V_0 = 200$ meV. From the Fig.~\ref{fig:iHG-n2-max} we see that for $\theta_{1}=0^{\circ}$, where the polarization of pump and signal modes are perpendicular, we obtain the smallest value of $n_{2} =0.013$ for all the mixing angels. The mean photon occupation $n_{2}(t)$ increases with increasing angles due to the fact that $2\hat{A}_{1}\hat{A}_{2}\sin(\theta_{1})$ becomes larger. For $\theta_{1}=90^{\circ}$ we find the highest down-conversion of photons as we obtain $n_{2} =0.0527$ since $\sin(\theta_{1})=1$ and the polarization of both modes are parallel with momenta contribution only in the y-axis (i.e., $\hat{p}_{y}\neq 0$ while $\hat{p}_{x}= 0$). The Mandel $Q_{2}(t)$ for $\theta_{1}=90^{\circ}$ shows the highest non-classical (anti-bunching) features of the down-converted photons as $Q_{2}=-0.0387$. The values of $n_{2}^{max}$ and $Q_{2}^{min}$ for angles $\theta_{1} = 0^{\circ}, 30^{\circ}, 45^{\circ}, 60^{\circ}, 90^{\circ}$ are given in Tab.~(\ref{tab:angles}). Both the 3-levels and Maxwell-Schr\"odinger approximations deviate from the ab-initio results, yet provide upper and lower bounds. Although the Maxwell-Schr\"odinger~\cite{Jestaedt2020} takes the mixing angle into account, it misses the induced correlations between the two modes, highlighting that an efficient down-conversion is driven in this strong-coupling case by the quantum correlations between the different modes and the matter system. The three-level approximation, on the other hand, overestimate these quantum correlations.

\begin{table}
\begin{center}
\begin{tabular}{ |  c| c| c| }
\hline
$\theta_{1} \; (^{\circ})$  &   Exact $n_{2}^{max}$ & Exact $Q_{2}^{min}$ \\\hline \hline
0   & 0.0130  & 0.0 \\\hline
30  & 0.0206  & 0.0 \\\hline
45  & 0.0301  & -0.0123 \\\hline
60  & 0.0413  & -0.0258 \\\hline
90  & 0.0527  & -0.0387 \\\hline
\end{tabular}
\end{center}
\caption{The values for maximum $n_{2}^{max}$ and minimum Mandel $Q_{2}^{min}$ for different mixing angles. Increasing $\theta_{1}$ increases $n_{2}^{max}$ and decreases $Q_{2}^{min}$.}
\label{tab:angles}
\end{table}

\subsection{Optimization of the matter spectrum}
~\label{subsec:anharmonicty}

\begin{figure}[t] 
\centerline{\includegraphics[width=0.5\textwidth]{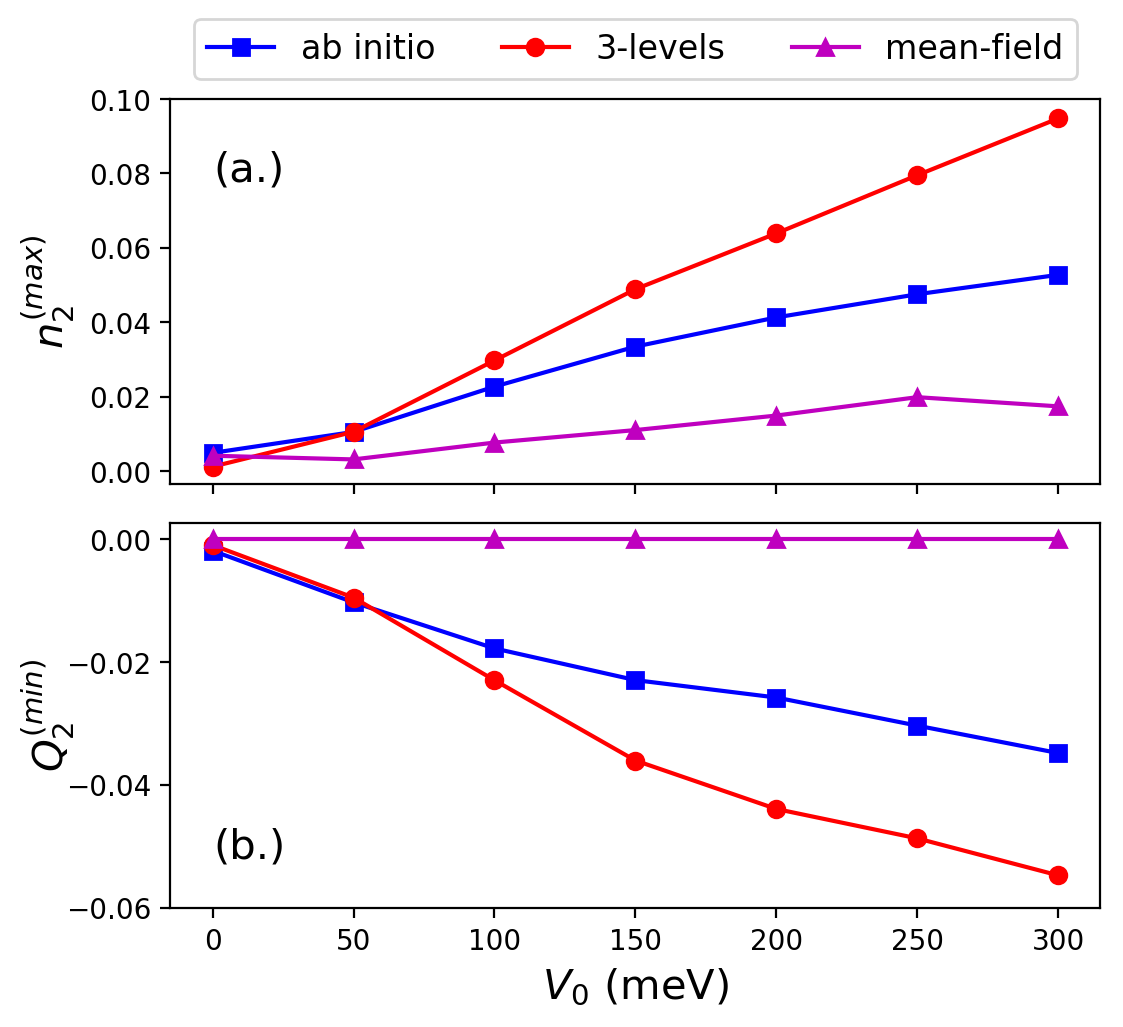}}
\caption{(a.) Maximum down-converted photons $n_{2}$ for increasing $V_{0}$ for the different levels of theory. (b.) Minimum Mandel $Q_{2}$ parameter for increasing $V_{0}$ for the different levels of theory.}
\label{fig:V0-n2-Q2}
\end{figure}

Next, we investigate the influence of the anharmonicity of the electronic subsystem on the down-conversion process. We choose mode 1 in a coherent state with on average 4 photons, assume strong coupling ($\lambda =0.017$) and choose a mixing angle $\theta_{1}=60^{\circ}$ such that both momentum components of the matter system are non-zero. We vary the parameter $V_{0} =0, 50, 100, 150, 200, 250, 300$ meV (see also App.~\ref{app:quantum-ring}) which changes the transition energies between states from harmonic (all have the same transition energy) to anharmonic (different transitions have different energies). Table.~(\ref{tab:anharmonic}) shows the transition energies from the electronic ground state to the first excited state that the pump energy is resonant to and its down-converted energies. In addition, increasing $V_{0}$ also increases the transition elements which leads to a stronger effective coupling between light and matter (see App.~(\ref{app:few-levels})). 

We find (Fig.~\ref{fig:V0-n2-Q2}) that with increasing $V_0$ the maximum photon occupation $n_2$ increases. Simultaneously, the state of the generated photon pairs becomes increasingly anti-bunched as $Q_{2}$ becomes more negative. This results from (i.) the increasing dipole moments due to the reduction of the quantum ring width for increasing $V_{0}$ and (ii.) the increase in the effective coupling strength due to the reduced scaling of the energies (see Tab.~(\ref{tab:anharmonic})) as the effective coupling is related to the frequency by $g_{\alpha}=\lambda\sqrt{\hbar/2\omega_{\alpha}}$ (see App.~\ref{app:quantum-ring}). We nicely see how the three-level approximation at $V_{0}=0$ meV fails, because many more than just three levels are important when all transitions have the same energy and we cannot separate a few specific transitions. Yet even when the system becomes strongly anharmonic and hence a separation of states is more reasonable, the few-level approximation overestimates the results strongly. Again, the Maxwell-Schr\"odinger approximation consistently under-estimates the results. 
\begin{table}
\begin{center}
\begin{tabular}{ | c | c | c| c| c| }
\hline
$V_{0}$ (meV) &  $\hbar\omega_{1}$ (meV) & $\hbar\omega_{2}$ (meV) & Exact $n_{2}^{max}$ & Exact $Q_{2}^{min}$ \\\hline \hline
0.0   & 10.00  & 5.00   & 0.0049  & -0.0019 \\\hline
50.0  & 3.121  & 1.560  & 0.0105  & -0.0103 \\\hline
100.0 & 1.924  & 0.962  & 0.0227  & -0.0178 \\\hline
150.0 & 1.580  & 0.790  & 0.0334  & -0.0229 \\\hline
200.0 & 1.413  & 0.706  & 0.0413  & -0.0258 \\\hline
250.0 & 1.311  & 0.655  & 0.0475  & -0.0303 \\\hline
300.0 & 1.239  & 0.619  & 0.0527  & -0.0348 \\\hline
\end{tabular}
\end{center}
\caption{Pump energies for  resonant coupling of the electron ground-state and first excited state and corresponding down-converted energies of the signal field for different values of $V_{0}$. The numerically exact real-space values for maximum $n_{2}^{(max)}$ and minimum $Q_{2}^{(min)}$.}
\label{tab:anharmonic}
\end{table}

\subsection{Optimization of the coupling and the input field}
~\label{subsec:strong}

\begin{figure}[t] 
\centerline{\includegraphics[width=0.5\textwidth]{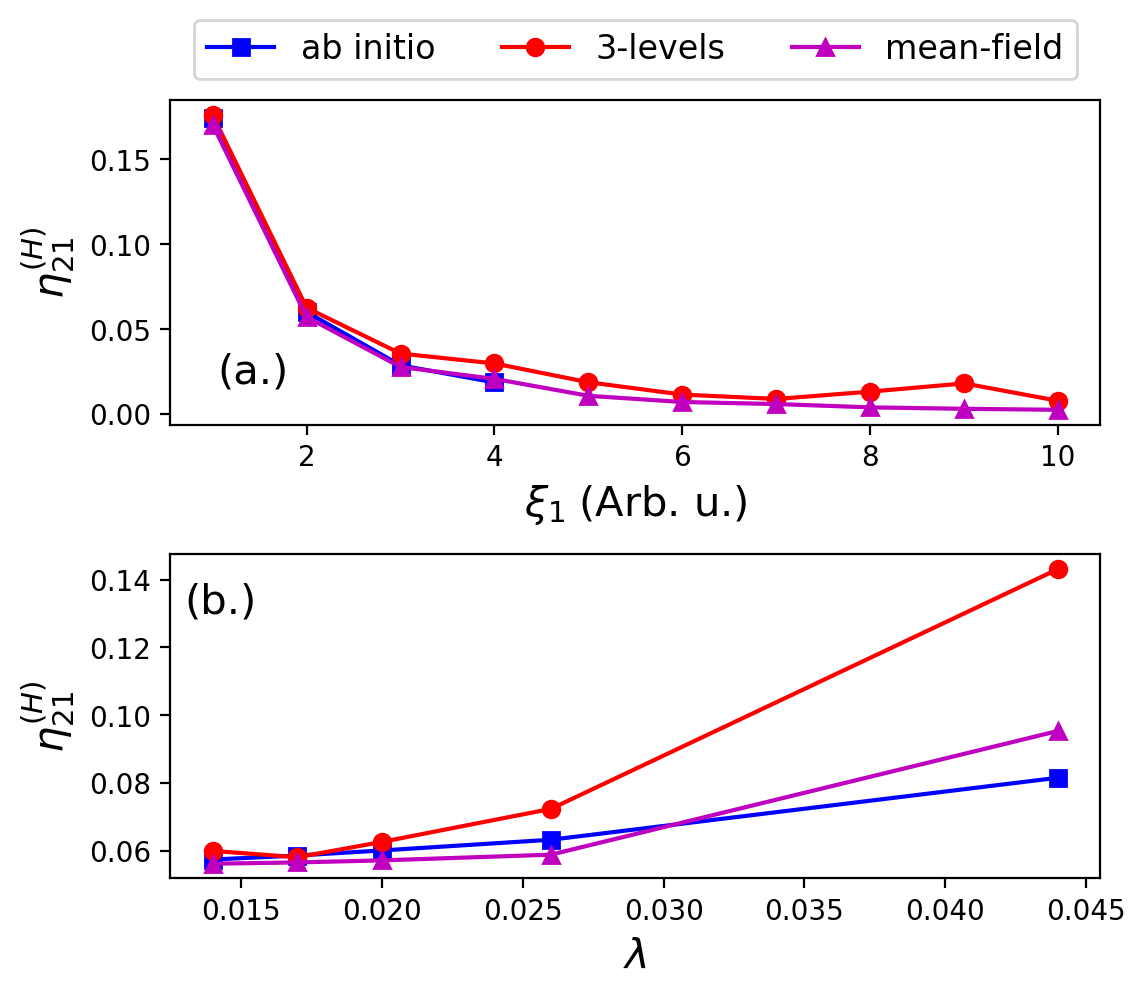}}
\caption{(a.) Comparison of the down-conversion efficiency for the 3-levels and mean-field approximations of the real-space numerically exact coupled system for increasing field amplitude $\xi_{1}$. For the exact system, we vary $\xi_{1}=1,2,3,4$ because of the large dimensionality. (b.) Increasing efficiency for increasing $\lambda$ for the 3-level and mean-field approximations of the real-space coupled system.}
\label{fig:strong-fields}
\end{figure}

Finally, we consider the influence of the coupling as well as the strength of the input field on the efficiency of the down-conversion process. It is clear that by increasing the number of photons in the input mode 1 we will generate more photons in mode 2. We therefore have to think about a better way to judge the efficiency. We choose here the ratio of the photon energy of down-converted photon to the photon energy of the pump field to judge the efficiency of the down-conversion, i.e., $\eta_{21}^{(H)} = \max(H_{2}(t)/H_{1}(t_{0}))$. In this way just increasing the input field will not automatically lead to a higher efficiency. We then first scan for a mixing angle of $\theta_{1}=60^{\circ}$, a binding potential of $V_0 =200$ meV and a coupling strength of $\lambda = 0.017$ different initial states for mode 1 by $\xi_{1}=1,2,3,...,10$ such that the the input field has $n_{1}=1,4,9,...,100$ photons at the initial time. In Fig.~\ref{fig:strong-fields} (a.) we display that by increasing the strength of the pump field through the amplitude $\xi_{1}$ the efficiency of the down-conversion decreases. Since the ab-initio simulation for such large number of photons becomes numerically very expensive, we extrapolate by the three-level and the Maxwell-Schr\"odinger approximation. As is expected, for more photons the Maxwell-Schr\"odinger theory becomes increasingly accurate and in the limit of very large photon numbers it should become exact. Since the quantum ring can only convert a finite amount of photons in the considered 40 ps, in this limit $\eta_{21}^{(H)}$ goes to zero. On the contrary, if we instead fix the amplitude of the input mode to $\xi_{1}=2$ with 4 photons, but we increase the coupling strength $\lambda=0.014,0.017,0.019,0.026,0.044$, we find (see Fig.~\ref{fig:strong-fields} (b.)) that the efficiency increases. Interestingly, both approximations overestimate the efficiency for large $\lambda$. For very large coupling strengths, it is no longer the quantum correlations that provide a higher down-conversion rate but the shear strength of the coupling terms dominates. By manipulating the photonic environment to reach the ultra-strong coupling limit, the inverse second-order harmonic generation is strongly enhanced.

\section{Summary and Outlook}
\label{sec:summary}

In this work we have highlighted how ab-initio non-relativistic quantum-electrodynamics (QED) simulations, that treat light and matter on equal footing, provide novel ways of engineering a parametric down-conversion (PDC) process. By coupling strongly to a photonic environment, hybrid light-matter states (polaritons) emerge that can efficiently mediate and control the down-conversion process. This can be done even for processes via non-dipole allowed or virtual states. We have focused on a simple yet instructive model of a GaAs quantum ring coupled to a photonic environment, e.g., an optical cavity. We have shown that for short times (on the order of several tens of picoseconds) the dissipation of a photonic bath can be ignored in our setup and we have focused on coherent simulations. Changing the coupling strength to the input and signal modes, e.g., by increasing the quality of the cavity, then allowed to control the timing of the down-conversion process and shift it to earlier times. Thus strong coupling can help to limit the detrimental effects of dissipation on the PDC process. We then highlighted that while the interpretation of the PDC process as a single-photon process becomes less straightforward in the case where we have a coherent initial state in the input mode or when we replace the input mode by a classical external pump, the main features of the down-conversion remain intact independent of the exact form of the input mode. This suggests that our results are relatively robust with respect to details of the bath, the initial states or the different pumping schemes. We have then optimized the down-conversion efficiency and the non-classicality with respect to the level-structure of the quantum ring, the coupling strength to the photonic environment and the polarization directions of its modes, and with respect to the number of input photons. We find that parallel modes to maximize the matter-mediated mode-mode interactions, strong anharmonicity to reduce the number of electronic states involved and ultra-strong coupling between light and matter maximize both targets. We find that common approximations are helpful to estimate limiting cases but often do not capture the qualitative details of the down-conversion process for strong-coupling situations.

Our results demonstrate the possibilities that become accessible with ab-initio light-matter methods in the context of engineering novel photon sources. Since these methods stay applicable from the weak to the ultra-strong coupling regime and connect few-level models to the Maxwell-Schr\"odinger picture of non-linear optics seamlessly, we can scan a wide range of parameters to find optimal conditions not only for efficiency but also for other objectives, such as the Mandel $Q$ parameter discussed here. While our simulations are still far from exact, since we do not model, for instance, the emission process or the phononic bath, we think our results demonstrate that using the full flexibility of the matter system and the photonic environment allows a detailed control of down-conversion process also in practice. We envision that using methods from polaritonic chemistry and material sciences to enhance the coupling between light and matter even at room temperature might provide new ways to access this full flexibility. Furthermore, since our description is not geared to only a two-photon down-conversion process but includes all processes, we can consider more intricate processes such as 3 or even more photon down-conversions, and can try to develop completely new ways to generate photons with specific properties on demand. With coupling strongly to many virtual states and the emergence of hybrid light-matter states at these energies not only inverse second-harmonic generation can be realized but also inverse high-harmonic generation seems feasible. This will be the subject of future work.

Since we have restricted to just a single active electron in our exploratory study, the total number of photons down-converted is relatively small. However, if we have an ensemble of quantum rings or a general many-electron system, the total number of down-converted photons can be efficiently increased. Such setups can be accessed with, for instance, QEDFT~\cite{ruggenthaler2014,ruggenthaler2015} or polaritonic coupled cluster~\cite{mordovina2020polaritonic, haugland2020coupled}, which can provide not only qualitative results but also a quantitative prediction how hybrid light-matter states can generate photons on demand. This in certain cases can even include the nanophotonic structure as part of the simulation as well as the full emission process~\cite{Jestaedt2020}. The results of this work demonstrate that the design of efficient photon sources is a very interesting working avenue for the emerging field of ab-initio light-matter interactions.


\section{Acknowledgements}
We would like to thank Frank Schlawin, Nicolas Tancogne-Dejean and Arunangshu Debnath for insightful discussions, and Sebastian Ohlmann for the help with the efficient massive parallel implementation of our code. We acknowledge financial support from the European Research Council (ERC-2015-AdG-694097) and the SFB925 "Light induced dynamics and control of correlated quantum systems". This work was supported by the Excellence Cluster "CUI: Advanced Imaging of Matter" of the Deutsche Forschungsgemeinschaft (DFG), EXC 2056, project ID 390715994.


\appendix

\section{2D semiconductor quantum ring}
~\label{app:quantum-ring}

\begin{figure}[!tbp]
\centerline{\includegraphics[width=0.5\textwidth]{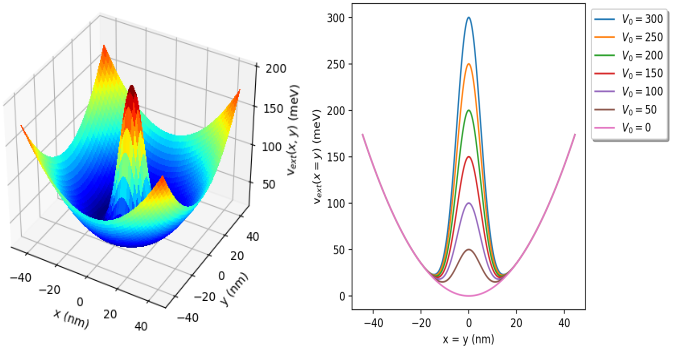}}
\caption{(a.) Real-space 2D potential of the quantum ring with potential strength parameter $V_{0}=200$ meV. (b.) Diagonal cut ($x=y$) of 2D potential showing increasing Gaussian peak for increasing $V_{0}$. Ring radius $r_{0}=44$ nm. }%
\label{fig:potential}%
\end{figure}

\begin{figure}[t] 
\centerline{\includegraphics[width=0.5\textwidth]{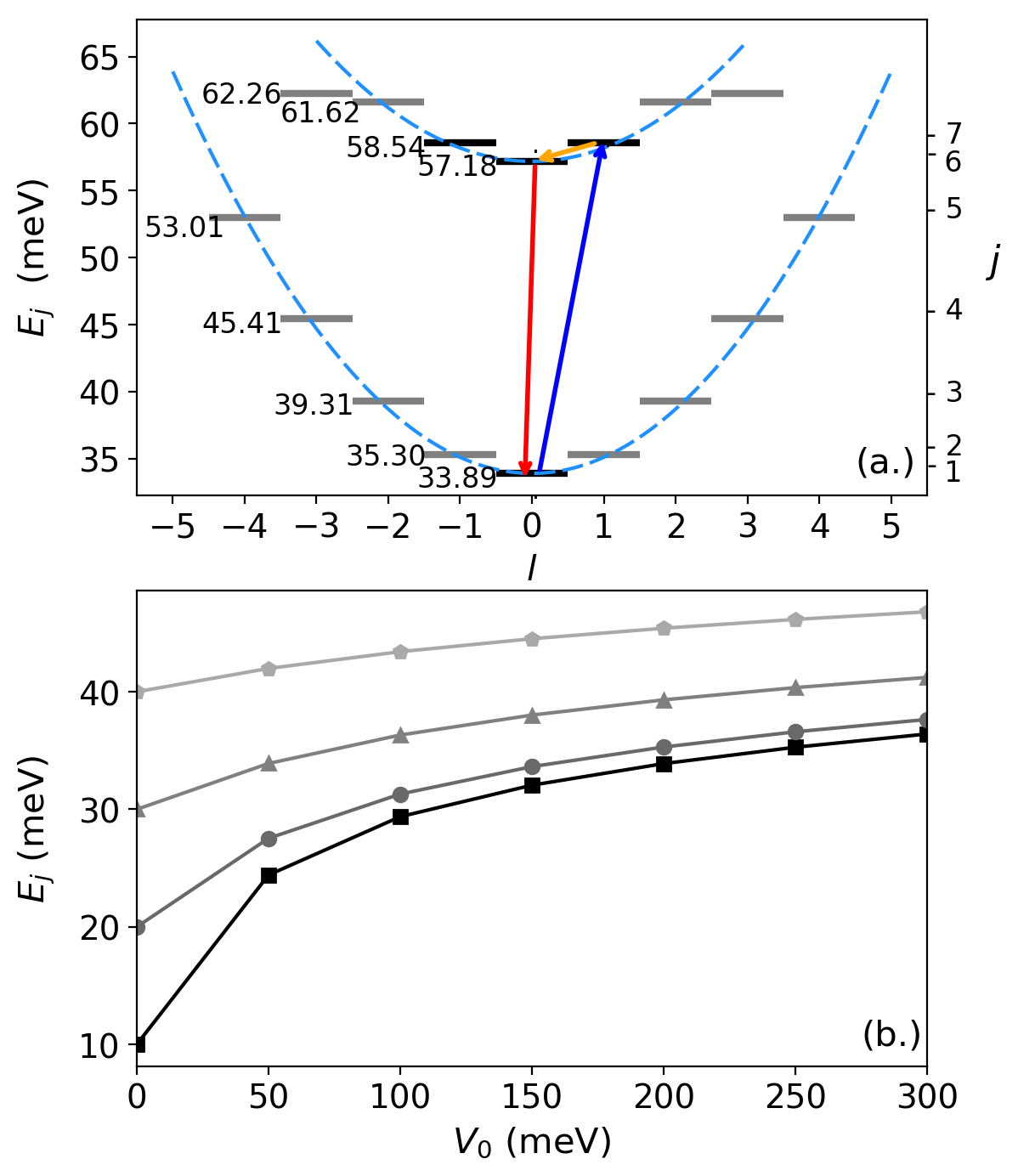}}
\caption{(a.) Energy spectrum of the 2D quantum ring showing the two lowest radial bands with degenerate and non-degenerate excited states and a non-degenerate ground-state for $V_{0}=200$ meV against angular momentum $l$. (b.) Variation of four lowest energies of (a.) for increasing potential strength parameter.}
\label{fig:spectrum-GaAs}
\end{figure}

The 2D semiconductor quantum ring of finite width features a single effective electron confined in two-dimensions in real-space ($\textbf{r} = x \textbf{e}_{x} + y \textbf{e}_{y}$). The bare electron Hamiltonian is given by
\begin{equation}
\hat{H}_{el} = -\frac{\hbar^{2}}{2m} \left(\frac{\partial^{2}}{\partial x^{2}} + \frac{\partial^{2}}{\partial y^{2}}\right)  + \frac{1}{2}m\omega_{0}^{2}\textbf{r}^{2} + V_{0} e^{-\textbf{r}^{2}/d^{2}} , \label{electron-hamiltonian}
\end{equation}
where the potential part with $\textbf{r}^{2} = x^{2} + y^{2}$ introduces a parabolic confinement and a Gaussian peak located at the center as shown in Fig.~(\ref{fig:potential}.a). The parameters of the potential are chosen in order to reflect the energy and length scales of quantum ring experiments~\cite{fuhrer2001,ihn2005} as $\hbar \omega_{0} = 10$ meV, $d = 10$ nm, $m = 0.067 m_{e}$, and $V_{0} =200$ meV (unless otherwise stated). The potential strength $V_{0}$ in meV can be varied and introduces the anharmonicity (nonlinearity) in the electronic system as in Fig.~(\ref{fig:potential}.b). This is particularly important because increasing $V_{0}$ increases the transition dipole amplitudes (see  Tab.~(\ref{tab:GaAs-dipole-matrix-elements}.b)) which leads to a stronger coupling since the coupling parameter is proportional to the transition dipoles as discussed in App.~(\ref{app:few-levels}).  For $V_{0} =0$ meV, Eq.~(\ref{electron-hamiltonian}) reduces to a two-dimensional isotropic harmonic oscillator with energies $E_{n} = \hbar\omega_{0}(2n +1)$ where $n=0,1,2,3,... \,$ and the degeneracy in energy is $(n+1)$. For $V_{0} >0$ meV (and we choose $V_{0}=200$ meV unless otherwise stated) the eigenstates $\varphi_{j}^{l}$ are labeled by the angular momentum $l=0,\pm 1, \pm 2, \pm 3, ... \,$ and the index $j=|l|+1$ enumerates over the energy levels (see Fig.~(\ref{fig:spectrum-GaAs}a)).  The ground state and excited state with $l=0$ are singlets, whereas the excited states with finite angular momentum differing from $l=0$ are doubly degenerate as shown in Fig.~(\ref{fig:spectrum-GaAs}a). Dipole-allowed transitions occur only between states with consecutive angular momenta~\cite{rasanen2007,flick2015}. For example, in the first radial band the allowed transitions are  $\varphi_{1}^{0}\leftrightarrow\varphi_{2}^{1,-1}\leftrightarrow\varphi_{3}^{2,-2} \leftrightarrow\varphi_{4}^{3,-3} \leftrightarrow\varphi_{5}^{4,-4}$ and for the two lowest radial bands, some of the allowed transitions are $\varphi_{1}^{0}\leftrightarrow\varphi_{2}^{1,-1}\leftrightarrow\varphi_{6}^{0} \leftrightarrow\varphi_{7}^{1,-1}$.  Making use of the dielectric constant $\epsilon=12.7\epsilon_{0}$, we work in scaled effective atomic units as in Ref.~\cite{rasanen2007} by defining effective units as $\text{Ha}^{*}=(m/\epsilon^{2})\text{Ha} \approx 11.30$ meV, $a_{B}^{*}=(m/\epsilon)a_{0}\approx10.03$ nm, and $u_{t}^{*}=\hbar/\text{Ha}^{*}\approx 58.23$ fs.

The choice of the quantum ring as the down-conversion medium is motivated by two properties. First, since it is a variable atom-like system, the electronic spectrum can be altered which is usually realized experimentally~\cite{fomin2018}. For example, the parabolic confinement potential of a quantum ring can be altered which changes the geometric properties by reducing the ring width (see Fig.~(\ref{fig:potential}.b)) thereby changing the electronic spectra as shown in Fig.~(\ref{fig:spectrum-GaAs}b). Secondly, transitions between states with the same angular momentum (i.e. $\varphi_{j}^{-l}$ to $\varphi_{j}^{l}$ and vice versa) have been shown to be accessible by driving the transitions with a coherent laser field~\cite{rasanen2007} breaking the inversion symmetry of the quantum ring. Here, we take another route by considering the Pauli-Fierz Hamiltonian in which the electronic subsystem is the GaAs quantum ring and show that the down-conversion is possible between non-dipole allowed transitions since the photon modes are quantized and treated inclusively as dynamical contributions of the coupled system. Specifically, photon emission into mode 3 in the cascaded process is possible for coupling resonantly to the non-dipole allowed transition $\varphi_{1}^{0}\leftrightarrow\varphi_{6}^{0}$. The transition dipoles in effective atomic units for the electronic states considered in this work are given in Tab.~(\ref{tab:GaAs-dipole-matrix-elements}.a).
\begin{table}[bth]
\begin{tabular}{ | c | c | }
	\hline
	{Transitions} & {Amplitudes}   \\\hline \hline
	$\langle\psi_{1}^{0}|\hat{x}|\psi_{7}^{1}\rangle$ & 0.2077 \\\hline
	$\langle\psi_{6}^{0}|\hat{x}|\psi_{7}^{1}\rangle$ & 1.2786 \\\hline
	$\langle\psi_{1}^{0}|\hat{x}|\psi_{2}^{1}\rangle$ & 1.0867 \\\hline
	$\langle\psi_{1}^{0}|\hat{x}|\psi_{6}^{0}\rangle$ & $4.0090\times 10^{-13}$ \\\hline
\end{tabular}
	\hfill
\begin{tabular}{ | c | c |  }
	\hline
	$V_{0}$ (meV) &  $\langle\psi_{1}^{0}|\hat{x}|\psi_{2}^{1}\rangle$  \\\hline \hline
	0 &  0.53159199 \\\hline
	50 &  0.84520553 \\\hline
	100 &  0.97645376  \\\hline
	150 &  1.04263107 \\\hline
	200 &  1.08705932  \\\hline
	250 &  1.11987106 \\\hline
	300 &  1.14420501  \\\hline
\end{tabular}
\caption{Left: The $x$-component of the transition dipole matrix elements for selected transitions shown in Fig.~(\ref{fig:potential}.a) for $V_{0}=200$ meV between the various states. Right: Increasing $x$-component of the dipole transition amplitudes between the ground-state and first degenerate excited states for increasing potential strength parameter $V_{0}$.}
\label{tab:GaAs-dipole-matrix-elements}
\end{table}

\section{External perturbation for down-conversion}
~\label{app:external-pump}

We consider the case in which the photons to be down-converted come from an external classical current that perturbs the pump mode of the coupled system. This is the case in Eq.~(\ref{external-cPDC}) where an external field $j_{1}(t)$ couples to the vector potential $\hat{A}_{1}$ of mode 1. This field injects photons into the system that interact and are down-converted into the vacant modes. On the other hand, we consider an external field that couples to the electrons, thus inducing down-conversion by driving the electronic system. This requires the pump field $\hat{A}_{1}$ in Eq.~(\ref{cspdc-hamiltonian}) to be an external classical field $A_{1}(t) = \lambda_{1}q_{1}(t)$. In order to compute the classical field $q_{1}(t)$, we consider that mode 1 is driven by an external classical field $j_{1}(t)$ given by
\begin{align}
\hat{H}_{1}(t) =  \frac{1}{2}  \left(\hat{p}_{1}^{2} + \omega_{1}^{2}\hat{q}_{1}^{2}  \right) + \hat{A}_{1}\cdot j_{1}(t) . \label{driven-HO}
\end{align}
The mode-resolved classical equation of motion of the photon coordinate $\hat{q}_{1}$  is given by $
\left(\partial^{2}/\partial t^{2} + \omega_{1}^{2}\right)q_{1}(t) =  -\lambda_{1} j_{1}(t)$, of which the solution of the mode-resolved Maxwell equation is given as
\begin{align}
q_{1}(t) &=  -\int_{0}^{t} dt' \frac{\lambda_{1}}{\omega_{1}}\sin\left(\omega_{1}(t-t') \right) j_{1}(t') \nonumber \\
& \quad + q_{1}^{(0)}\cos(\omega_{1}t) + \frac{\dot{q}_{1}^{(0)}}{\omega_{1}}\sin(\omega_{1}t) . \label{solution}
\end{align}
With the solution of the classical trajectory $q_{1}(t)$, the time-dependent external pump of Eq.(\ref{external-PDC}) is given by
\begin{align}
\hat{H}_{ext}(t) &= - \frac{e}{m}A_{1}(t)\;\hat{p}_{x} \nonumber \\
& + \frac{e^{2}}{2m}\left(A_{1}^{2}(t) - 2A_{1}(t)\hat{A}_{2}\sin\theta_{2} - 2A_{1}(t)\hat{A}_{3}\cos\theta_{3} \right) . \label{spdc-pump}
\end{align}
where $A_{1}(t) = \lambda_{1}q_{1}(t)$ and the down-conversion Hamiltonian is
\begin{align}
\hat{H}_{S}' &= \hat{H}_{el} + \hat{H}_{2} + \hat{H}_{3}  \nonumber\\
& \quad  - \frac{e}{m}\left[\hat{A}_{2}\left(-\hat{p}_{x}\sin\theta_{2} + \hat{p}_{y}\cos\theta_{2}\right) \right. \nonumber \\
& \left. \qquad\qquad   + \hat{A}_{3}\left(\hat{p}_{x}\sin\theta_{3} + \hat{p}_{y}\cos\theta_{3}\right)\right]  \nonumber\\
& \quad + \frac{e^{2}}{2m}\left[ \hat{A}_{2}^{2} + \hat{A}_{3}^{2}  + 2\hat{A}_{3}\hat{A}_{2}\cos(\theta_{2}+\theta_{3})\right] . \label{spdc-hamiltonian}
\end{align}
In Fig.~(\ref{fig:ini_vs_ext_ns}) for strong coupling, we show the comparison for the evolution dynamics of the down-conversion with an external current $j_{1}(t)$ (as in Subsec.~(\ref{sec:methods})) and with an initial factorizable state as in Sec.~(\ref{sec:ab-initio}). At $t=0.23$ ps, the external pump $j_{1}(t)$ injects $n_{1}=4$ photons and this qualitatively correspond to the case of an initial state which at the initial time have $n_{1}=4$ photons. The profiles for the photon occupation are qualitatively the same for the evolved time. The apparent slight deviation in the profiles is attributed to the values of the observables at the initial time. For example, at the initial time, $n_{2}=0$ for the initial factorizable state and $n_{2}=0.0018$ for the external current with a correlated ground-state. Therefore, the profiles for the evolved observable quantities are independent of the initial state chosen.

\section{Approximations of the electron-photon coupled system}

\subsection{Few level approximation}
~\label{app:few-levels}

The few-level approximation of the real-space electron-photon coupled system approximates Eq.~(\ref{cspdc-hamiltonian-SB}) to
\begin{align}
\hat{H}_{\text{FL}} &= \sum_{i=1}^{N}E_{i}|\varphi_{i}\rangle\langle\varphi_{i}| + \hat{H}_{1} + \hat{H}_{2} + \hat{H}_{3} \nonumber\\
& \quad - \frac{e}{m}\sum_{i,j}^{N}\left[ \hat{A}_{2}\left(-\langle\hat{p}_{x}\rangle_{ij} \cos\theta_{2} + \langle\hat{p}_{y}\rangle_{ij} \sin\theta_{2} \right) \right.  \nonumber\\
& \left. \qquad\qquad\qquad + \hat{A}_{3}\left(\langle\hat{p}_{x}\rangle_{ij} \sin\theta_{3} + \langle\hat{p}_{y}\rangle_{ij} \cos\theta_{3} \right) \right.  \nonumber\\
& \left. \qquad\qquad\qquad + \hat{A}_{1}\langle\hat{p}_{x}\rangle_{ij} \right]|\varphi_{i}\rangle\langle\varphi_{j}|  \nonumber\\
& \quad + \frac{e^{2}}{2m}\left[\hat{A}_{1}^{2} + \hat{A}_{2}^{2} + \hat{A}_{3}^{2}  -2\hat{A}_{2}\hat{A}_{1}\sin(\theta_{2})  \right. \nonumber\\ 
& \left. \qquad\qquad\quad - 2\hat{A}_{3}\hat{A}_{1} \sin(\theta_{3}) + 2\hat{A}_{3}\hat{A}_{2}\cos(\theta_{2}+\theta_{3})\right]  , \nonumber
\end{align}
where $\langle\hat{p}_{x}\rangle_{ij} = \langle\varphi_{i}|\hat{p}_{x}|\varphi_{j}\rangle$, $\langle\hat{p}_{y}\rangle_{ij} = \langle\varphi_{i}|\hat{p}_{y}|\varphi_{j}\rangle$ represent the transition momenta matrix elements between relevant electronic states of the coupling to the photons. This approximation assumes that the few lowest energy levels of the electronic spectrum are separated in energy from the higher lying energy levels. Assuming this approximation holds, we truncate the Hilbert space to the $N-$lowest energy levels of interest~\cite{stefano2019}. Due to the degeneracy in the electronic spectrum, for the non-degenerate down-conversion we considered the four electronic states $\varphi_{1}^{0}, \varphi_{6}^{0}, \varphi_{7}^{1,-1} $ and for the degenerate down-conversion we considered the three electronic states $\varphi_{1}^{0},  \varphi_{2}^{1,-1} $.

From Eq.~(\ref{cspdc-hamiltonian}), we deduce the electron-photon coupling from the bilinear term which can be written as $\hat{H}_{int}=-\frac{e}{m^{*}}\sum_{i,j}^{N}\sum_{\alpha}^{M}\boldsymbol{\lambda}_{\alpha}\hat{q}_{\alpha}\langle\left(\textbf{e}_{\alpha}\cdot\hat{\textbf{p}}\right)\rangle_{ij}|\varphi_{i}\rangle\langle\varphi_{j}|$. The interaction term can be expressed as $\hat{H}_{int} =i\sum_{i,j}^{N} \sum_{\alpha}^{M} g_{\alpha}^{(ij)} \left(\hat{a}_{\alpha}+\hat{a}_{\alpha}^{\dagger}\right)$ where the coupling term is $g_{\alpha}^{(ij)} =\frac{\omega_{ij}}{\omega_{\alpha}} \left(\hbar\omega_{\alpha} /\epsilon_{0}L\right)^{1/2}d_{ij}$ and the dipole matrix elements is $d_{ij}= \langle\left(\textbf{e}_{\alpha}\cdot\hat{\textbf{r}}\right)\rangle_{ij}$ which is related to the momentum matrix elements by $\langle\left(\textbf{e}_{\alpha}\cdot\hat{\textbf{p}}\right)\rangle_{ij}=i\, \omega_{ij} \, m \, d_{ij}$. For free space radiation, the electron photon coupling is related to the radiative decay rate as $g_{\alpha}^{(ij)}= \left(\frac{3\hbar^{2}c^{3}}{\omega_{\alpha}\omega_{ij}\epsilon_{0}L}\gamma_{ij}\right)^{1/2}$. Here we used the relation between the radiative decay rate and the transition dipole matrix elements $\gamma_{ij}=\frac{\omega_{ij}^{3} |d_{ij}|^{2}}{3\pi\epsilon_{0}c^{3}} \,$.

\subsection{Maxwell-Schr\"{o}dinger approximation}
~\label{app:maxwell-schroedinger}

\begin{figure}[t] 
	\centerline{\includegraphics[width=0.5\textwidth]{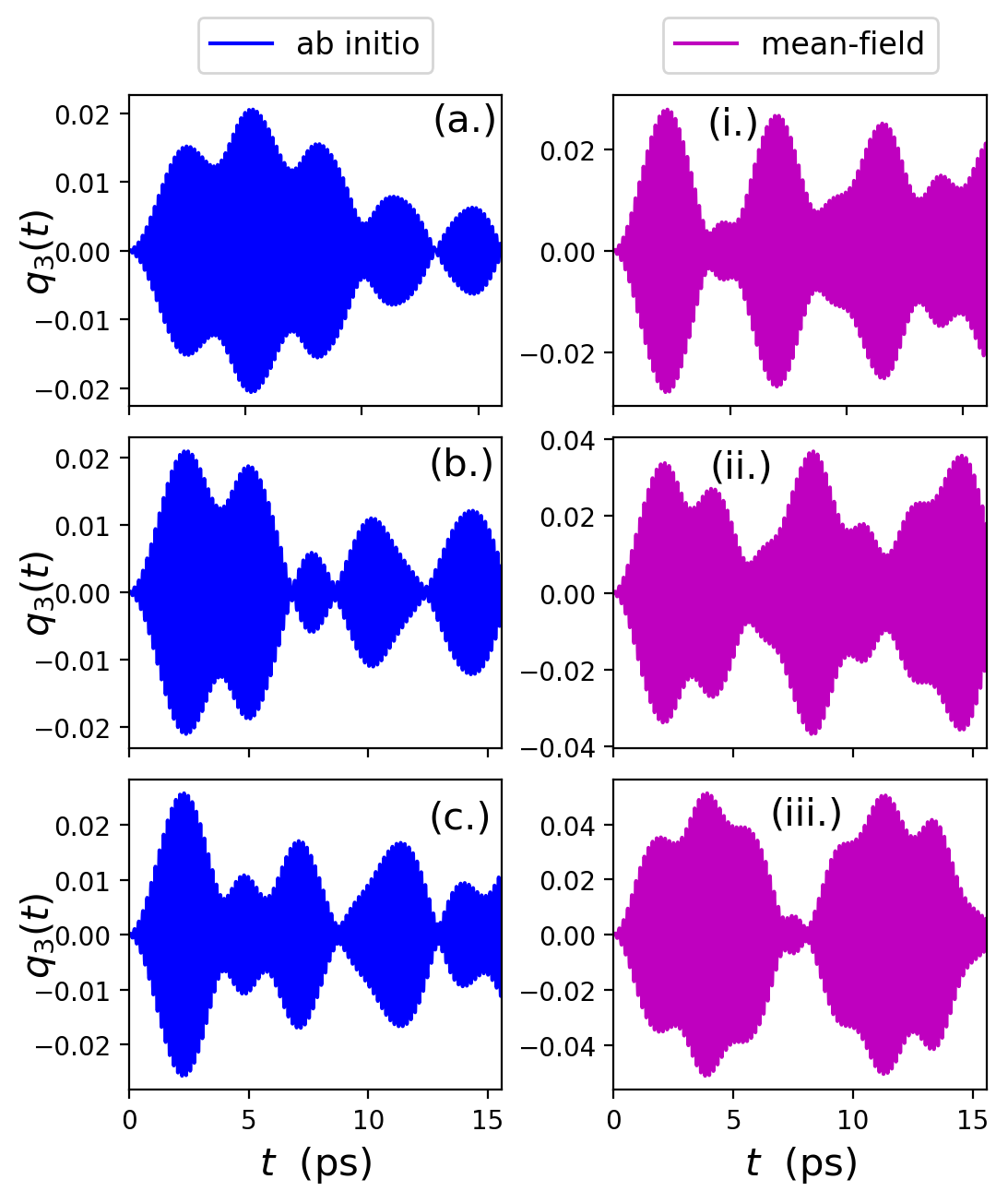}}
	\caption{Comparison of the field $q_{3}(t)$ for both real-space and mean-field.  The comparisons (a.) \& (i), (b.) \& (ii), (c.) \& (iii) are respectively the cases where the amplitude of the pump field is $\xi_{1}=2,3,4$ with respective photon number $n_{1}=4,9,16$ at the initial time. The amplitude of $q_{3}(t)$ for the mean-field increases over the real-space for increasing $\xi_{1}$.}
	\label{fig:photon-q-1}
\end{figure}

\begin{figure}[t] 
	\centerline{\includegraphics[width=0.5\textwidth]{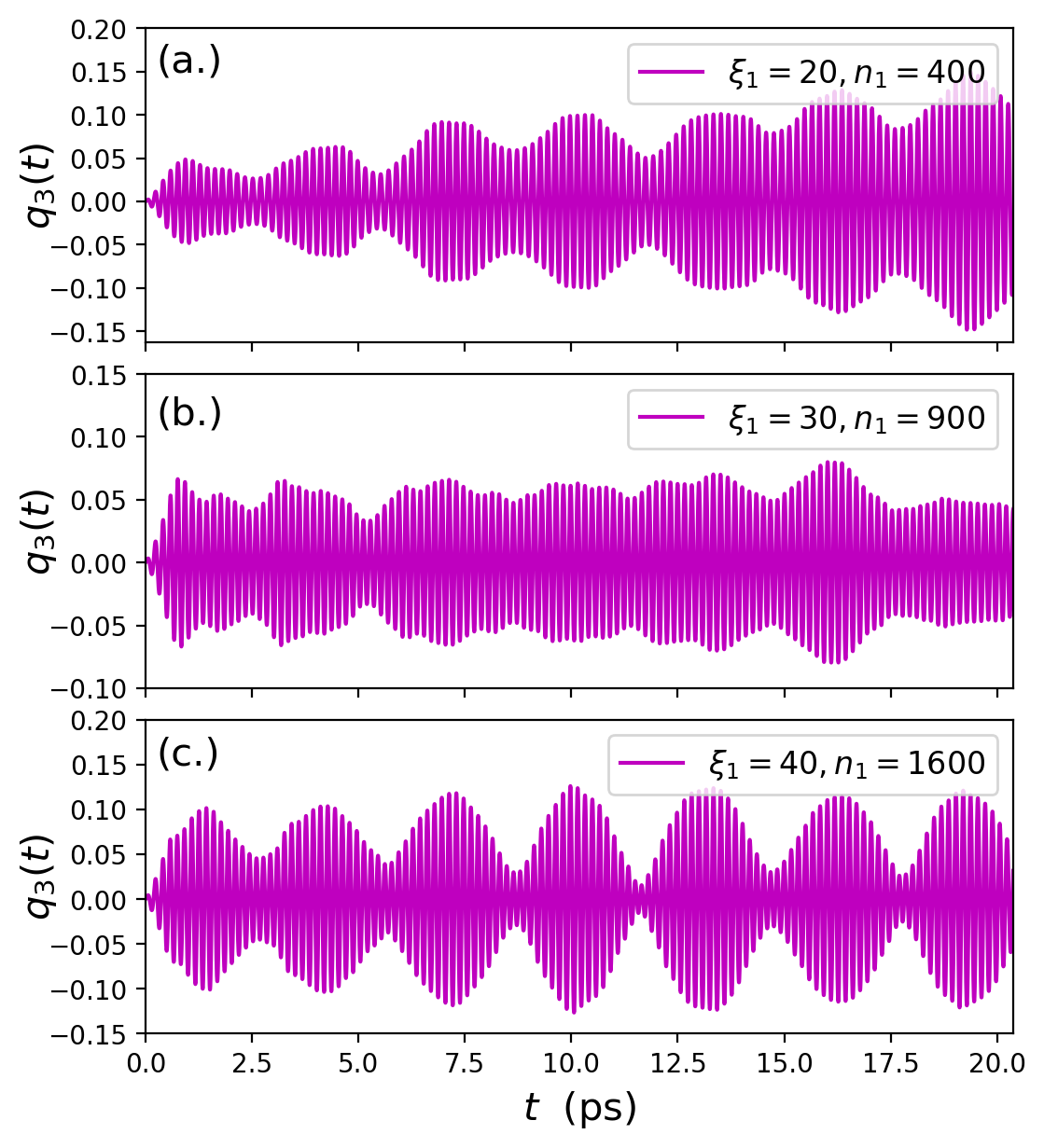}}
	\caption{The photon field $q_{3}(t)$ for mode 3 for different amplitude $\xi_{1}$ of the coherent pump field with corresponding photon number $n_{1}=|\xi_{1}|^{2}$. The more intense the pump field is, results to a slight increase in the amplitude of $q_{3}(t)$ for time-evolution shown. The time-step is $\Delta t = 0.0058$ fs. }
	\label{fig:photon-q-2}
\end{figure}

By making a mean-field ansatz of the coupled electron-photon system, we can approximate the correlated electron-photon wavefunction as $\Psi(\textbf{r},q_{\alpha}) \approx \varphi(\textbf{r})\otimes\phi(q_{1})\otimes\phi(q_{2})\otimes...\otimes\phi(q_{\alpha})$~\cite{ruggenthaler2017b}. This ansatz simplifies the dynamics of the time-evolved correlated problem of Eqs.~(\ref{schroed}) and (\ref{maxwell}) where the self-consistent Hamiltonian and currents are 
\begin{align}
\hat{H}_{\text{MS}}([q_{\alpha}];t) &= \hat{H}_{el}  - \frac{e\lambda_{1}}{m^{*}}q_{1}(t)\hat{p}_{x} \nonumber\\
& \quad  - \frac{e}{m^{*}}\left[\lambda_{2}q_{2}(t)\left(-\hat{p}_{x}\sin\theta_{2} + \hat{p}_{y}\cos\theta_{2}\right) \right. \nonumber \\
& \left. \qquad\qquad   + \lambda_{3}q_{3}(t)\left(\hat{p}_{x}\sin\theta_{3} + \hat{p}_{y}\cos\theta_{3}\right)\right]   . \label{MS-hamiltonian}\\
j_{1}([\textbf{p},q_{\alpha}]) &=  \frac{e\lambda_{1}}{m^{*}}p_{x}(t) \nonumber\\
& \quad + \frac{e^{2}}{m^{*}}\left[-\lambda_{1}^{2}q_{1}(t) + \lambda_{2}\lambda_{1}q_{2}(t)\sin\theta_{2}  \right. \nonumber\\ 
& \left. \qquad\qquad + \lambda_{3}\lambda_{1}q_{3}(t) \sin\theta_{3} \right] . \label{j1-current} \\
j_{2}([\textbf{p},q_{\alpha}]) &=  \frac{e\lambda_{2}}{m^{*}}(-p_{x}(t)\sin\theta_{2} + p_{y}(t)\cos\theta_{2}) \nonumber\\
& \quad + \frac{e^{2}}{m^{*}}\left[-\lambda_{2}^{2}q_{2}(t) + \lambda_{2}\lambda_{1}q_{1}(t)\sin\theta_{2}  \right. \nonumber\\ 
& \left. \qquad\qquad - \lambda_{2}\lambda_{3}q_{3}(t) \cos(\theta_{2}+\theta_{3}) \right] . \label{j2-current} \\
j_{3}([\textbf{p},q_{\alpha}]) &=  \frac{e\lambda_{3}}{m^{*}}(p_{x}(t)\sin\theta_{3} + p_{y}(t)\cos\theta_{3}) \nonumber\\
& \quad + \frac{e^{2}}{m^{*}}\left[-\lambda_{3}^{2}q_{3}(t) + \lambda_{3}\lambda_{1}q_{1}(t)\sin\theta_{2}  \right. \nonumber\\ 
& \left. \qquad\qquad - \lambda_{3}\lambda_{2}q_{2}(t) \cos(\theta_{2}+\theta_{3}) \right] . \label{j3-current}
\end{align}
Here, the $q_{\alpha}$'s are the photon coordinates of the respective modes $\alpha=1,2,3$ and $\textbf{p}=(p_{x},p_{y})$ are the x- and y-momenta of the electron. 

We compute the photon occupation by first computing the Hamiltonians of the photonic subsystem $H_{\alpha} =\frac{1}{2}\left(p_{\alpha}^{2}+\omega_{\alpha}^{2}q_{\alpha}^{2}\right)+\frac{\hbar\omega_{\alpha}}{2}$, where we added the zero-point energy to account for the energy shift which does not appear in the mean-field treatment. From the photonic Hamiltonian, we deduce the photon occupation to be $n_{\alpha}=H_{\alpha}/\hbar\omega_{\alpha}-0.5$. Equally, we determine the photonic observable of Eq.~(\ref{mandel-q}) by defining a mean-field creation and annihilation operators
\begin{align}
a_{\alpha} = \frac{1}{\sqrt{2\hbar\omega_{\alpha}}}\left(\omega_{\alpha}q_{\alpha}+ip_{\alpha}\right), \nonumber\\
a_{\alpha}^{\dagger} = \frac{1}{\sqrt{2\hbar\omega_{\alpha}}}\left(\omega_{\alpha}q_{\alpha}-ip_{\alpha}\right). \nonumber
\end{align}
For the choice of the initial states in Sec.~(\ref{subsec:evolution}), the initial values of the photon coordinate are $q_{1} =\langle\xi_{1}|\hat{q}_{1}|\xi_{1}\rangle=|\xi_{1}|\sqrt{2\hbar/\omega_{1}}$, $q_{2} =\langle0_{2}|\hat{q}_{2}|0_{2}\rangle=0$, $q_{3} =\langle0_{3}|\hat{q}_{3}|0_{3}\rangle=0$ and momenta are $p_{1} =\langle\xi_{1}|\hat{p}_{1}|\xi_{1}\rangle=|\xi_{1}|\sqrt{2\hbar\omega_{1}}$, $p_{2} =\langle0_{2}|\hat{p}_{2}|0_{2}\rangle=0$, $p_{3} =\langle0_{3}|\hat{p}_{3}|0_{3}\rangle=0$.

In Fig.~(\ref{fig:photon-q-1}) we show a comparison between the Maxwell-Schr\"{o}dinger approximation (mean-field) and the real-space result for the non-degenerate case in weak coupling for the generated field $q_{3}(t)$ of mode 3 when the pump field have different amplitudes $\xi_{1}=2,3,4$ with respective photon numbers $n_{1}=|\xi_{1}|^{2}=4,9,16$ at the initial time. Considering the maximum amplitude of $q_{3}(t)$ of the evolution, the mean-field is larger than the real-space for increasing amplitude of the pump field. Also, in Fig.~(\ref{fig:photon-q-2}) we show for the mean-field case that by increasing the amplitude of the pump mode by an order of magnitude, i.e., $\xi_{1}=20,30,40$, the amplitude of $q_{3}(t)$ slightly increases. This is in accord with non-linear optics which for intense fields (large number of photons), the down-conversion have a low efficiency~\cite{burnham1970}.

\section{Numerical details}
~\label{app:numerical-details}

We outline details of the representation of the photon subspace in Fock number basis for the different input fields and the description of the many mode case including a restricted sampled photon bath. The numerical details follows chronologically as in the successive sections in the paper:
\begin{itemize}
	\item dissipation and coherence in Sec.~\ref{sec:dissipation-and-coherence}: for each of the modes 1, 2, 3, we included three photon Fock states which are the vacuum, one-photon and two-photon states. In order to be able to treat the photon bath consisting of $M-3=M_{70}=70$ modes numerically exactly, we truncate the Fock space and consider only the vacuum state, the $M_{70}$ one-photon states, and the $(M_{70}^{2}+M_{70})/2$ two-photon states as in Ref.~\cite{flick2017}. For only this case, we consider all the 12 electronic states in Fig.~(\ref{fig:spectrum-GaAs}a) up to the state with energy $58.54$ meV.
	
	\item temporal control Sec.~\ref{sec:input-fock-state}: we sample 20 photon Fock states for each of the modes 1, 2, and 3.
	
	\item in Secs.~\ref{sec:methods} till the end, we sample 30 photon Fock states for the individual modes 1, 2, and 3. The choice of 30 photon Fock states is to well represent the coherent state.
\end{itemize}

\section{Density matrix of the coupled system}
~\label{app:density-matrix}

We quantify features of entanglement in the electron-photon coupled system by defining appropriate one-body reduced density matrices~\cite{schaefer2019,buchholz2019} for the individual subsystems
\begin{align}
\gamma_{\text{M}}(\textbf{r},\textbf{r}') &= \iiint dq_{1}dq_{2}dq_{3}\Psi(\textbf{r},q_{1},q_{2},q_{3})\Psi^{\ast}(\textbf{r}',q_{1},q_{2},q_{3})  , \nonumber\\
\gamma_{1}(q_{1},q_{1}') &= \iiint d\textbf{r}dq_{2}dq_{3}\Psi(\textbf{r},q_{1},q_{2},q_{3})\Psi^{\ast}(\textbf{r},q_{1}',q_{2},q_{3})  ,\nonumber\\
\gamma_{2}(q_{2},q_{2}') &= \iiint d\textbf{r}dq_{1}dq_{3}\Psi(\textbf{r},q_{1},q_{2},q_{3})\Psi^{\ast}(\textbf{r},q_{1},q_{2}',q_{3})  ,\nonumber\\
\gamma_{3}(q_{3},q_{3}') &= \iiint d\textbf{r}dq_{1}dq_{2}\Psi(\textbf{r},q_{1},q_{2},q_{3})\Psi^{\ast}(\textbf{r},q_{1},q_{2},q_{3}')  .\nonumber
\end{align}
Where $\gamma_{\text{M}}(\textbf{r},\textbf{r}')$, $\gamma_{1}(q_{1},q_{1}')$, $\gamma_{2}(q_{2},q_{2}')$, $\gamma_{3}(q_{3},q_{3}')$ are respectively the one-body reduced density matrices of the electronic and photonic subsystems. We choose a normalization of these reduced density matrices to one, such that the following holds
\begin{align}
\textrm{Tr}(|\Psi\rangle\langle\Psi^{\ast}|) = \textrm{Tr}(\gamma_{\text{M}}) = \textrm{Tr}(\gamma_{1}) = \textrm{Tr}(\gamma_{2}) = \textrm{Tr}(\gamma_{3}) = 1. \nonumber
\end{align}
From this normalization, the reduced density matrices can be used to compute the purity of the subsystems by requiring
\begin{align}
\textrm{Tr}(\gamma_{\text{M}}^{2}) = \textrm{Tr}(\gamma_{1}^{2}) = \textrm{Tr}(\gamma_{2}^{2}) = \textrm{Tr}(\gamma_{3}^{2}) = 1. \label{norm-1RDM}
\end{align}
We consider only the evolution of the purity of the photonic subsystems.

\section{Fock state occupations for different input fields}
~\label{app:fock-coherent-occup}

\begin{figure}[bth]
	\centerline{\includegraphics[width=0.5\textwidth]{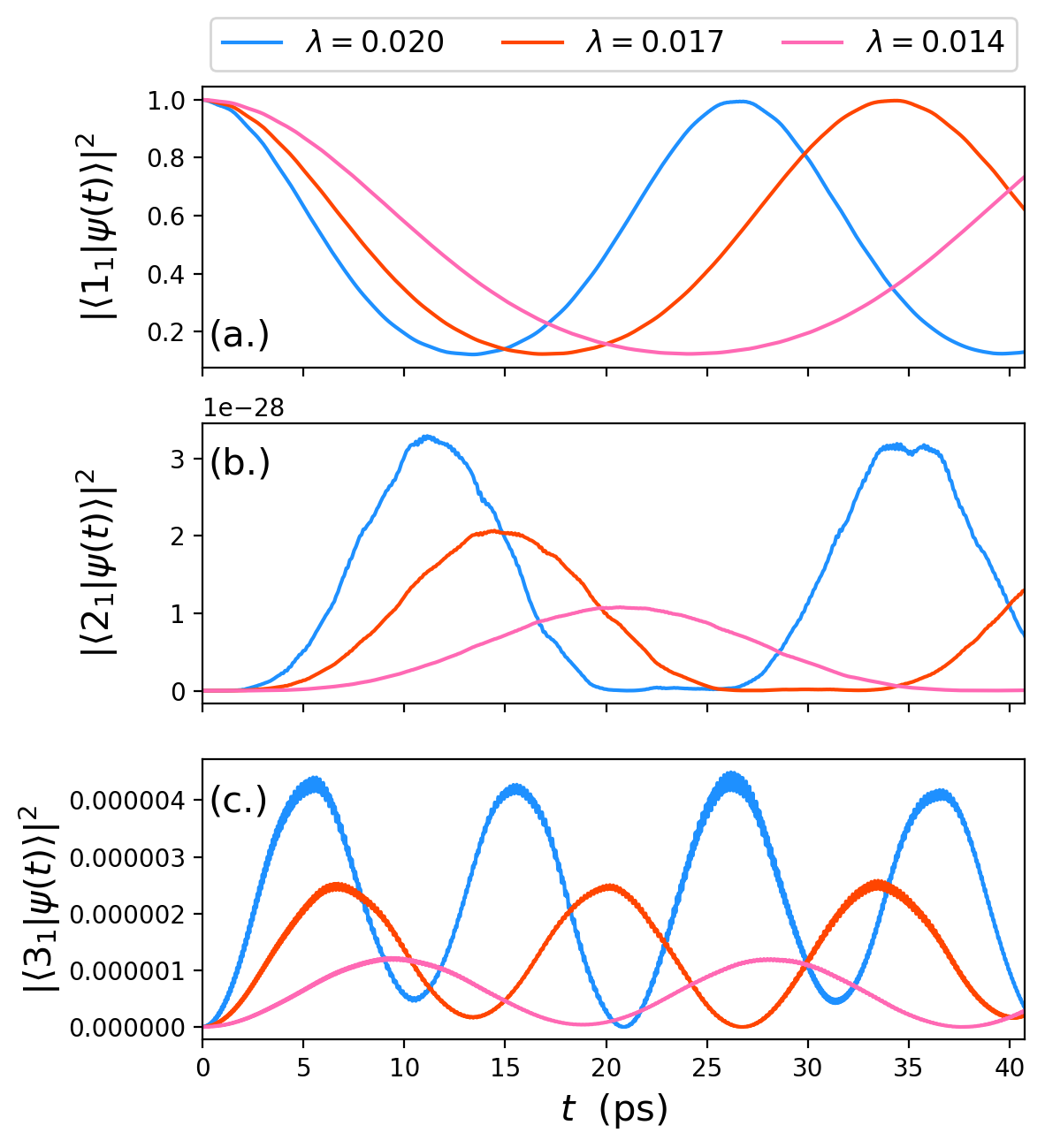}}
	\caption{Real-time Fock state occupations of the pump mode 1 in a single-photon Fock state from weak to ultra-strong coupling.}
	\label{fig:fock-occupation-mode-1}
\end{figure}
\begin{figure}[bth]
	\centerline{\includegraphics[width=0.5\textwidth]{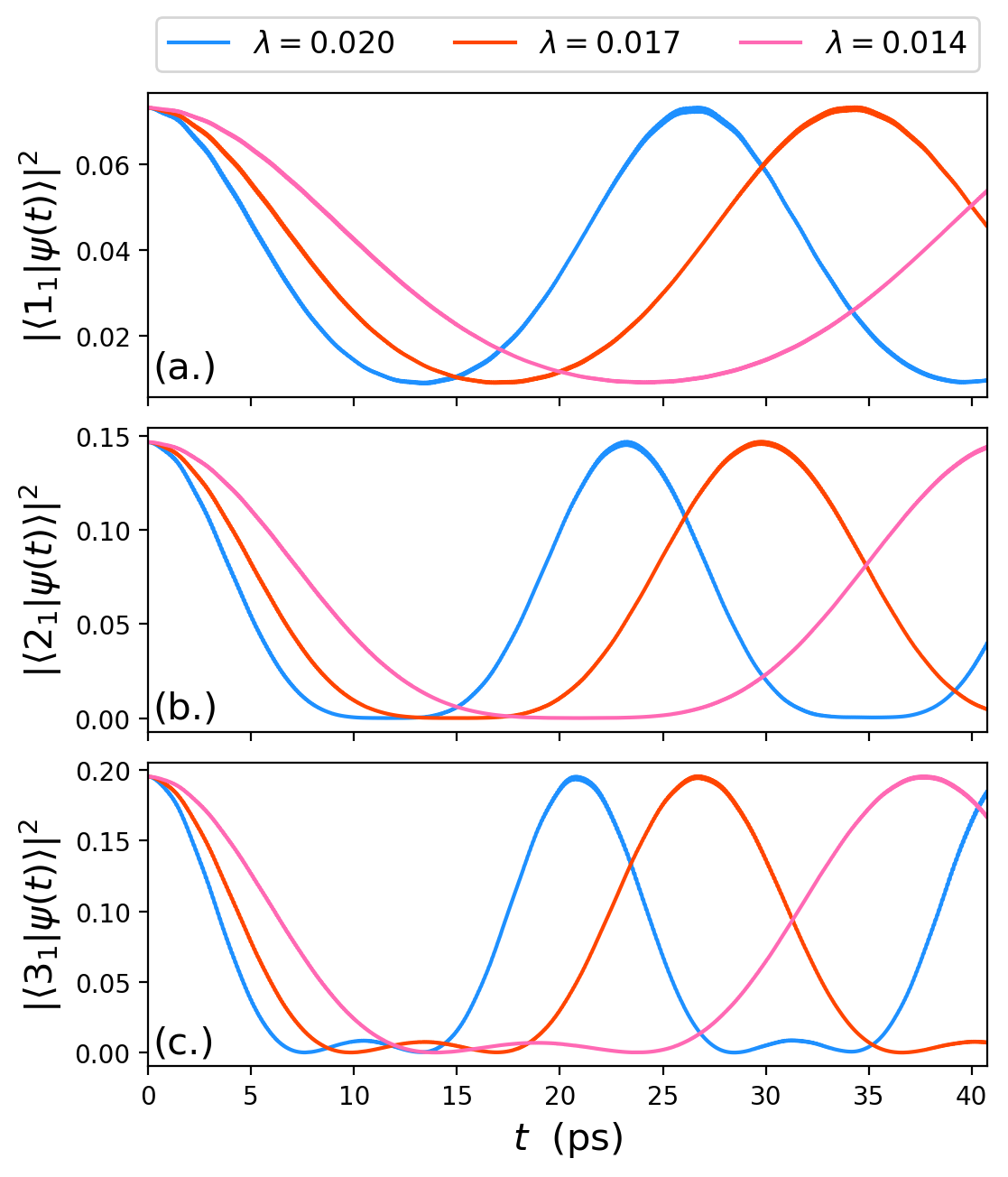}}
	\caption{Real-time Fock state occupations of the pump mode 1 in a coherent state from weak to ultra-strong coupling.}
	\label{fig:coherent-fock-occupation-mode-1}
\end{figure}
\begin{figure}[t] 
	\centerline{\includegraphics[width=0.5\textwidth]{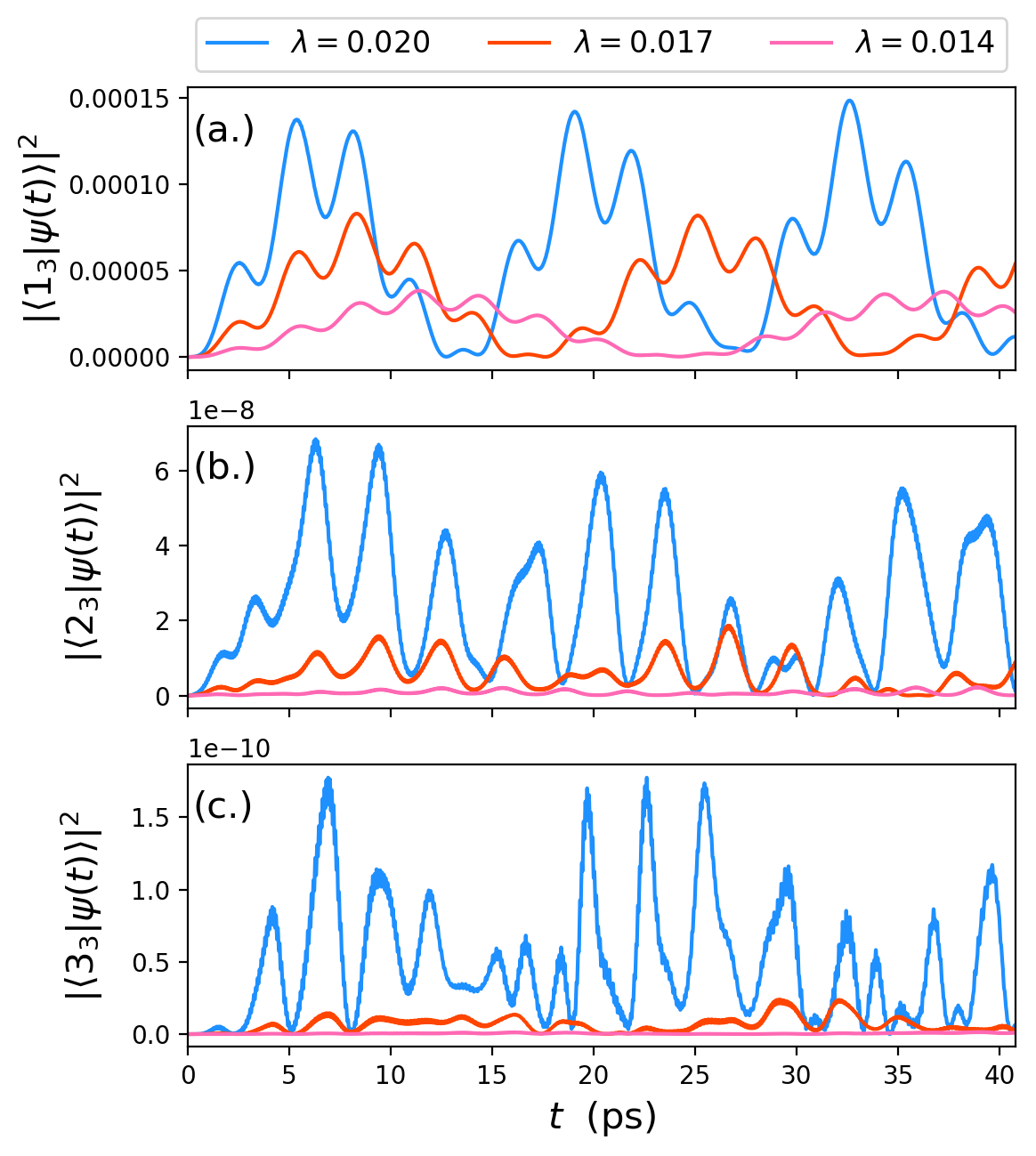}}
	\caption{Real-time Fock state occupations of the signal mode 3 from weak to ultra-strong coupling for the input coherent state. A similar trend of the one-photon Fock state in (a.) being mostly populated in comparison to the two- and three-photon Fock states in (b.) and (c.), respectively.}
	\label{fig:coherent-fock-occupation-mode-3}
\end{figure}

In this section, we show the Fock state occupations for the different fields, i.e., single-photon input field and the coherent state with $n_{1}(0)=4$ photons. 

In Fig.~(\ref{fig:fock-occupation-mode-1}) with the input field in a single-photon Fock state, the population of the two-photon Fock state is zero through out the time-evolution while that of the three-photon Fock state is weakly populated. In accordance with the Fock state populations in Figs.~(\ref{fig:fock-occupation-mode-2}) and (\ref{fig:fock-occupation-mode-3}), this shows that the down-converted photons are of a single-photon nature.

In the case when the input field is a coherent state, the one-, two- and three-photon Fock states are populated with increasing amplitude from $|1_{1}\rangle$, $|2_{1}\rangle$ and $|3_{1}\rangle$. This is true for the input mode Fig.~\ref{fig:coherent-fock-occupation-mode-1} but also for the signal modes Figs.~\ref{fig:coherent-fock-occupation-mode-2} and \ref{fig:coherent-fock-occupation-mode-3}.


\vspace{10em}

\bibliography{01_light_matter_coupling} 

\begin{thebibliography}{67}%
\makeatletter
\providecommand \@ifxundefined [1]{%
 \@ifx{#1\undefined}
}%
\providecommand \@ifnum [1]{%
 \ifnum #1\expandafter \@firstoftwo
 \else \expandafter \@secondoftwo
 \fi
}%
\providecommand \@ifx [1]{%
 \ifx #1\expandafter \@firstoftwo
 \else \expandafter \@secondoftwo
 \fi
}%
\providecommand \natexlab [1]{#1}%
\providecommand \enquote  [1]{``#1''}%
\providecommand \bibnamefont  [1]{#1}%
\providecommand \bibfnamefont [1]{#1}%
\providecommand \citenamefont [1]{#1}%
\providecommand \href@noop [0]{\@secondoftwo}%
\providecommand \href [0]{\begingroup \@sanitize@url \@href}%
\providecommand \@href[1]{\@@startlink{#1}\@@href}%
\providecommand \@@href[1]{\endgroup#1\@@endlink}%
\providecommand \@sanitize@url [0]{\catcode `\\12\catcode `\$12\catcode
  `\&12\catcode `\#12\catcode `\^12\catcode `\_12\catcode `\%12\relax}%
\providecommand \@@startlink[1]{}%
\providecommand \@@endlink[0]{}%
\providecommand \url  [0]{\begingroup\@sanitize@url \@url }%
\providecommand \@url [1]{\endgroup\@href {#1}{\urlprefix }}%
\providecommand \urlprefix  [0]{URL }%
\providecommand \Eprint [0]{\href }%
\providecommand \doibase [0]{http://dx.doi.org/}%
\providecommand \selectlanguage [0]{\@gobble}%
\providecommand \bibinfo  [0]{\@secondoftwo}%
\providecommand \bibfield  [0]{\@secondoftwo}%
\providecommand \translation [1]{[#1]}%
\providecommand \BibitemOpen [0]{}%
\providecommand \bibitemStop [0]{}%
\providecommand \bibitemNoStop [0]{.\EOS\space}%
\providecommand \EOS [0]{\spacefactor3000\relax}%
\providecommand \BibitemShut  [1]{\csname bibitem#1\endcsname}%
\let\auto@bib@innerbib\@empty
\bibitem [{\citenamefont {O’Brien}\ \emph {et~al.}(2009)\citenamefont
  {O’Brien}, \citenamefont {Furusawa},\ and\ \citenamefont
  {Vu\v{c}kovi\'{c}}}]{obrien2009}%
  \BibitemOpen
  \bibfield  {author} {\bibinfo {author} {\bibfnamefont {Jeremy~L.}\
  \bibnamefont {O’Brien}}, \bibinfo {author} {\bibfnamefont {Akira}\
  \bibnamefont {Furusawa}}, \ and\ \bibinfo {author} {\bibfnamefont {Jelena}\
  \bibnamefont {Vu\v{c}kovi\'{c}}},\ }\bibfield  {title} {\enquote {\bibinfo
  {title} {Photonic quantum technologies},}\ }\href {\doibase
  10.1038/nphoton.2009.229} {\bibfield  {journal} {\bibinfo  {journal} {Nat.
  Photonics}\ }\textbf {\bibinfo {volume} {3}},\ \bibinfo {pages} {687--695}
  (\bibinfo {year} {2009})}\BibitemShut {NoStop}%
\bibitem [{\citenamefont {Zheng}\ and\ \citenamefont {Guo}(2000)}]{zheng2000}%
  \BibitemOpen
  \bibfield  {author} {\bibinfo {author} {\bibfnamefont {Shi-Biao}\
  \bibnamefont {Zheng}}\ and\ \bibinfo {author} {\bibfnamefont {Guang-Can}\
  \bibnamefont {Guo}},\ }\bibfield  {title} {\enquote {\bibinfo {title}
  {Efficient scheme for two-atom entanglement and quantum information
  processing in cavity qed},}\ }\href {\doibase 10.1103/PhysRevLett.85.2392}
  {\bibfield  {journal} {\bibinfo  {journal} {Phys. Rev. Lett.}\ }\textbf
  {\bibinfo {volume} {85}},\ \bibinfo {pages} {2392} (\bibinfo {year}
  {2000})}\BibitemShut {NoStop}%
\bibitem [{\citenamefont {Jennewein}\ \emph {et~al.}(2000)\citenamefont
  {Jennewein}, \citenamefont {Simon}, \citenamefont {Weihs}, \citenamefont
  {Weinfurter},\ and\ \citenamefont {Zeilinger}}]{jennewein2000}%
  \BibitemOpen
  \bibfield  {author} {\bibinfo {author} {\bibfnamefont {Thomas}\ \bibnamefont
  {Jennewein}}, \bibinfo {author} {\bibfnamefont {Christoph}\ \bibnamefont
  {Simon}}, \bibinfo {author} {\bibfnamefont {Gregor}\ \bibnamefont {Weihs}},
  \bibinfo {author} {\bibfnamefont {Harald}\ \bibnamefont {Weinfurter}}, \ and\
  \bibinfo {author} {\bibfnamefont {Anton}\ \bibnamefont {Zeilinger}},\
  }\bibfield  {title} {\enquote {\bibinfo {title} {Quantum cryptography with
  entangled photons},}\ }\href {\doibase 10.1103/PhysRevLett.84.4729}
  {\bibfield  {journal} {\bibinfo  {journal} {Phys. Rev. Lett.}\ }\textbf
  {\bibinfo {volume} {84}},\ \bibinfo {pages} {4729} (\bibinfo {year}
  {2000})}\BibitemShut {NoStop}%
\bibitem [{\citenamefont {Boschi}\ \emph {et~al.}(1998)\citenamefont {Boschi},
  \citenamefont {Brancaa}, \citenamefont {Martini}, \citenamefont {Hardy},\
  and\ \citenamefont {Popescu}}]{boschi1998}%
  \BibitemOpen
  \bibfield  {author} {\bibinfo {author} {\bibfnamefont {D.}~\bibnamefont
  {Boschi}}, \bibinfo {author} {\bibfnamefont {S.}~\bibnamefont {Brancaa}},
  \bibinfo {author} {\bibfnamefont {F.~De}\ \bibnamefont {Martini}}, \bibinfo
  {author} {\bibfnamefont {L.}~\bibnamefont {Hardy}}, \ and\ \bibinfo {author}
  {\bibfnamefont {S.}~\bibnamefont {Popescu}},\ }\bibfield  {title} {\enquote
  {\bibinfo {title} {Experimental realization of teleporting an unknown pure
  quantum state via dual classical and einstein-podolsky-rosen channels},}\
  }\href {\doibase 10.1103/PhysRevLett.80.1121} {\bibfield  {journal} {\bibinfo
   {journal} {Phys. Rev. Lett.}\ }\textbf {\bibinfo {volume} {80}},\ \bibinfo
  {pages} {1121} (\bibinfo {year} {1998})}\BibitemShut {NoStop}%
\bibitem [{\citenamefont {Stevenson}\ \emph {et~al.}(2012)\citenamefont
  {Stevenson}, \citenamefont {Salter}, \citenamefont {Nilsson}, \citenamefont
  {Bennett}, \citenamefont {Ward}, \citenamefont {Farrer}, \citenamefont
  {Ritchie},\ and\ \citenamefont {Shields}}]{stevenson2012}%
  \BibitemOpen
  \bibfield  {author} {\bibinfo {author} {\bibfnamefont {R.~M.}\ \bibnamefont
  {Stevenson}}, \bibinfo {author} {\bibfnamefont {C.~L.}\ \bibnamefont
  {Salter}}, \bibinfo {author} {\bibfnamefont {J.}~\bibnamefont {Nilsson}},
  \bibinfo {author} {\bibfnamefont {A.~J.}\ \bibnamefont {Bennett}}, \bibinfo
  {author} {\bibfnamefont {M.~B.}\ \bibnamefont {Ward}}, \bibinfo {author}
  {\bibfnamefont {I.}~\bibnamefont {Farrer}}, \bibinfo {author} {\bibfnamefont
  {D.~A.}\ \bibnamefont {Ritchie}}, \ and\ \bibinfo {author} {\bibfnamefont
  {A.~J.}\ \bibnamefont {Shields}},\ }\bibfield  {title} {\enquote {\bibinfo
  {title} {Indistinguishable entangled photons generated by a light-emitting
  diode},}\ }\href {\doibase 10.1103/PhysRevLett.108.040503} {\bibfield
  {journal} {\bibinfo  {journal} {Phys. Rev. Lett.}\ }\textbf {\bibinfo
  {volume} {108}},\ \bibinfo {pages} {040503} (\bibinfo {year}
  {2012})}\BibitemShut {NoStop}%
\bibitem [{\citenamefont {Callsen}\ \emph {et~al.}(2013)\citenamefont
  {Callsen}, \citenamefont {Carmele}, \citenamefont {H\"{o}nig}, \citenamefont
  {Kindel}, \citenamefont {Brunnmeier}, \citenamefont {Wagner}, \citenamefont
  {Stock}, \citenamefont {Reparaz}, \citenamefont {Schliwa}, \citenamefont
  {Reitzenstein}, \citenamefont {Knorr},\ and\ \citenamefont
  {Hoffmann}}]{callsen2013}%
  \BibitemOpen
  \bibfield  {author} {\bibinfo {author} {\bibfnamefont {G.}~\bibnamefont
  {Callsen}}, \bibinfo {author} {\bibfnamefont {A.}~\bibnamefont {Carmele}},
  \bibinfo {author} {\bibfnamefont {G.}~\bibnamefont {H\"{o}nig}}, \bibinfo
  {author} {\bibfnamefont {C.}~\bibnamefont {Kindel}}, \bibinfo {author}
  {\bibfnamefont {J.}~\bibnamefont {Brunnmeier}}, \bibinfo {author}
  {\bibfnamefont {M.~R.}\ \bibnamefont {Wagner}}, \bibinfo {author}
  {\bibfnamefont {E.}~\bibnamefont {Stock}}, \bibinfo {author} {\bibfnamefont
  {J.~S.}\ \bibnamefont {Reparaz}}, \bibinfo {author} {\bibfnamefont
  {A.}~\bibnamefont {Schliwa}}, \bibinfo {author} {\bibfnamefont
  {S.}~\bibnamefont {Reitzenstein}}, \bibinfo {author} {\bibfnamefont
  {A.}~\bibnamefont {Knorr}}, \ and\ \bibinfo {author} {\bibfnamefont
  {A.}~\bibnamefont {Hoffmann}},\ }\bibfield  {title} {\enquote {\bibinfo
  {title} {Steering photon statistics in single quantum dots: From one- to
  two-photon emission},}\ }\href {\doibase 10.1103/PhysRevB.87.245314}
  {\bibfield  {journal} {\bibinfo  {journal} {Phys. Rev. B.}\ }\textbf
  {\bibinfo {volume} {87}},\ \bibinfo {pages} {245314} (\bibinfo {year}
  {2013})}\BibitemShut {NoStop}%
\bibitem [{\citenamefont {Boyd}(1992)}]{boyd1992}%
  \BibitemOpen
  \bibfield  {author} {\bibinfo {author} {\bibfnamefont {Robert~W.}\
  \bibnamefont {Boyd}},\ }\href@noop {} {\emph {\bibinfo {title} {Nonlinear
  Optics}}}\ (\bibinfo  {publisher} {Academic, New York},\ \bibinfo {year}
  {1992})\BibitemShut {NoStop}%
\bibitem [{\citenamefont {Ou}\ and\ \citenamefont {Lu}(1999)}]{ou1999}%
  \BibitemOpen
  \bibfield  {author} {\bibinfo {author} {\bibfnamefont {Z.~Y.}\ \bibnamefont
  {Ou}}\ and\ \bibinfo {author} {\bibfnamefont {Y.~J.}\ \bibnamefont {Lu}},\
  }\bibfield  {title} {\enquote {\bibinfo {title} {Cavity enhanced spontaneous
  parametric down-conversion for the prolongation of correlation time between
  conjugate photons},}\ }\href {\doibase 10.1103/PhysRevLett.83.2556}
  {\bibfield  {journal} {\bibinfo  {journal} {Phys. Rev. Lett.}\ }\textbf
  {\bibinfo {volume} {83}},\ \bibinfo {pages} {2556} (\bibinfo {year}
  {1999})}\BibitemShut {NoStop}%
\bibitem [{\citenamefont {Dousse}\ \emph {et~al.}(2019)\citenamefont {Dousse},
  \citenamefont {Suffczy\'{n}ski}, \citenamefont {Beveratos}, \citenamefont
  {Krebs}, \citenamefont {Lema\^{i}tre}, \citenamefont {Sagnes}, \citenamefont
  {Bloch}, \citenamefont {Voisin},\ and\ \citenamefont
  {Senellart}}]{dousse2010}%
  \BibitemOpen
  \bibfield  {author} {\bibinfo {author} {\bibfnamefont {Adrien}\ \bibnamefont
  {Dousse}}, \bibinfo {author} {\bibfnamefont {Jan}\ \bibnamefont
  {Suffczy\'{n}ski}}, \bibinfo {author} {\bibfnamefont {Alexios}\ \bibnamefont
  {Beveratos}}, \bibinfo {author} {\bibfnamefont {Olivier}\ \bibnamefont
  {Krebs}}, \bibinfo {author} {\bibfnamefont {Aristide}\ \bibnamefont
  {Lema\^{i}tre}}, \bibinfo {author} {\bibfnamefont {Isabelle}\ \bibnamefont
  {Sagnes}}, \bibinfo {author} {\bibfnamefont {Jacqueline}\ \bibnamefont
  {Bloch}}, \bibinfo {author} {\bibfnamefont {Paul}\ \bibnamefont {Voisin}}, \
  and\ \bibinfo {author} {\bibfnamefont {Pascale}\ \bibnamefont {Senellart}},\
  }\bibfield  {title} {\enquote {\bibinfo {title} {Ultrabright source of
  entangled photon pairs},}\ }\href {\doibase 10.1038/nature09148} {\bibfield
  {journal} {\bibinfo  {journal} {Nature}\ }\textbf {\bibinfo {volume} {466}},\
  \bibinfo {pages} {217--220} (\bibinfo {year} {2019})}\BibitemShut {NoStop}%
\bibitem [{\citenamefont {S\'{a}nchez-Burillo}\ \emph
  {et~al.}(2016)\citenamefont {S\'{a}nchez-Burillo}, \citenamefont
  {Mart\'{i}n-Moreno}, \citenamefont {Garc\'{i}a-Ripoll},\ and\ \citenamefont
  {Zueco}}]{burillo2016}%
  \BibitemOpen
  \bibfield  {author} {\bibinfo {author} {\bibfnamefont {E.}~\bibnamefont
  {S\'{a}nchez-Burillo}}, \bibinfo {author} {\bibfnamefont {L.}~\bibnamefont
  {Mart\'{i}n-Moreno}}, \bibinfo {author} {\bibfnamefont {J.~J.}\ \bibnamefont
  {Garc\'{i}a-Ripoll}}, \ and\ \bibinfo {author} {\bibfnamefont
  {D.}~\bibnamefont {Zueco}},\ }\bibfield  {title} {\enquote {\bibinfo {title}
  {Full two-photon down-conversion of a single photon},}\ }\href {\doibase
  10.1103/PhysRevA.94.053814} {\bibfield  {journal} {\bibinfo  {journal} {Phys.
  Rev. A.}\ }\textbf {\bibinfo {volume} {94}},\ \bibinfo {pages} {053814}
  (\bibinfo {year} {2016})}\BibitemShut {NoStop}%
\bibitem [{\citenamefont {Chang}\ \emph {et~al.}(2016)\citenamefont {Chang},
  \citenamefont {Gonz\'{a}lez-Tudela}, \citenamefont {Munoz}, \citenamefont
  {Navarrete-Benlloch},\ and\ \citenamefont {Shi}}]{chang2016}%
  \BibitemOpen
  \bibfield  {author} {\bibinfo {author} {\bibfnamefont {Yue}\ \bibnamefont
  {Chang}}, \bibinfo {author} {\bibfnamefont {Alejandro}\ \bibnamefont
  {Gonz\'{a}lez-Tudela}}, \bibinfo {author} {\bibfnamefont
  {Carlos~S\'{a}nchez}\ \bibnamefont {Munoz}}, \bibinfo {author} {\bibfnamefont
  {Carlos}\ \bibnamefont {Navarrete-Benlloch}}, \ and\ \bibinfo {author}
  {\bibfnamefont {Tao}\ \bibnamefont {Shi}},\ }\bibfield  {title} {\enquote
  {\bibinfo {title} {Deterministic down-converter and continuous photon-pair
  source within the bad-cavity limit},}\ }\href {\doibase
  10.1103/PhysRevLett.117.203602} {\bibfield  {journal} {\bibinfo  {journal}
  {Phys. Rev. Lett.}\ }\textbf {\bibinfo {volume} {117}},\ \bibinfo {pages}
  {203602} (\bibinfo {year} {2016})}\BibitemShut {NoStop}%
\bibitem [{\citenamefont {Evans}\ \emph {et~al.}(2010)\citenamefont {Evans},
  \citenamefont {Bennink}, \citenamefont {Grice}, \citenamefont {Humble},\ and\
  \citenamefont {Schaake}}]{evans2010}%
  \BibitemOpen
  \bibfield  {author} {\bibinfo {author} {\bibfnamefont {P.}~\bibnamefont
  {Evans}}, \bibinfo {author} {\bibfnamefont {R.}~\bibnamefont {Bennink}},
  \bibinfo {author} {\bibfnamefont {W.}~\bibnamefont {Grice}}, \bibinfo
  {author} {\bibfnamefont {T.}~\bibnamefont {Humble}}, \ and\ \bibinfo {author}
  {\bibfnamefont {J.}~\bibnamefont {Schaake}},\ }\bibfield  {title} {\enquote
  {\bibinfo {title} {Bright source of spectrally uncorrelated
  polarization-entangled photons with nearly single-mode emission},}\ }\href
  {\doibase 10.1103/PhysRevLett.105.253601} {\bibfield  {journal} {\bibinfo
  {journal} {Phys. Rev. Lett.}\ }\textbf {\bibinfo {volume} {105}},\ \bibinfo
  {pages} {253601} (\bibinfo {year} {2010})}\BibitemShut {NoStop}%
\bibitem [{\citenamefont {Bruno}\ \emph {et~al.}(2014)\citenamefont {Bruno},
  \citenamefont {Martin}, \citenamefont {Guerreiro}, \citenamefont
  {Sanguinetti},\ and\ \citenamefont {Thew}}]{bruno2014}%
  \BibitemOpen
  \bibfield  {author} {\bibinfo {author} {\bibfnamefont {N.}~\bibnamefont
  {Bruno}}, \bibinfo {author} {\bibfnamefont {A.}~\bibnamefont {Martin}},
  \bibinfo {author} {\bibfnamefont {T.}~\bibnamefont {Guerreiro}}, \bibinfo
  {author} {\bibfnamefont {B.}~\bibnamefont {Sanguinetti}}, \ and\ \bibinfo
  {author} {\bibfnamefont {R.~T.}\ \bibnamefont {Thew}},\ }\bibfield  {title}
  {\enquote {\bibinfo {title} {Pulsed source of spectrally uncorrelated and
  indistinguishable photons at telecom wavelengths},}\ }\href {\doibase
  10.1364/OE.22.017246} {\bibfield  {journal} {\bibinfo  {journal} {Optics
  Express}\ }\textbf {\bibinfo {volume} {22}},\ \bibinfo {pages} {17246--17253}
  (\bibinfo {year} {2014})}\BibitemShut {NoStop}%
\bibitem [{\citenamefont {Law}\ and\ \citenamefont {Eberly}(1996)}]{law1996}%
  \BibitemOpen
  \bibfield  {author} {\bibinfo {author} {\bibfnamefont {C.~K.}\ \bibnamefont
  {Law}}\ and\ \bibinfo {author} {\bibfnamefont {J.~H.}\ \bibnamefont
  {Eberly}},\ }\bibfield  {title} {\enquote {\bibinfo {title} {Arbitrary
  control of a quantum electromagnetic field},}\ }\href {\doibase
  10.1103/PhysRevLett.76.1055} {\bibfield  {journal} {\bibinfo  {journal}
  {Phys. Rev. Lett.}\ }\textbf {\bibinfo {volume} {76}},\ \bibinfo {pages}
  {1055} (\bibinfo {year} {1996})}\BibitemShut {NoStop}%
\bibitem [{\citenamefont {Thompson}\ \emph {et~al.}(2006)\citenamefont
  {Thompson}, \citenamefont {Simon}, \citenamefont {Loh},\ and\ \citenamefont
  {Vuleti\'{c}}}]{thompson2006}%
  \BibitemOpen
  \bibfield  {author} {\bibinfo {author} {\bibfnamefont {James~K.}\
  \bibnamefont {Thompson}}, \bibinfo {author} {\bibfnamefont {Jonathan}\
  \bibnamefont {Simon}}, \bibinfo {author} {\bibfnamefont {Huanqian}\
  \bibnamefont {Loh}}, \ and\ \bibinfo {author} {\bibfnamefont {Vladan}\
  \bibnamefont {Vuleti\'{c}}},\ }\bibfield  {title} {\enquote {\bibinfo {title}
  {A high-brightness source of narrowband, identical-photon pairs},}\ }\href
  {\doibase 10.1126/science.1127676} {\bibfield  {journal} {\bibinfo  {journal}
  {Science}\ }\textbf {\bibinfo {volume} {313}},\ \bibinfo {pages} {74--77}
  (\bibinfo {year} {2006})}\BibitemShut {NoStop}%
\bibitem [{\citenamefont {Akopian}\ \emph {et~al.}(2006)\citenamefont
  {Akopian}, \citenamefont {Lindner}, \citenamefont {Poem}, \citenamefont
  {Berlatzky}, \citenamefont {Avron}, \citenamefont {Gershoni}, \citenamefont
  {Gerardot},\ and\ \citenamefont {Petroff}}]{akopian2006}%
  \BibitemOpen
  \bibfield  {author} {\bibinfo {author} {\bibfnamefont {N.}~\bibnamefont
  {Akopian}}, \bibinfo {author} {\bibfnamefont {N.~H.}\ \bibnamefont
  {Lindner}}, \bibinfo {author} {\bibfnamefont {E.}~\bibnamefont {Poem}},
  \bibinfo {author} {\bibfnamefont {Y.}~\bibnamefont {Berlatzky}}, \bibinfo
  {author} {\bibfnamefont {J.}~\bibnamefont {Avron}}, \bibinfo {author}
  {\bibfnamefont {D.}~\bibnamefont {Gershoni}}, \bibinfo {author}
  {\bibfnamefont {B.~D.}\ \bibnamefont {Gerardot}}, \ and\ \bibinfo {author}
  {\bibfnamefont {P.~M.}\ \bibnamefont {Petroff}},\ }\bibfield  {title}
  {\enquote {\bibinfo {title} {Entangled photon pairs from semiconductor
  quantum dots},}\ }\href {\doibase 10.1103/PhysRevLett.96.130501} {\bibfield
  {journal} {\bibinfo  {journal} {Phys. Rev. Lett.}\ }\textbf {\bibinfo
  {volume} {96}},\ \bibinfo {pages} {130501} (\bibinfo {year}
  {2006})}\BibitemShut {NoStop}%
\bibitem [{\citenamefont {Pathak}\ and\ \citenamefont
  {Hughes}(2011)}]{pathak2011}%
  \BibitemOpen
  \bibfield  {author} {\bibinfo {author} {\bibfnamefont {P.~K.}\ \bibnamefont
  {Pathak}}\ and\ \bibinfo {author} {\bibfnamefont {S.}~\bibnamefont
  {Hughes}},\ }\bibfield  {title} {\enquote {\bibinfo {title} {Coherent
  generation of time-bin entangled photon pairs using the biexciton cascade and
  cavity-assisted piecewise adiabatic passage},}\ }\href {\doibase
  10.1103/PhysRevB.83.245301} {\bibfield  {journal} {\bibinfo  {journal} {Phys.
  Rev. B}\ }\textbf {\bibinfo {volume} {83}},\ \bibinfo {pages} {245301}
  (\bibinfo {year} {2011})}\BibitemShut {NoStop}%
\bibitem [{\citenamefont {M\"{u}ller}\ \emph {et~al.}(2014)\citenamefont
  {M\"{u}ller}, \citenamefont {Bounouar}, \citenamefont {J\"{o}ns},
  \citenamefont {Gl\"{a}ssl},\ and\ \citenamefont {Michler}}]{muller2014}%
  \BibitemOpen
  \bibfield  {author} {\bibinfo {author} {\bibfnamefont {M.}~\bibnamefont
  {M\"{u}ller}}, \bibinfo {author} {\bibfnamefont {S.}~\bibnamefont
  {Bounouar}}, \bibinfo {author} {\bibfnamefont {K.~D.}\ \bibnamefont
  {J\"{o}ns}}, \bibinfo {author} {\bibfnamefont {M.}~\bibnamefont
  {Gl\"{a}ssl}}, \ and\ \bibinfo {author} {\bibfnamefont {P.}~\bibnamefont
  {Michler}},\ }\bibfield  {title} {\enquote {\bibinfo {title} {On-demand
  generation of indistinguishable polarization-entangled photon pairs},}\
  }\href {\doibase 10.1038/nphoton.2013.377} {\bibfield  {journal} {\bibinfo
  {journal} {Nat. Photonics}\ }\textbf {\bibinfo {volume} {8}},\ \bibinfo
  {pages} {224--228} (\bibinfo {year} {2014})}\BibitemShut {NoStop}%
\bibitem [{\citenamefont {P\'{e}rez-S\'{a}nchez}\ and\ \citenamefont
  {Yuen-Zhou}(2020)}]{juan2020}%
  \BibitemOpen
  \bibfield  {author} {\bibinfo {author} {\bibfnamefont {Juan~B.}\ \bibnamefont
  {P\'{e}rez-S\'{a}nchez}}\ and\ \bibinfo {author} {\bibfnamefont {Joel}\
  \bibnamefont {Yuen-Zhou}},\ }\bibfield  {title} {\enquote {\bibinfo {title}
  {Polariton assisted down-conversion of photons via nonadiabatic molecular
  dynamics: A molecular dynamical casimir effect},}\ }\href {\doibase
  10.1021/acs.jpclett.9b02870} {\bibfield  {journal} {\bibinfo  {journal} {J.
  Phys. Chem. Lett.}\ }\textbf {\bibinfo {volume} {11}},\ \bibinfo {pages}
  {152--159} (\bibinfo {year} {2020})}\BibitemShut {NoStop}%
\bibitem [{\citenamefont {Abdo}\ \emph {et~al.}(2013)\citenamefont {Abdo},
  \citenamefont {Sliwa}, \citenamefont {Schackert}, \citenamefont {Bergeal},
  \citenamefont {Hatridge}, \citenamefont {Frunzio}, \citenamefont {Stone},\
  and\ \citenamefont {Devoret}}]{abdo2013}%
  \BibitemOpen
  \bibfield  {author} {\bibinfo {author} {\bibfnamefont {Baleegh}\ \bibnamefont
  {Abdo}}, \bibinfo {author} {\bibfnamefont {Katrina}\ \bibnamefont {Sliwa}},
  \bibinfo {author} {\bibfnamefont {Flavius}\ \bibnamefont {Schackert}},
  \bibinfo {author} {\bibfnamefont {Nicolas}\ \bibnamefont {Bergeal}}, \bibinfo
  {author} {\bibfnamefont {Michael}\ \bibnamefont {Hatridge}}, \bibinfo
  {author} {\bibfnamefont {Luigi}\ \bibnamefont {Frunzio}}, \bibinfo {author}
  {\bibfnamefont {A.~Douglas}\ \bibnamefont {Stone}}, \ and\ \bibinfo {author}
  {\bibfnamefont {Michel}\ \bibnamefont {Devoret}},\ }\bibfield  {title}
  {\enquote {\bibinfo {title} {Full coherent frequency conversion between two
  propagating microwave modes},}\ }\href {\doibase
  10.1103/PhysRevLett.110.173902} {\bibfield  {journal} {\bibinfo  {journal}
  {Phys. Rev. Lett.}\ }\textbf {\bibinfo {volume} {110}},\ \bibinfo {pages}
  {173902} (\bibinfo {year} {2013})}\BibitemShut {NoStop}%
\bibitem [{\citenamefont {Kamal}\ \emph {et~al.}(2014)\citenamefont {Kamal},
  \citenamefont {Roy}, \citenamefont {Clarke},\ and\ \citenamefont
  {Devoret}}]{kamal2014}%
  \BibitemOpen
  \bibfield  {author} {\bibinfo {author} {\bibfnamefont {Archana}\ \bibnamefont
  {Kamal}}, \bibinfo {author} {\bibfnamefont {Ananda}\ \bibnamefont {Roy}},
  \bibinfo {author} {\bibfnamefont {John}\ \bibnamefont {Clarke}}, \ and\
  \bibinfo {author} {\bibfnamefont {Michel~H.}\ \bibnamefont {Devoret}},\
  }\bibfield  {title} {\enquote {\bibinfo {title} {Asymmetric frequency
  conversion in nonlinear systems driven by a biharmonic pump},}\ }\href
  {\doibase 10.1103/PhysRevLett.113.247003} {\bibfield  {journal} {\bibinfo
  {journal} {Phys. Rev. Lett.}\ }\textbf {\bibinfo {volume} {113}},\ \bibinfo
  {pages} {247003} (\bibinfo {year} {2014})}\BibitemShut {NoStop}%
\bibitem [{\citenamefont {Burnham}\ and\ \citenamefont
  {Weinberg}(1970)}]{burnham1970}%
  \BibitemOpen
  \bibfield  {author} {\bibinfo {author} {\bibfnamefont {David~C.}\
  \bibnamefont {Burnham}}\ and\ \bibinfo {author} {\bibfnamefont {Donald~L.}\
  \bibnamefont {Weinberg}},\ }\bibfield  {title} {\enquote {\bibinfo {title}
  {Observation of simultaneity in parametric production of optical photon
  pairs},}\ }\href {\doibase 10.1103/PhysRevLett.25.84} {\bibfield  {journal}
  {\bibinfo  {journal} {Phys. Rev. Lett.}\ }\textbf {\bibinfo {volume} {25}},\
  \bibinfo {pages} {84} (\bibinfo {year} {1970})}\BibitemShut {NoStop}%
\bibitem [{\citenamefont {Stevenson}\ \emph {et~al.}(2006)\citenamefont
  {Stevenson}, \citenamefont {Young}, \citenamefont {Atkinson}, \citenamefont
  {Cooper}, \citenamefont {Ritchie},\ and\ \citenamefont
  {Shields}}]{stevenson2006}%
  \BibitemOpen
  \bibfield  {author} {\bibinfo {author} {\bibfnamefont {R.~M.}\ \bibnamefont
  {Stevenson}}, \bibinfo {author} {\bibfnamefont {R.~J.}\ \bibnamefont
  {Young}}, \bibinfo {author} {\bibfnamefont {P.}~\bibnamefont {Atkinson}},
  \bibinfo {author} {\bibfnamefont {K.}~\bibnamefont {Cooper}}, \bibinfo
  {author} {\bibfnamefont {D.~A.}\ \bibnamefont {Ritchie}}, \ and\ \bibinfo
  {author} {\bibfnamefont {A.~J.}\ \bibnamefont {Shields}},\ }\bibfield
  {title} {\enquote {\bibinfo {title} {A semiconductor source of triggered
  entangled photon pairs},}\ }\href {\doibase 10.1038/nature04446} {\bibfield
  {journal} {\bibinfo  {journal} {Nature}\ }\textbf {\bibinfo {volume} {439}},\
  \bibinfo {pages} {179--182} (\bibinfo {year} {2006})}\BibitemShut {NoStop}%
\bibitem [{\citenamefont {ans C.~A.~Mastrandrea}\ \emph
  {et~al.}(2009)\citenamefont {ans C.~A.~Mastrandrea}, \citenamefont
  {Vinattieri}, \citenamefont {Sanguinetti}, \citenamefont {Mano},
  \citenamefont {Kuroda}, \citenamefont {Koguchi}, \citenamefont {Sakoda},\
  and\ \citenamefont {Gurioli}}]{abbarchi2009}%
  \BibitemOpen
  \bibfield  {author} {\bibinfo {author} {\bibfnamefont {M.~Abbarchi}\
  \bibnamefont {ans C.~A.~Mastrandrea}}, \bibinfo {author} {\bibfnamefont
  {A.}~\bibnamefont {Vinattieri}}, \bibinfo {author} {\bibfnamefont
  {S.}~\bibnamefont {Sanguinetti}}, \bibinfo {author} {\bibfnamefont
  {T.}~\bibnamefont {Mano}}, \bibinfo {author} {\bibfnamefont {T.}~\bibnamefont
  {Kuroda}}, \bibinfo {author} {\bibfnamefont {N.}~\bibnamefont {Koguchi}},
  \bibinfo {author} {\bibfnamefont {K.}~\bibnamefont {Sakoda}}, \ and\ \bibinfo
  {author} {\bibfnamefont {M.}~\bibnamefont {Gurioli}},\ }\bibfield  {title}
  {\enquote {\bibinfo {title} {Photon antibunching in double quantum ring
  structures},}\ }\href {\doibase 10.1103/PhysRevB.79.085308} {\bibfield
  {journal} {\bibinfo  {journal} {Phys. Rev. B}\ }\textbf {\bibinfo {volume}
  {79}},\ \bibinfo {pages} {085308} (\bibinfo {year} {2009})}\BibitemShut
  {NoStop}%
\bibitem [{\citenamefont {Fomin}(2018)}]{fomin2018}%
  \BibitemOpen
  \bibfield  {author} {\bibinfo {author} {\bibfnamefont {Vladimir~M.}\
  \bibnamefont {Fomin}},\ }\href {\doibase 10.1007/978-3-642-39197-2} {\emph
  {\bibinfo {title} {Physics of Quantum Rings}}}\ (\bibinfo  {publisher}
  {Springer-Verlag Berlin Heidelberg},\ \bibinfo {year} {2018})\BibitemShut
  {NoStop}%
\bibitem [{\citenamefont {Couteau}(2018)}]{couteau2018}%
  \BibitemOpen
  \bibfield  {author} {\bibinfo {author} {\bibfnamefont {Christophe}\
  \bibnamefont {Couteau}},\ }\bibfield  {title} {\enquote {\bibinfo {title}
  {Spontaneous parametric down-conversion},}\ }\href {\doibase
  10.1080/00107514.2018.1488463} {\bibfield  {journal} {\bibinfo  {journal}
  {Contemporary Physics}\ }\textbf {\bibinfo {volume} {59}},\ \bibinfo {pages}
  {291--304} (\bibinfo {year} {2018})}\BibitemShut {NoStop}%
\bibitem [{\citenamefont {Kockum}\ \emph {et~al.}(2017)\citenamefont {Kockum},
  \citenamefont {Macr\`{i}}, \citenamefont {Garziano}, \citenamefont
  {Savasta},\ and\ \citenamefont {Nori}}]{kockum2017}%
  \BibitemOpen
  \bibfield  {author} {\bibinfo {author} {\bibfnamefont {Anton~Frisk}\
  \bibnamefont {Kockum}}, \bibinfo {author} {\bibfnamefont {Vincenzo}\
  \bibnamefont {Macr\`{i}}}, \bibinfo {author} {\bibfnamefont {Luigi}\
  \bibnamefont {Garziano}}, \bibinfo {author} {\bibfnamefont {Salvatore}\
  \bibnamefont {Savasta}}, \ and\ \bibinfo {author} {\bibfnamefont {Franco}\
  \bibnamefont {Nori}},\ }\bibfield  {title} {\enquote {\bibinfo {title}
  {Frequency conversion in ultrastrong cavity qed},}\ }\href {\doibase
  10.1038/s41598-017-04225-3} {\bibfield  {journal} {\bibinfo  {journal}
  {Scientific Reports}\ }\textbf {\bibinfo {volume} {7}},\ \bibinfo {pages}
  {5313} (\bibinfo {year} {2017})}\BibitemShut {NoStop}%
\bibitem [{\citenamefont {Villas-B\^{o}as}\ \emph {et~al.}(2003)\citenamefont
  {Villas-B\^{o}as}, \citenamefont {de~Almeida}, \citenamefont {Serra},\ and\
  \citenamefont {Moussa}}]{villas2003}%
  \BibitemOpen
  \bibfield  {author} {\bibinfo {author} {\bibfnamefont {C.~J.}\ \bibnamefont
  {Villas-B\^{o}as}}, \bibinfo {author} {\bibfnamefont {N.~G.}\ \bibnamefont
  {de~Almeida}}, \bibinfo {author} {\bibfnamefont {R.~M.}\ \bibnamefont
  {Serra}}, \ and\ \bibinfo {author} {\bibfnamefont {M.~H.~Y.}\ \bibnamefont
  {Moussa}},\ }\bibfield  {title} {\enquote {\bibinfo {title} {Squeezing
  arbitrary cavity-field states through their interaction with a single driven
  atom},}\ }\href {\doibase 10.1103/PhysRevA.68.061801} {\bibfield  {journal}
  {\bibinfo  {journal} {Phys. Rev. A.}\ }\textbf {\bibinfo {volume} {68}},\
  \bibinfo {pages} {061801(R)} (\bibinfo {year} {2003})}\BibitemShut {NoStop}%
\bibitem [{\citenamefont {Serra}\ \emph {et~al.}(2005)\citenamefont {Serra},
  \citenamefont {Villas-B\^{o}as}, \citenamefont {de~Almeida},\ and\
  \citenamefont {Moussa}}]{villas2005}%
  \BibitemOpen
  \bibfield  {author} {\bibinfo {author} {\bibfnamefont {R.~M.}\ \bibnamefont
  {Serra}}, \bibinfo {author} {\bibfnamefont {C.~J.}\ \bibnamefont
  {Villas-B\^{o}as}}, \bibinfo {author} {\bibfnamefont {N.~G.}\ \bibnamefont
  {de~Almeida}}, \ and\ \bibinfo {author} {\bibfnamefont {M.~H.~Y.}\
  \bibnamefont {Moussa}},\ }\bibfield  {title} {\enquote {\bibinfo {title}
  {Frequency up- and down-conversions in two-mode cavity quantum
  electrodynamics},}\ }\href {\doibase 10.1103/PhysRevA.71.045802} {\bibfield
  {journal} {\bibinfo  {journal} {Phys. Rev. A}\ }\textbf {\bibinfo {volume}
  {71}},\ \bibinfo {pages} {045802} (\bibinfo {year} {2005})}\BibitemShut
  {NoStop}%
\bibitem [{\citenamefont {Duque}\ \emph {et~al.}(2012)\citenamefont {Duque},
  \citenamefont {Mora-Ramos},\ and\ \citenamefont {Duque}}]{duque2012}%
  \BibitemOpen
  \bibfield  {author} {\bibinfo {author} {\bibfnamefont {Carlos~M.}\
  \bibnamefont {Duque}}, \bibinfo {author} {\bibfnamefont {Miguel~E.}\
  \bibnamefont {Mora-Ramos}}, \ and\ \bibinfo {author} {\bibfnamefont
  {Carlos~A.}\ \bibnamefont {Duque}},\ }\bibfield  {title} {\enquote {\bibinfo
  {title} {Quantum disc plus inverse square potential. an analytical model for
  two-dimensional quantum rings: Study of nonlinear optical properties},}\
  }\href {\doibase 10.1002/andp.201200055} {\bibfield  {journal} {\bibinfo
  {journal} {Ann. Phys.}\ }\textbf {\bibinfo {volume} {524}},\ \bibinfo {pages}
  {327--337} (\bibinfo {year} {2012})}\BibitemShut {NoStop}%
\bibitem [{\citenamefont {R\"as\"anen}\ \emph {et~al.}(2007)\citenamefont
  {R\"as\"anen}, \citenamefont {Castro}, \citenamefont {Werschnik},
  \citenamefont {Rubio},\ and\ \citenamefont {Gross}}]{rasanen2007}%
  \BibitemOpen
  \bibfield  {author} {\bibinfo {author} {\bibfnamefont {E.}~\bibnamefont
  {R\"as\"anen}}, \bibinfo {author} {\bibfnamefont {A.}~\bibnamefont {Castro}},
  \bibinfo {author} {\bibfnamefont {J.}~\bibnamefont {Werschnik}}, \bibinfo
  {author} {\bibfnamefont {A.}~\bibnamefont {Rubio}}, \ and\ \bibinfo {author}
  {\bibfnamefont {E.~K.~U.}\ \bibnamefont {Gross}},\ }\bibfield  {title}
  {\enquote {\bibinfo {title} {Optimal control of quantum rings by terahertz
  laser pulses},}\ }\href {\doibase 10.1103/PhysRevLett.98.157404} {\bibfield
  {journal} {\bibinfo  {journal} {Phys. Rev. Lett.}\ }\textbf {\bibinfo
  {volume} {98}},\ \bibinfo {pages} {157404} (\bibinfo {year}
  {2007})}\BibitemShut {NoStop}%
\bibitem [{\citenamefont {Spohn}(2004)}]{spohn2004}%
  \BibitemOpen
  \bibfield  {author} {\bibinfo {author} {\bibfnamefont {Herbert}\ \bibnamefont
  {Spohn}},\ }\href
  {https://www.cambridge.org/core/books/dynamics-of-charged-particles-and-their-radiation-field/92D241C60F65E559EB1B7AFCD9E47F43}
  {\emph {\bibinfo {title} {Dynamics of charged particles and their radiation
  field}}}\ (\bibinfo  {publisher} {Cambridge university press},\ \bibinfo
  {year} {2004})\BibitemShut {NoStop}%
\bibitem [{\citenamefont {Ruggenthaler}\ \emph {et~al.}(2018)\citenamefont
  {Ruggenthaler}, \citenamefont {Tancogne-Dejean}, \citenamefont {Flick},
  \citenamefont {Appel},\ and\ \citenamefont {Rubio}}]{ruggenthaler2017b}%
  \BibitemOpen
  \bibfield  {author} {\bibinfo {author} {\bibfnamefont {Michael}\ \bibnamefont
  {Ruggenthaler}}, \bibinfo {author} {\bibfnamefont {Nicolas}\ \bibnamefont
  {Tancogne-Dejean}}, \bibinfo {author} {\bibfnamefont {Johannes}\ \bibnamefont
  {Flick}}, \bibinfo {author} {\bibfnamefont {Heiko}\ \bibnamefont {Appel}}, \
  and\ \bibinfo {author} {\bibfnamefont {Angel}\ \bibnamefont {Rubio}},\
  }\bibfield  {title} {\enquote {\bibinfo {title} {From a
  quantum-electrodynamical light{\textendash}matter description to novel
  spectroscopies},}\ }\href {\doibase 10.1038/s41570-018-0118} {\bibfield
  {journal} {\bibinfo  {journal} {Nature Reviews Chemistry}\ }\textbf {\bibinfo
  {volume} {2}},\ \bibinfo {pages} {0118} (\bibinfo {year} {2018})}\BibitemShut
  {NoStop}%
\bibitem [{\citenamefont {Jest\"{a}dt}\ \emph {et~al.}(2020)\citenamefont
  {Jest\"{a}dt}, \citenamefont {Ruggenthaler}, \citenamefont {Oliveira},
  \citenamefont {Rubio},\ and\ \citenamefont {Appel}}]{Jestaedt2020}%
  \BibitemOpen
  \bibfield  {author} {\bibinfo {author} {\bibfnamefont {Ren\'{e}}\
  \bibnamefont {Jest\"{a}dt}}, \bibinfo {author} {\bibfnamefont {Michael}\
  \bibnamefont {Ruggenthaler}}, \bibinfo {author} {\bibfnamefont {Micael
  J.~T.}\ \bibnamefont {Oliveira}}, \bibinfo {author} {\bibfnamefont {Angel}\
  \bibnamefont {Rubio}}, \ and\ \bibinfo {author} {\bibfnamefont {Heiko}\
  \bibnamefont {Appel}},\ }\bibfield  {title} {\enquote {\bibinfo {title}
  {Light-matter interactions within the ehrenfest-maxwell-pauli-kohn-sham
  framework: fundamentals, implementation, and nano-optical applications},}\
  }\href {https://doi.org/10.1080/00018732.2019.1695875} {\bibfield  {journal}
  {\bibinfo  {journal} {Advances in Phys.}\ }\textbf {\bibinfo {volume} {68}},\
  \bibinfo {pages} {225--333} (\bibinfo {year} {2020})}\BibitemShut {NoStop}%
\bibitem [{\citenamefont {Flick}\ \emph {et~al.}(2017)\citenamefont {Flick},
  \citenamefont {Ruggenthaler}, \citenamefont {Appel},\ and\ \citenamefont
  {Rubio}}]{flick2017}%
  \BibitemOpen
  \bibfield  {author} {\bibinfo {author} {\bibfnamefont {Johannes}\
  \bibnamefont {Flick}}, \bibinfo {author} {\bibfnamefont {Michael}\
  \bibnamefont {Ruggenthaler}}, \bibinfo {author} {\bibfnamefont {Heiko}\
  \bibnamefont {Appel}}, \ and\ \bibinfo {author} {\bibfnamefont {Angel}\
  \bibnamefont {Rubio}},\ }\bibfield  {title} {\enquote {\bibinfo {title}
  {Atoms and molecules in cavities, from weak to strong coupling in
  quantum-electrodynamics (qed) chemistry},}\ }\href {\doibase
  10.1073/pnas.1615509114} {\bibfield  {journal} {\bibinfo  {journal}
  {Proceedings of the National Academy of Sciences}\ }\textbf {\bibinfo
  {volume} {114}},\ \bibinfo {pages} {3026--3034} (\bibinfo {year}
  {2017})}\BibitemShut {NoStop}%
\bibitem [{\citenamefont {Rokaj}\ \emph {et~al.}(2018)\citenamefont {Rokaj},
  \citenamefont {Welakuh}, \citenamefont {Ruggenthaler},\ and\ \citenamefont
  {Rubio}}]{rokaj2017}%
  \BibitemOpen
  \bibfield  {author} {\bibinfo {author} {\bibfnamefont {Vasil}\ \bibnamefont
  {Rokaj}}, \bibinfo {author} {\bibfnamefont {Davis~M}\ \bibnamefont
  {Welakuh}}, \bibinfo {author} {\bibfnamefont {Michael}\ \bibnamefont
  {Ruggenthaler}}, \ and\ \bibinfo {author} {\bibfnamefont {Angel}\
  \bibnamefont {Rubio}},\ }\bibfield  {title} {\enquote {\bibinfo {title}
  {Light–matter interaction in the long-wavelength limit: no ground-state
  without dipole self-energy},}\ }\href
  {http://stacks.iop.org/0953-4075/51/i=3/a=034005} {\bibfield  {journal}
  {\bibinfo  {journal} {Journal of Physics B: Atomic, Molecular and Optical
  Physics}\ }\textbf {\bibinfo {volume} {51}},\ \bibinfo {pages} {034005}
  (\bibinfo {year} {2018})}\BibitemShut {NoStop}%
\bibitem [{\citenamefont {Sch\"{a}fer}\ \emph {et~al.}(2018)\citenamefont
  {Sch\"{a}fer}, \citenamefont {Ruggenthaler},\ and\ \citenamefont
  {Rubio}}]{schaefer2018}%
  \BibitemOpen
  \bibfield  {author} {\bibinfo {author} {\bibfnamefont {Christian}\
  \bibnamefont {Sch\"{a}fer}}, \bibinfo {author} {\bibfnamefont {Michael}\
  \bibnamefont {Ruggenthaler}}, \ and\ \bibinfo {author} {\bibfnamefont
  {Angel}\ \bibnamefont {Rubio}},\ }\bibfield  {title} {\enquote {\bibinfo
  {title} {Ab initio nonrelativistic quantum electrodynamics: Bridging quantum
  chemistry and quantum optics from weak to strong coupling},}\ }\href
  {\doibase 10.1103/PhysRevA.98.043801} {\bibfield  {journal} {\bibinfo
  {journal} {Phys. Rev. A}\ }\textbf {\bibinfo {volume} {98}},\ \bibinfo
  {pages} {043801} (\bibinfo {year} {2018})}\BibitemShut {NoStop}%
\bibitem [{\citenamefont {Sch\"{a}fer}\ \emph {et~al.}(2020)\citenamefont
  {Sch\"{a}fer}, \citenamefont {Ruggenthaler}, \citenamefont {Rokaj},\ and\
  \citenamefont {Rubio}}]{schaefer2020}%
  \BibitemOpen
  \bibfield  {author} {\bibinfo {author} {\bibfnamefont {Christian}\
  \bibnamefont {Sch\"{a}fer}}, \bibinfo {author} {\bibfnamefont {Michael}\
  \bibnamefont {Ruggenthaler}}, \bibinfo {author} {\bibfnamefont {Vasil}\
  \bibnamefont {Rokaj}}, \ and\ \bibinfo {author} {\bibfnamefont {Angel}\
  \bibnamefont {Rubio}},\ }\bibfield  {title} {\enquote {\bibinfo {title}
  {Relevance of the quadratic diamagnetic and self-polarization terms in cavity
  quantum electrodynamics},}\ }\href {\doibase 10.1021/acsphotonics.9b01649}
  {\bibfield  {journal} {\bibinfo  {journal} {ACS Photonics}\ }\textbf
  {\bibinfo {volume} {7}},\ \bibinfo {pages} {975--990} (\bibinfo {year}
  {2020})}\BibitemShut {NoStop}%
\bibitem [{\citenamefont {Flick}\ \emph {et~al.}(2019)\citenamefont {Flick},
  \citenamefont {Welakuh}, \citenamefont {Ruggenthaler}, \citenamefont
  {Appel},\ and\ \citenamefont {Rubio}}]{flick2019}%
  \BibitemOpen
  \bibfield  {author} {\bibinfo {author} {\bibfnamefont {Johannes}\
  \bibnamefont {Flick}}, \bibinfo {author} {\bibfnamefont {Davis~M.}\
  \bibnamefont {Welakuh}}, \bibinfo {author} {\bibfnamefont {Michael}\
  \bibnamefont {Ruggenthaler}}, \bibinfo {author} {\bibfnamefont {Heiko}\
  \bibnamefont {Appel}}, \ and\ \bibinfo {author} {\bibfnamefont {Angel}\
  \bibnamefont {Rubio}},\ }\bibfield  {title} {\enquote {\bibinfo {title}
  {Light-matter response in nonrelativistic quantum electrodynamics},}\ }\href
  {\doibase 10.1021/acsphotonics.9b00768} {\bibfield  {journal} {\bibinfo
  {journal} {ACS Photonics}\ }\textbf {\bibinfo {volume} {6}},\ \bibinfo
  {pages} {2757--2778} (\bibinfo {year} {2019})}\BibitemShut {NoStop}%
\bibitem [{\citenamefont {Rokaj}\ \emph {et~al.}(2020)\citenamefont {Rokaj},
  \citenamefont {Ruggenthaler}, \citenamefont {Eich},\ and\ \citenamefont
  {Rubio}}]{rokaj2020}%
  \BibitemOpen
  \bibfield  {author} {\bibinfo {author} {\bibfnamefont {Vasil}\ \bibnamefont
  {Rokaj}}, \bibinfo {author} {\bibfnamefont {Michael}\ \bibnamefont
  {Ruggenthaler}}, \bibinfo {author} {\bibfnamefont {Florian~G.}\ \bibnamefont
  {Eich}}, \ and\ \bibinfo {author} {\bibfnamefont {Angel}\ \bibnamefont
  {Rubio}},\ }\bibfield  {title} {\enquote {\bibinfo {title} {The free electron
  gas in cavity quantum electrodynamics},}\ }\href@noop {} {\bibfield
  {journal} {\bibinfo  {journal} {arXiv preprint arXiv:2006.09236}\ } (\bibinfo
  {year} {2020})}\BibitemShut {NoStop}%
\bibitem [{\citenamefont {Greiner}\ and\ \citenamefont
  {Reinhardt}(1996)}]{greiner1996}%
  \BibitemOpen
  \bibfield  {author} {\bibinfo {author} {\bibfnamefont {Walter}\ \bibnamefont
  {Greiner}}\ and\ \bibinfo {author} {\bibfnamefont {Joachim}\ \bibnamefont
  {Reinhardt}},\ }\href@noop {} {\emph {\bibinfo {title} {Field
  quantization}}}\ (\bibinfo  {publisher} {Springer},\ \bibinfo {year}
  {1996})\BibitemShut {NoStop}%
\bibitem [{\citenamefont {Flick}\ \emph {et~al.}(2015)\citenamefont {Flick},
  \citenamefont {Ruggenthaler}, \citenamefont {Appel},\ and\ \citenamefont
  {Rubio}}]{flick2015}%
  \BibitemOpen
  \bibfield  {author} {\bibinfo {author} {\bibfnamefont {Johannes}\
  \bibnamefont {Flick}}, \bibinfo {author} {\bibfnamefont {Michael}\
  \bibnamefont {Ruggenthaler}}, \bibinfo {author} {\bibfnamefont {Heiko}\
  \bibnamefont {Appel}}, \ and\ \bibinfo {author} {\bibfnamefont {Angel}\
  \bibnamefont {Rubio}},\ }\bibfield  {title} {\enquote {\bibinfo {title}
  {Kohn-sham approach to quantum electrodynamical density-functional theory:
  Exact time-dependent effective potentials in real space},}\ }\href {\doibase
  10.1073/pnas.1518224112} {\bibfield  {journal} {\bibinfo  {journal} {Proc.
  Natl. Acad. Sci. U. S. A.}\ }\textbf {\bibinfo {volume} {112}},\ \bibinfo
  {pages} {15285--15290} (\bibinfo {year} {2015})}\BibitemShut {NoStop}%
\bibitem [{\citenamefont {Ruggenthaler}\ \emph {et~al.}(2014)\citenamefont
  {Ruggenthaler}, \citenamefont {Flick}, \citenamefont {Pellegrini},
  \citenamefont {Appel}, \citenamefont {Tokatly},\ and\ \citenamefont
  {Rubio}}]{ruggenthaler2014}%
  \BibitemOpen
  \bibfield  {author} {\bibinfo {author} {\bibfnamefont {Michael}\ \bibnamefont
  {Ruggenthaler}}, \bibinfo {author} {\bibfnamefont {Johannes}\ \bibnamefont
  {Flick}}, \bibinfo {author} {\bibfnamefont {Camilla}\ \bibnamefont
  {Pellegrini}}, \bibinfo {author} {\bibfnamefont {Heiko}\ \bibnamefont
  {Appel}}, \bibinfo {author} {\bibfnamefont {Ilya~V.}\ \bibnamefont
  {Tokatly}}, \ and\ \bibinfo {author} {\bibfnamefont {Angel}\ \bibnamefont
  {Rubio}},\ }\bibfield  {title} {\enquote {\bibinfo {title}
  {Quantum-electrodynamical density-functional theory: Bridging quantum optics
  and electronic-structure theory},}\ }\href {\doibase
  10.1103/PhysRevA.90.012508} {\bibfield  {journal} {\bibinfo  {journal} {Phys.
  Rev. A}\ }\textbf {\bibinfo {volume} {90}},\ \bibinfo {pages} {012508}
  (\bibinfo {year} {2014})}\BibitemShut {NoStop}%
\bibitem [{\citenamefont {Pellegrini}\ \emph {et~al.}(2015)\citenamefont
  {Pellegrini}, \citenamefont {Flick}, \citenamefont {Tokatly}, \citenamefont
  {Appel},\ and\ \citenamefont {Rubio}}]{pellegrini2015}%
  \BibitemOpen
  \bibfield  {author} {\bibinfo {author} {\bibfnamefont {Camilla}\ \bibnamefont
  {Pellegrini}}, \bibinfo {author} {\bibfnamefont {Johannes}\ \bibnamefont
  {Flick}}, \bibinfo {author} {\bibfnamefont {Ilya~V.}\ \bibnamefont
  {Tokatly}}, \bibinfo {author} {\bibfnamefont {Heiko}\ \bibnamefont {Appel}},
  \ and\ \bibinfo {author} {\bibfnamefont {Angel}\ \bibnamefont {Rubio}},\
  }\bibfield  {title} {\enquote {\bibinfo {title} {Optimized effective
  potential for quantum electrodynamical time-dependent density functional
  theory},}\ }\href {\doibase 10.1103/PhysRevLett.115.093001} {\bibfield
  {journal} {\bibinfo  {journal} {Phys. Rev. Lett.}\ }\textbf {\bibinfo
  {volume} {115}},\ \bibinfo {pages} {093001} (\bibinfo {year}
  {2015})}\BibitemShut {NoStop}%
\bibitem [{\citenamefont {Flick}\ \emph {et~al.}(2018)\citenamefont {Flick},
  \citenamefont {Sch\"{a}fer}, \citenamefont {Ruggenthaler}, \citenamefont
  {Appel},\ and\ \citenamefont {Rubio}}]{flick2017c}%
  \BibitemOpen
  \bibfield  {author} {\bibinfo {author} {\bibfnamefont {Johannes}\
  \bibnamefont {Flick}}, \bibinfo {author} {\bibfnamefont {Christian}\
  \bibnamefont {Sch\"{a}fer}}, \bibinfo {author} {\bibfnamefont {Michael}\
  \bibnamefont {Ruggenthaler}}, \bibinfo {author} {\bibfnamefont {Heiko}\
  \bibnamefont {Appel}}, \ and\ \bibinfo {author} {\bibfnamefont {Angel}\
  \bibnamefont {Rubio}},\ }\bibfield  {title} {\enquote {\bibinfo {title} {Ab
  initio optimized effective potentials for real molecules in optical cavities:
  Photon contributions to the molecular ground state},}\ }\href {\doibase
  10.1021/acsphotonics.7b01279} {\bibfield  {journal} {\bibinfo  {journal} {ACS
  Photonics}\ }\textbf {\bibinfo {volume} {5}},\ \bibinfo {pages} {992--1005}
  (\bibinfo {year} {2018})}\BibitemShut {NoStop}%
\bibitem [{\citenamefont {Mordovina}\ \emph {et~al.}(2020)\citenamefont
  {Mordovina}, \citenamefont {Bungey}, \citenamefont {Appel}, \citenamefont
  {Knowles}, \citenamefont {Rubio},\ and\ \citenamefont
  {Manby}}]{mordovina2020polaritonic}%
  \BibitemOpen
  \bibfield  {author} {\bibinfo {author} {\bibfnamefont {Uliana}\ \bibnamefont
  {Mordovina}}, \bibinfo {author} {\bibfnamefont {Callum}\ \bibnamefont
  {Bungey}}, \bibinfo {author} {\bibfnamefont {Heiko}\ \bibnamefont {Appel}},
  \bibinfo {author} {\bibfnamefont {Peter~J}\ \bibnamefont {Knowles}}, \bibinfo
  {author} {\bibfnamefont {Angel}\ \bibnamefont {Rubio}}, \ and\ \bibinfo
  {author} {\bibfnamefont {Frederick~R}\ \bibnamefont {Manby}},\ }\bibfield
  {title} {\enquote {\bibinfo {title} {Polaritonic coupled-cluster theory},}\
  }\href {\doibase 10.1103/PhysRevResearch.2.023262} {\bibfield  {journal}
  {\bibinfo  {journal} {Physical Review Research}\ }\textbf {\bibinfo {volume}
  {2}},\ \bibinfo {pages} {023262} (\bibinfo {year} {2020})}\BibitemShut
  {NoStop}%
\bibitem [{\citenamefont {Haugland}\ \emph {et~al.}(2020)\citenamefont
  {Haugland}, \citenamefont {Ronca}, \citenamefont {Kj{\o}nstad}, \citenamefont
  {Rubio},\ and\ \citenamefont {Koch}}]{haugland2020coupled}%
  \BibitemOpen
  \bibfield  {author} {\bibinfo {author} {\bibfnamefont {Tor~S}\ \bibnamefont
  {Haugland}}, \bibinfo {author} {\bibfnamefont {Enrico}\ \bibnamefont
  {Ronca}}, \bibinfo {author} {\bibfnamefont {Eirik~F}\ \bibnamefont
  {Kj{\o}nstad}}, \bibinfo {author} {\bibfnamefont {Angel}\ \bibnamefont
  {Rubio}}, \ and\ \bibinfo {author} {\bibfnamefont {Henrik}\ \bibnamefont
  {Koch}},\ }\bibfield  {title} {\enquote {\bibinfo {title} {Coupled cluster
  theory for molecular polaritons: Changing ground and excited states},}\
  }\href@noop {} {\bibfield  {journal} {\bibinfo  {journal} {arXiv preprint
  arXiv:2005.04477}\ } (\bibinfo {year} {2020})}\BibitemShut {NoStop}%
\bibitem [{\citenamefont {Hartmann}\ \emph {et~al.}(2019)\citenamefont
  {Hartmann}, \citenamefont {Otten}, \citenamefont {Fedin}, \citenamefont
  {Talapin}, \citenamefont {Cygorek}, \citenamefont {Hawrylak}, \citenamefont
  {Korkusinski}, \citenamefont {Gray}, \citenamefont {Hartschuh},\ and\
  \citenamefont {Ma}}]{hartmann2019}%
  \BibitemOpen
  \bibfield  {author} {\bibinfo {author} {\bibfnamefont {Nicolai~F.}\
  \bibnamefont {Hartmann}}, \bibinfo {author} {\bibfnamefont {Matthew}\
  \bibnamefont {Otten}}, \bibinfo {author} {\bibfnamefont {Igor}\ \bibnamefont
  {Fedin}}, \bibinfo {author} {\bibfnamefont {Dmitri}\ \bibnamefont {Talapin}},
  \bibinfo {author} {\bibfnamefont {Moritz}\ \bibnamefont {Cygorek}}, \bibinfo
  {author} {\bibfnamefont {Pawel}\ \bibnamefont {Hawrylak}}, \bibinfo {author}
  {\bibfnamefont {Marek}\ \bibnamefont {Korkusinski}}, \bibinfo {author}
  {\bibfnamefont {Stephen}\ \bibnamefont {Gray}}, \bibinfo {author}
  {\bibfnamefont {Achim}\ \bibnamefont {Hartschuh}}, \ and\ \bibinfo {author}
  {\bibfnamefont {Xuedan}\ \bibnamefont {Ma}},\ }\bibfield  {title} {\enquote
  {\bibinfo {title} {Uniaxial transition dipole moments in semiconductor
  quantum rings caused by broken rotational symmetry},}\ }\href {\doibase
  10.1038/s41467-019-11225-6} {\bibfield  {journal} {\bibinfo  {journal}
  {Nature Communications}\ }\textbf {\bibinfo {volume} {10}},\ \bibinfo {pages}
  {3253} (\bibinfo {year} {2019})}\BibitemShut {NoStop}%
\bibitem [{\citenamefont {Vinasco}\ \emph {et~al.}(2018)\citenamefont
  {Vinasco}, \citenamefont {Radu}, \citenamefont {Kasapoglu}, \citenamefont
  {Restrepo}, \citenamefont {Morales}, \citenamefont {Feddi}, \citenamefont
  {Mora-Ramos},\ and\ \citenamefont {Duque}}]{vinasco2018}%
  \BibitemOpen
  \bibfield  {author} {\bibinfo {author} {\bibfnamefont {J.~A.}\ \bibnamefont
  {Vinasco}}, \bibinfo {author} {\bibfnamefont {A.}~\bibnamefont {Radu}},
  \bibinfo {author} {\bibfnamefont {E.}~\bibnamefont {Kasapoglu}}, \bibinfo
  {author} {\bibfnamefont {R.~L.}\ \bibnamefont {Restrepo}}, \bibinfo {author}
  {\bibfnamefont {A.~L.}\ \bibnamefont {Morales}}, \bibinfo {author}
  {\bibfnamefont {E.}~\bibnamefont {Feddi}}, \bibinfo {author} {\bibfnamefont
  {M.~E.}\ \bibnamefont {Mora-Ramos}}, \ and\ \bibinfo {author} {\bibfnamefont
  {C.~A.}\ \bibnamefont {Duque}},\ }\bibfield  {title} {\enquote {\bibinfo
  {title} {Effects of geometry on the electronic properties of semiconductor
  elliptical quantum rings},}\ }\href {\doibase 10.1038/s41598-018-31512-4}
  {\bibfield  {journal} {\bibinfo  {journal} {Scientific Reports}\ }\textbf
  {\bibinfo {volume} {8}},\ \bibinfo {pages} {13299} (\bibinfo {year}
  {2018})}\BibitemShut {NoStop}%
\bibitem [{\citenamefont {Hochbruck}\ and\ \citenamefont
  {Lubich}(1997)}]{hochbruck1997}%
  \BibitemOpen
  \bibfield  {author} {\bibinfo {author} {\bibfnamefont {Marlis}\ \bibnamefont
  {Hochbruck}}\ and\ \bibinfo {author} {\bibfnamefont {Christian}\ \bibnamefont
  {Lubich}},\ }\bibfield  {title} {\enquote {\bibinfo {title} {On krylov
  subspace approximations to the matrix exponential operator},}\ }\href
  {\doibase 10.1137/S0036142995280572} {\bibfield  {journal} {\bibinfo
  {journal} {SIAM J. Numer. Anal.}\ }\textbf {\bibinfo {volume} {34}},\
  \bibinfo {pages} {1911--1925.} (\bibinfo {year} {1997})}\BibitemShut
  {NoStop}%
\bibitem [{\citenamefont {Mandel}(1979)}]{mandel1979}%
  \BibitemOpen
  \bibfield  {author} {\bibinfo {author} {\bibfnamefont {L.}~\bibnamefont
  {Mandel}},\ }\bibfield  {title} {\enquote {\bibinfo {title} {Sub-poissonian
  photon statistics in resonance fluorescence},}\ }\href {\doibase
  10.1364/OL.4.000205} {\bibfield  {journal} {\bibinfo  {journal} {Opt. Lett.}\
  }\textbf {\bibinfo {volume} {4}},\ \bibinfo {pages} {205--207} (\bibinfo
  {year} {1979})}\BibitemShut {NoStop}%
\bibitem [{\citenamefont {Loudon}(2000)}]{loudon2000}%
  \BibitemOpen
  \bibfield  {author} {\bibinfo {author} {\bibfnamefont {Rodney}\ \bibnamefont
  {Loudon}},\ }\href@noop {} {\emph {\bibinfo {title} {The Quantum Theory of
  Light}}}\ (\bibinfo  {publisher} {Oxford University Press},\ \bibinfo {year}
  {2000})\BibitemShut {NoStop}%
\bibitem [{\citenamefont {Gerry}\ and\ \citenamefont
  {Knight}(2005)}]{gerry2005}%
  \BibitemOpen
  \bibfield  {author} {\bibinfo {author} {\bibfnamefont {C.~C.}\ \bibnamefont
  {Gerry}}\ and\ \bibinfo {author} {\bibfnamefont {P.~L.}\ \bibnamefont
  {Knight}},\ }\href@noop {} {\emph {\bibinfo {title} {Introductory Quantum
  Optics}}}\ (\bibinfo  {publisher} {Cambridge University Press, Cambridge},\
  \bibinfo {year} {2005})\BibitemShut {NoStop}%
\bibitem [{\citenamefont {Kalaga}\ \emph {et~al.}(2016)\citenamefont {Kalaga},
  \citenamefont {Kowalewska-Kudłaszyk}, \citenamefont {Leo\'{n}ski},\ and\
  \citenamefont {Barasi\'{n}ski}}]{kalaga2016}%
  \BibitemOpen
  \bibfield  {author} {\bibinfo {author} {\bibfnamefont {J.~K.}\ \bibnamefont
  {Kalaga}}, \bibinfo {author} {\bibfnamefont {A.}~\bibnamefont
  {Kowalewska-Kudłaszyk}}, \bibinfo {author} {\bibfnamefont {W.}~\bibnamefont
  {Leo\'{n}ski}}, \ and\ \bibinfo {author} {\bibfnamefont {A.}~\bibnamefont
  {Barasi\'{n}ski}},\ }\bibfield  {title} {\enquote {\bibinfo {title} {Quantum
  correlations and entanglement in a model comprised of a short chain of
  nonlinear oscillators},}\ }\href {\doibase 10.1103/PhysRevA.94.032304}
  {\bibfield  {journal} {\bibinfo  {journal} {Phys. Rev. A}\ }\textbf {\bibinfo
  {volume} {94}},\ \bibinfo {pages} {032304} (\bibinfo {year}
  {2016})}\BibitemShut {NoStop}%
\bibitem [{\citenamefont {Bocchieri}\ and\ \citenamefont
  {Loinger}(1957)}]{PhysRev.107.337}%
  \BibitemOpen
  \bibfield  {author} {\bibinfo {author} {\bibfnamefont {P.}~\bibnamefont
  {Bocchieri}}\ and\ \bibinfo {author} {\bibfnamefont {A.}~\bibnamefont
  {Loinger}},\ }\bibfield  {title} {\enquote {\bibinfo {title} {Quantum
  recurrence theorem},}\ }\href {\doibase 10.1103/PhysRev.107.337} {\bibfield
  {journal} {\bibinfo  {journal} {Phys. Rev.}\ }\textbf {\bibinfo {volume}
  {107}},\ \bibinfo {pages} {337--338} (\bibinfo {year} {1957})}\BibitemShut
  {NoStop}%
\bibitem [{\citenamefont {Breuer}\ and\ \citenamefont
  {Petruccione}(2007)}]{breuer2007}%
  \BibitemOpen
  \bibfield  {author} {\bibinfo {author} {\bibfnamefont {Heinz-Peter}\
  \bibnamefont {Breuer}}\ and\ \bibinfo {author} {\bibfnamefont {Francesco}\
  \bibnamefont {Petruccione}},\ }\href {\doibase
  10.1093/acprof:oso/9780199213900.001.0001} {\emph {\bibinfo {title} {The
  Theory of Open Quantum Systems}}}\ (\bibinfo  {publisher} {Oxford University
  Press},\ \bibinfo {year} {2007})\BibitemShut {NoStop}%
\bibitem [{fig()}]{figcomment}%
  \BibitemOpen
  \href@noop {} {\emph {\bibinfo {title} {At $t=0$ ps, the initial states are
  factorizable and the numerator of Eq.~(\ref{g-2-coherence}) is separable as a
  product of $\langle\hat{n}_{\alpha}\rangle\langle\hat{n}_{\beta}\rangle$.
  Analytically, $g_{\alpha\beta}^{(2)}=1$, however, numerically this is
  infinite since
  $\langle\hat{n}_{\alpha}\rangle=\langle\hat{n}_{\beta}\rangle=0$ at $t=0$ and
  $g_{\alpha\beta}^{(2)}(t=0)$ is not a number. Therefore, we make a cut for
  the time interval between $t=0$ and $t=0.58$ ps to take care of the
  divergence.}}}\BibitemShut {Stop}%
\bibitem [{\citenamefont {Muthukrishnan}\ \emph {et~al.}(2004)\citenamefont
  {Muthukrishnan}, \citenamefont {Agarwal},\ and\ \citenamefont
  {Scully}}]{muthukrishnan2004}%
  \BibitemOpen
  \bibfield  {author} {\bibinfo {author} {\bibfnamefont {Ashok}\ \bibnamefont
  {Muthukrishnan}}, \bibinfo {author} {\bibfnamefont {Girish~S.}\ \bibnamefont
  {Agarwal}}, \ and\ \bibinfo {author} {\bibfnamefont {Marlan~O.}\ \bibnamefont
  {Scully}},\ }\bibfield  {title} {\enquote {\bibinfo {title} {Inducing
  disallowed two-atom transitions with temporally entangled photons},}\ }\href
  {\doibase 10.1103/PhysRevLett.93.093002} {\bibfield  {journal} {\bibinfo
  {journal} {Phys. Rev. Lett.}\ }\textbf {\bibinfo {volume} {93}},\ \bibinfo
  {pages} {093002} (\bibinfo {year} {2004})}\BibitemShut {NoStop}%
\bibitem [{\citenamefont {Villar}\ \emph {et~al.}(2006)\citenamefont {Villar},
  \citenamefont {Martinelli}, \citenamefont {Fabre},\ and\ \citenamefont
  {Nussenzveig}}]{villar2006}%
  \BibitemOpen
  \bibfield  {author} {\bibinfo {author} {\bibfnamefont {A.~S.}\ \bibnamefont
  {Villar}}, \bibinfo {author} {\bibfnamefont {M.}~\bibnamefont {Martinelli}},
  \bibinfo {author} {\bibfnamefont {C.}~\bibnamefont {Fabre}}, \ and\ \bibinfo
  {author} {\bibfnamefont {P.}~\bibnamefont {Nussenzveig}},\ }\bibfield
  {title} {\enquote {\bibinfo {title} {Direct production of tripartite
  pump-signal-idler entanglement in the above-threshold optical parametric
  oscillator},}\ }\href {\doibase 10.1103/PhysRevLett.97.140504} {\bibfield
  {journal} {\bibinfo  {journal} {Phys. Rev. Lett.}\ }\textbf {\bibinfo
  {volume} {97}},\ \bibinfo {pages} {140504} (\bibinfo {year}
  {2006})}\BibitemShut {NoStop}%
\bibitem [{\citenamefont {xi~Liu}\ \emph {et~al.}(2005)\citenamefont {xi~Liu},
  \citenamefont {You}, \citenamefont {Wei}, \citenamefont {Sun},\ and\
  \citenamefont {Nori}}]{liu2005}%
  \BibitemOpen
  \bibfield  {author} {\bibinfo {author} {\bibfnamefont {Yu}~\bibnamefont
  {xi~Liu}}, \bibinfo {author} {\bibfnamefont {J.~Q.}\ \bibnamefont {You}},
  \bibinfo {author} {\bibfnamefont {L.~F.}\ \bibnamefont {Wei}}, \bibinfo
  {author} {\bibfnamefont {C.~P.}\ \bibnamefont {Sun}}, \ and\ \bibinfo
  {author} {\bibfnamefont {Franco}\ \bibnamefont {Nori}},\ }\bibfield  {title}
  {\enquote {\bibinfo {title} {Optical selection rules and phase-dependent
  adiabatic state control in a superconducting quantum circuit},}\ }\href
  {\doibase 10.1103/PhysRevLett.95.087001} {\bibfield  {journal} {\bibinfo
  {journal} {Phys. Rev. Lett.}\ }\textbf {\bibinfo {volume} {95}},\ \bibinfo
  {pages} {087001} (\bibinfo {year} {2005})}\BibitemShut {NoStop}%
\bibitem [{\citenamefont {xi~Liu}\ \emph {et~al.}(2014)\citenamefont {xi~Liu},
  \citenamefont {Sun}, \citenamefont {Peng}, \citenamefont {Miranowicz},
  \citenamefont {Tsai},\ and\ \citenamefont {Nori}}]{liu2014}%
  \BibitemOpen
  \bibfield  {author} {\bibinfo {author} {\bibfnamefont {Yu}~\bibnamefont
  {xi~Liu}}, \bibinfo {author} {\bibfnamefont {Hui-Chen}\ \bibnamefont {Sun}},
  \bibinfo {author} {\bibfnamefont {Z.~H.}\ \bibnamefont {Peng}}, \bibinfo
  {author} {\bibfnamefont {Adam}\ \bibnamefont {Miranowicz}}, \bibinfo {author}
  {\bibfnamefont {J.~S.}\ \bibnamefont {Tsai}}, \ and\ \bibinfo {author}
  {\bibfnamefont {Franco}\ \bibnamefont {Nori}},\ }\bibfield  {title} {\enquote
  {\bibinfo {title} {Controllable microwave three-wave mixing via a single
  three-level superconducting quantum circuit},}\ }\href {\doibase
  10.1038/srep07289} {\bibfield  {journal} {\bibinfo  {journal} {Scientific
  Reports}\ }\textbf {\bibinfo {volume} {4}},\ \bibinfo {pages} {7289}
  (\bibinfo {year} {2014})}\BibitemShut {NoStop}%
\bibitem [{\citenamefont {Stefano}\ \emph {et~al.}(2019)\citenamefont
  {Stefano}, \citenamefont {Settineri}, \citenamefont {Macr\`{i}},
  \citenamefont {Garziano}, \citenamefont {Stassi}, \citenamefont {Savasta},\
  and\ \citenamefont {Nori}}]{stefano2019}%
  \BibitemOpen
  \bibfield  {author} {\bibinfo {author} {\bibfnamefont {Omar~Di}\ \bibnamefont
  {Stefano}}, \bibinfo {author} {\bibfnamefont {Alessio}\ \bibnamefont
  {Settineri}}, \bibinfo {author} {\bibfnamefont {Vincenzo}\ \bibnamefont
  {Macr\`{i}}}, \bibinfo {author} {\bibfnamefont {Luigi}\ \bibnamefont
  {Garziano}}, \bibinfo {author} {\bibfnamefont {Roberto}\ \bibnamefont
  {Stassi}}, \bibinfo {author} {\bibfnamefont {Salvatore}\ \bibnamefont
  {Savasta}}, \ and\ \bibinfo {author} {\bibfnamefont {Franco}\ \bibnamefont
  {Nori}},\ }\bibfield  {title} {\enquote {\bibinfo {title} {Resolution of
  gauge ambiguities in ultrastrong-coupling cavity quantum electrodynamics},}\
  }\href {\doibase 10.1038/s41567-019-0534-4} {\bibfield  {journal} {\bibinfo
  {journal} {Nature Physics}\ }\textbf {\bibinfo {volume} {15}},\ \bibinfo
  {pages} {803--808} (\bibinfo {year} {2019})}\BibitemShut {NoStop}%
\bibitem [{\citenamefont {{Ruggenthaler}}(2015)}]{ruggenthaler2015}%
  \BibitemOpen
  \bibfield  {author} {\bibinfo {author} {\bibfnamefont {Michael}\ \bibnamefont
  {{Ruggenthaler}}},\ }\bibfield  {title} {\enquote {\bibinfo {title}
  {{Ground-State Quantum-Electrodynamical Density-Functional Theory}},}\
  }\href@noop {} {\bibfield  {journal} {\bibinfo  {journal} {ArXiv e-prints}\ }
  (\bibinfo {year} {2015})},\ \Eprint {http://arxiv.org/abs/1509.01417}
  {arXiv:1509.01417 [quant-ph]} \BibitemShut {NoStop}%
\bibitem [{\citenamefont {Fuhrer}\ \emph {et~al.}(2001)\citenamefont {Fuhrer},
  \citenamefont {L\"{u}scher}, \citenamefont {Ihn}, \citenamefont {Heinzel},
  \citenamefont {Ensslin}, \citenamefont {Wegscheider},\ and\ \citenamefont
  {Bichler}}]{fuhrer2001}%
  \BibitemOpen
  \bibfield  {author} {\bibinfo {author} {\bibfnamefont {A.}~\bibnamefont
  {Fuhrer}}, \bibinfo {author} {\bibfnamefont {S.}~\bibnamefont {L\"{u}scher}},
  \bibinfo {author} {\bibfnamefont {T.}~\bibnamefont {Ihn}}, \bibinfo {author}
  {\bibfnamefont {T.}~\bibnamefont {Heinzel}}, \bibinfo {author} {\bibfnamefont
  {K.}~\bibnamefont {Ensslin}}, \bibinfo {author} {\bibfnamefont
  {W.}~\bibnamefont {Wegscheider}}, \ and\ \bibinfo {author} {\bibfnamefont
  {M.}~\bibnamefont {Bichler}},\ }\bibfield  {title} {\enquote {\bibinfo
  {title} {Energy spectra of quantum rings},}\ }\href {\doibase
  10.1038/35101552} {\bibfield  {journal} {\bibinfo  {journal} {Nature}\
  }\textbf {\bibinfo {volume} {413}},\ \bibinfo {pages} {822--825} (\bibinfo
  {year} {2001})}\BibitemShut {NoStop}%
\bibitem [{\citenamefont {Ihn}\ \emph {et~al.}(2005)\citenamefont {Ihn},
  \citenamefont {Fuhrer}, \citenamefont {Meier}, \citenamefont {Sigrist},\ and\
  \citenamefont {Ensslin}}]{ihn2005}%
  \BibitemOpen
  \bibfield  {author} {\bibinfo {author} {\bibfnamefont {Thomas}\ \bibnamefont
  {Ihn}}, \bibinfo {author} {\bibfnamefont {Andreas}\ \bibnamefont {Fuhrer}},
  \bibinfo {author} {\bibfnamefont {Lorenz}\ \bibnamefont {Meier}}, \bibinfo
  {author} {\bibfnamefont {Martin}\ \bibnamefont {Sigrist}}, \ and\ \bibinfo
  {author} {\bibfnamefont {Klaus}\ \bibnamefont {Ensslin}},\ }\bibfield
  {title} {\enquote {\bibinfo {title} {Quantum physics in quantum rings},}\
  }\href {\doibase 10.1051/epn:2005302} {\bibfield  {journal} {\bibinfo
  {journal} {Europhysics News}\ }\textbf {\bibinfo {volume} {36}},\ \bibinfo
  {pages} {78--71} (\bibinfo {year} {2005})}\BibitemShut {NoStop}%
\bibitem [{\citenamefont {Sch\"{a}fer}\ \emph {et~al.}(2019)\citenamefont
  {Sch\"{a}fer}, \citenamefont {Ruggenthaler}, \citenamefont {Appel},\ and\
  \citenamefont {Rubio}}]{schaefer2019}%
  \BibitemOpen
  \bibfield  {author} {\bibinfo {author} {\bibfnamefont {Christian}\
  \bibnamefont {Sch\"{a}fer}}, \bibinfo {author} {\bibfnamefont {Michael}\
  \bibnamefont {Ruggenthaler}}, \bibinfo {author} {\bibfnamefont {Heiko}\
  \bibnamefont {Appel}}, \ and\ \bibinfo {author} {\bibfnamefont {Angel}\
  \bibnamefont {Rubio}},\ }\bibfield  {title} {\enquote {\bibinfo {title}
  {Modification of excitation and charge transfer in cavity
  quantum-electrodynamical chemistry},}\ }\href {\doibase
  10.1073/pnas.1814178116} {\bibfield  {journal} {\bibinfo  {journal} {PNAS}\
  }\textbf {\bibinfo {volume} {116}},\ \bibinfo {pages} {4883--4892} (\bibinfo
  {year} {2019})}\BibitemShut {NoStop}%
\bibitem [{\citenamefont {Buchholz}\ \emph {et~al.}(2019)\citenamefont
  {Buchholz}, \citenamefont {Theophilou}, \citenamefont {Nielsen},
  \citenamefont {Ruggenthaler},\ and\ \citenamefont {Rubio}}]{buchholz2019}%
  \BibitemOpen
  \bibfield  {author} {\bibinfo {author} {\bibfnamefont {Florian}\ \bibnamefont
  {Buchholz}}, \bibinfo {author} {\bibfnamefont {Iris}\ \bibnamefont
  {Theophilou}}, \bibinfo {author} {\bibfnamefont {Soeren E.~B.}\ \bibnamefont
  {Nielsen}}, \bibinfo {author} {\bibfnamefont {Michael}\ \bibnamefont
  {Ruggenthaler}}, \ and\ \bibinfo {author} {\bibfnamefont {Angel}\
  \bibnamefont {Rubio}},\ }\bibfield  {title} {\enquote {\bibinfo {title}
  {Reduced density-matrix approach to strong matter-photon interaction},}\
  }\href {\doibase 10.1021/acsphotonics.9b00648} {\bibfield  {journal}
  {\bibinfo  {journal} {ACS Photonics}\ }\textbf {\bibinfo {volume} {6}},\
  \bibinfo {pages} {2694--2711} (\bibinfo {year} {2019})}\BibitemShut {NoStop}%
\end{thebibliography}%
\newpage

\end{document}